%% file: main.tex
\documentclass[aps,pre,twocolumn,floats,floatfix,eqsecnum,longbibliography,nobibnotes,nofootinbib,superscriptaddress]{revtex4-1}
\usepackage{soul}
\usepackage{hyperref}
\hypersetup{colorlinks = true,linkcolor = blue, anchorcolor = blue, citecolor = blue, pdfstartpage=1, filecolor = red,urlcolor = black}
\usepackage{graphicx}
\usepackage{amsmath}
\usepackage{amssymb}
\usepackage{amsfonts}
\usepackage{sidecap}
\usepackage[pagewise, displaymath, mathlines,  columnwise]{lineno}
\linenumbers
\usepackage{times}
\usepackage{epigraph}
\usepackage[dvipsnames]{xcolor}
\usepackage{xspace}
\def\nin{\noindent}
\def\non{\nonumber}

\newcommand{\units}[1]{{\mathrm{ #1}}}
\def\be{\begin{equation}}
\def\ee{\end{equation}}
\def\bea{\begin{eqnarray}}
\def\eea{\end{eqnarray}}
\def\r{{\bf r}}
\def\k{{\bf k}}
\def\q{{\bf q}}

\def\g{{\bf \hat g}}
\def\n{{\bf \hat n}}
\def\G{{\cal G}}
\def\D{{\cal D}}
\def\P{{\cal P}}

\def\K{{\cal K}}
\def\O{{\cal O}}
\def\v{{\bf v}}
\def\lp{\left(}
\def\rp{\right)}
\def\lb{\left[}
\def\rb{\right]}
\def\la{\left<}
\def\ra{\right>}

\newcommand{\ts}[1]{\textstyle{#1}}
\def\etal{{\it et al.} }
\def\Gbare{G^{(0)}}

\def\Dinf{D_{\infty}}
\def\Dinst{D_{\rm inst}}

\def\w{\omega}
\def\d{\mbox{d}}
\def\sgn{\,{\rm sgn}\,}
\def\erf{\,\mathrm{erf}\,}
\def\Re{\mbox{\,Re\,}}
\def\Im{\mbox{\,Im\,}}

\def\const{\xspace{\rm const}\xspace}

\newcommand{\eq}[1]{(\ref{#1})}

\newcommand{\Da}{{D_{a}}}
\newcommand{\Depar}{{D_{e}^{\parallel}}}
\newcommand{\Deperp}{{D_{e}^{\perp}}}

\newcommand{\Dbar}{{\bar D}}

\newcommand{\lmax}{l_{\rm max}}

\usepackage[fulladjust]{marginnote}
\newcommand{\mpar}[1]{\!\xspace}

\newcommand{\del}[1]{\!\xspace}\newcommand{\new}{\!\xspace}\newcommand{\newe}{\!\xspace}\newcommand{\newv}{\!\xspace}\newcommand{\news}{\!\xspace}

\newcommand{\keep}{\color{black}}

\makeatletter

\renewcommand*{\p@subsection}{}

\renewcommand*{\p@subsubsection}{}
\makeatother

\begin{document}


\title{Quantifying brain microstructure with diffusion MRI: Theory and parameter estimation}

\author{Dmitry S. Novikov}
\email{dima@alum.mit.edu}
\affiliation{Center for Biomedical Imaging, Department of Radiology,
NYU School of Medicine, New York, NY, USA}

\author{Els Fieremans}
\email{els.fieremans@nyumc.org}
\affiliation{Center for Biomedical Imaging, Department of Radiology,
NYU School of Medicine, New York, NY, USA}

\author{Sune N. Jespersen}
\email{sune@cfin.au.dk}
\affiliation{CFIN/MINDLab, Department of Clinical Medicine and Department of Physics and Astronomy, Aarhus University, Aarhus, Denmark}

\author{Valerij G. Kiselev} 
\email{kiselev@ukl.uni-freiburg.de}
\affiliation{Medical Physics, Deptartment of Radiology, 
Faculty of Medicine, University of Freiburg, Germany}

\date{\today}

\begin{abstract}
\nin
We review, systematize and discuss models of diffusion in neuronal tissue, by putting them into an overarching physical context of  coarse-graining over an increasing diffusion length scale. From this perspective, we view  research on  quantifying brain microstructure as occurring along the three major avenues. 
The first avenue focusses on the transient, or time-dependent, effects in diffusion. These effects signify the gradual coarse-graining of tissue structure, which occurs qualitatively differently in different brain tissue compartments. We show that studying the transient effects has the potential to quantify the relevant length scales for  neuronal tissue, such as the packing correlation length for  neuronal fibers, the degree of neuronal beading, and compartment sizes. 
The second avenue corresponds to the long-time limit, when the observed signal can be approximated as a sum of multiple non-exchanging anisotropic Gaussian components. Here the challenge lies in parameter estimation and in resolving its hidden degeneracies. 
The third avenue employs  multiple diffusion encoding techniques, able to access information not contained in the conventional diffusion propagator.
We conclude with our outlook on the future  directions which can open exciting possibilities for designing quantitative markers of tissue physiology and pathology, based on  methods of studying mesoscopic transport in disordered systems. 
\end{abstract}

\maketitle



\renewcommand{\baselinestretch}{0.97}\normalsize
{\small \tableofcontents }
\renewcommand{\baselinestretch}{1.0}\normalsize

\input{Section_I}

\input{Section_II}

\input{Section_III}

\clearpage
\input{Section_IV}

\input{Section_V}

\begin{acknowledgments}

\nin
It is a pleasure to thank our numerous colleagues and members of our research groups for stimulating discussions and collaborations 
which are reflected in our Review.  
%
In particular, we thank Christian Beaulieu and Gene Kim for discussions on experimental issues, and  
Dan Wu and Jiangyang Zhang for sharing their OGSE data displayed in Fig.~\ref{fig:wu-does}. 
\new We also thank Ivana Drobnjak, Ileana Jelescu, Markus Nilsson and Marco Palombo, as well as the referees, 
for their thoughtful and constructive comments on the manuscript. \keep

Photo credit to Tom Deerinck and Mark Ellisman (National Center for Microscopy and Imaging Research) for the histology image illustrating a fiber fascicle in Figs.~\ref{fig:meso} and \ref{fig:SM}.

E.F. and D.S.N. were supported by 
the National Institute of Neurological Disorders and Stroke of the NIH under award number R01NS088040.
SNJ was supported by the Danish Ministry of Science, Technology and Innovations University Investment Grant (MINDLab, Grant no. 0601-01354B), and the Lundbeck Foundation R83-A7548.

\end{acknowledgments}



\bibliography{main.bbl}
\newpage \appendix \input{app}

\end{document}

%% file: Section_I.tex

\section{Diffusion MRI through a bird's eye} 
\label{sec:intro}


\epigraph{One of the most astonishing things about the world in which we live is that there seems to be interesting physics at all scales. $\la ... \ra$ 
To do physics amid this remarkable richness, it is convenient to be able to isolate a set of phenomena from all the rest, so that we can describe it without having to understand everything. Fortunately, this is often possible. We can divide the parameter space of the world into different regions, in each of which there is a different appropriate description of the important physics. Such an appropriate description of the important physics is an ``effective theory"}
{H. Georgi, {\it Effective Field Theory} \cite{georgi-EFT}}

\nin
Diffusion MRI (dMRI) is a macroscopic physical measurement of the voxel-averaged \new stochastic\keep\mpar{R2} motion of nuclear-spin-carrying molecules (typically water). This measurement occurs in a structurally complex tissue microenvironment such as the brain. Diffusion in complex media has been studied for about a century in a variety of fields, and is part of a broad class of transport phenomena in disordered systems. 

Our goal in this review article is to place biophysical  dMRI modeling into a broader physical context. Our overarching theme will be that of coarse-graining and effective theory, which will allow us to present and discuss neuronal tissue models of diffusion from a unifying perspective.

\subsection{Mesoscopic Bloch-Torrey equation as an effective theory}
\label{sec:BT}

\nin
One of the key 20$^{\rm th}$ century  advances in understanding the physics of complex systems was achieved by the development of {\it effective theory}, a paradigm to describe dynamics that involves only a handful of the so-called {\it relevant degrees of freedom}, or {\it relevant parameters}, thereby ignoring myriads of other, ``irrelevant" ones \cite{anderson1972,wilson1983,cardy-book}. 
This way of thinking was spurred by attempts to describe systems with an ever greater number of degrees of freedom, and  a subsequent realization that it is plain impossible to keep track of all of them at once.  

The more complex the system, the more the challenge of building an adequate theory shifts towards identifying which (few) parameters to keep, and which ones (almost all!) to ignore. Over time, selecting relevant parameters and formulating an adequate effective theory has become synonymous with the notion of understanding  the system's behavior. 

Having NMR as an example, the quantum-mechanical couplings of a very complex multi-spin Hamiltonian, together with all molecular degrees of freedom describing rotations, vibrations and translations, relevant at the nm and ps level, average out to produce effective parameters such as the relaxation rates $R_1$ and $R_2$, and the diffusion coefficient $D$, at least for the most common NMR measurements.  The parameters $R_1$ and $R_2$ emerge in Bloembergen--Purcell--Pound theory and enter the Bloch equations describing the semiclassical evolution of macroscopic magnetization \cite{BPP,Abragam}. Reducing a myriad of variables describing  molecular microenvironment to just a few relevant parameters has been a major scientific achievement of the 1940s--1960s NMR, and has formed the basis of effective theory of nuclear magnetization in  liquids.  

The step from NMR in uniform liquids to biological tissues has brought along a new challenge, which our community is only beginning to fully embrace. This challenge is associated with the above effective parameters  $R_1(\r)$, $R_2(\r)$ and $D(\r)$ acquiring {\it spatial dependence} at the scale  $\sim 0.1 - 10\, \mu$m,  set by the cellular architecture, much coarser than molecular dimensions. 
These spatial variations become relevant at the corresponding $\sim 1 - 1000\,$ms time scales of dMRI, --- much slower than the ps time scales on which the local relaxation rates and the diffusion coefficient emerge. 

\begin{figure}[b!]
\includegraphics[width=3.5in]{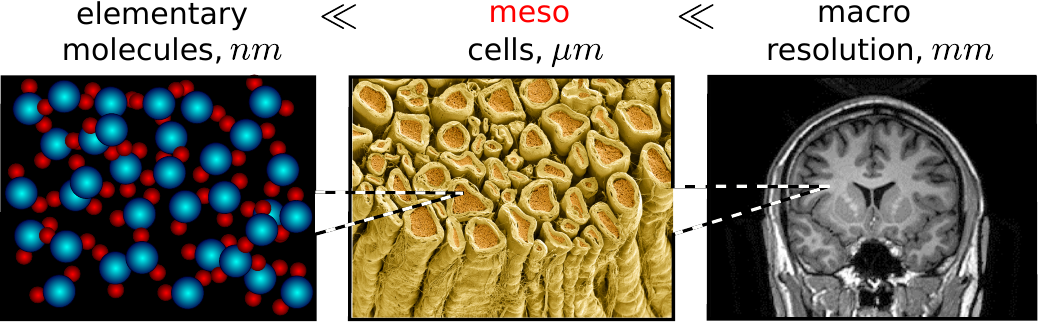}%
\caption{
The mesoscopic scale in brain dMRI, as an intermediate scale between the elementary (molecular) and the macroscopic (resolution). 
}
\label{fig:meso}
\end{figure}

From the physics standpoint, the spatial variations of $R_1(\r)$, $R_2(\r)$ and $D(\r)$ (with the latter including boundary conditions associated with cell membranes), occur at the {\it mesoscopic scale}, Fig.~\ref{fig:meso}.  
The term ``mesoscopic" originated in condensed matter physics some decades ago \cite{imry-book}, signifying focussing on the intermediate scales (``meso"),  in-between  elementary (say, atomic or molecular), and macroscopic (associated with the sample size or the measurement resolution). By design, this term is relative, depending on which spatio-temporal scales are deemed small and large. 

For dMRI, the mesoscopic scale corresponds to tissue heterogeneities at the scale defined by the MRI-controlled diffusion length, $L(t) \sim \sqrt{Dt} \sim 1 - 50\,\mu$m, which is the \new root mean squared \keep molecular displacement for times 
$t\sim 1 - 1000\,$ms. In the dMRI literature, it is  commonly referred to as the microstructure scale. 
This scale is commensurate with immense structural complexity of tissue architecture. 

At the mesoscale, quantum degrees of freedom become irrelevant (at least for the dMRI purposes), and the dynamics of transverse magnetization $m(t,\r)$ (a two dimensional vector, \mpar{R1}\new represented by \keep a complex number) can be captured by the  {\it mesoscopic Bloch-Torrey equation}
\be \label{BT}
\partial_t m(t,\r) = \partial_\r \lb D(\r) \partial_\r m(t,\r) \rb - \lb R_2(\r)  + i\Omega(t,\r) \rb m(t,\r) .
\ee
Here $\Omega(t,\r)$ is the Larmor frequency offset that may include externally applied diffusion-sensitizing \new \mpar{R2} Larmor frequency \keep gradients ${\bf g}(t)$, $\Omega(t,\r) = \Omega(\r) + {\bf g}(t)\r$, and the static $\Omega(\r)$ arises from the intrinsic mesoscopic magnetic structure of tissues due to paramagnetic ions such as iron, \new myelin susceptibility in the white matter, \keep or due to added contrast agent. 
\new 
While we focus on the transverse magnetization in what follows, the full version of the above equation includes the longitudinal magnetization components with $m$ being a three-dimensional vector. Further extension can incorporate multicomponent $m$ to describe the interplay between different proton pools, e.g., to describe magnetization transfer \cite{Henkelman1993,Henkelman2001}. 
\keep

The mesoscopic Bloch-Torrey equation (\ref{BT}) is an adequate effective description at the $\mu$m level, commensurate with typical diffusion length scales probed with dMRI. It is a mesoscopic equation in the sense that it involves scales in-between the quantum-mechanical molecular dynamics on the nm scale and the measurable signal in mm-sized MRI voxels. While the averaging up to the mesoscopic scale is already performed in its $\r$-dependent parameters, it is our task to perform the remaining averaging over a macroscopic voxel $V$ inherent to the observed (complex-valued) signal $S[t,{\bf g}(t)]\propto \int_V \! \d\r\, m(t,\r)$,  for which the $\mu$m-level spatially varying relaxation rates and diffusive properties produce the observable deviations from mono-exponential relaxation and Gaussian diffusion. It is because of this averaging that addressing the mesoscopic tissue complexity requires bringing the tools and intuition from condensed matter and statistical physics, in contrast to the quantum-mechanical description at the molecular level \cite{BPP,Abragam} and classical electrodynamics-based considerations used in designing  MR hardware.  

The overarching goals of ``microstructural", or ``mesoscopic" MRI modeling  are 
\renewcommand{\theenumi}{\bf (\roman{enumi})}
\begin{enumerate}
\item 
To identify the relevant tissue-specific parameters, which contribute to $R_1(\r)$, $R_2(\r)$, $D(\r)$, $\Omega(\r)$, 
and survive in the voxel-averaged signal (i.e., to build an appropriate effective theory for the macroscopic signal);  

\item To suggest optimal ways to probe them (i.e., to solve the corresponding parameter estimation problem). 
\end{enumerate}
Notice that to keep our terminology reasonably rigorous, we separated {\it modeling} into {\it theory} and {\it parameter estimation} (sometimes called ``fitting"); hence our title. These two  facets of modeling require very different tools and ways of thinking, as we will see below in Sections \ref{sec:Dt} and \ref{sec:gauss}, respectively.

\subsection{Coarse-graining and emergent phenomena}
\label{sec:emergent}

\nin
Equation (\ref{BT}) is an example of an {\it effective theory} --- i.e., an approximate description that emerges by averaging out the dynamics at the smaller spatial and temporal scales. 
It illustrates a general principle: pretty much every dynamical equation in physics is an effective theory (governed by an effective Hamiltonian or an effective action), i.e., it has {\it emerged} by identifying ``collective"  phenomena involving many-particle interactions at a more elementary level 
\cite{georgi-EFT,anderson1972,wilson1983,cardy-book}. 

Over the past century, physicists have come to realize that, at each level of complexity,  the effective theory and its relevant parameters can look very different \cite{anderson1972}, giving rise to the hierarchy of scales and of the corresponding emergent phenomena, from the most microscopic to the most macroscopic. Interactions between quarks and gluons give rise to protons and neutrons, so that their charge and mass can be viewed as effective parameters emerging by averaging over the quark/gluon degrees of freedom. Interactions between protons and neutrons forms a nucleus; interactions between nuclei and 
electrons give rise to all of chemistry, whereby the details of interactions between protons and neutrons inside nuclei become irrelevant. Interactions between molecules, coarse-grained over nm scale, give rise to  hydrodynamics, statistical mechanics and eventually, to biology, and so on.

It is remarkable that, for instance, there is not a hint of  classical hydrodynamics at the level of  the Schr\"odinger and Dirac equations describing the atomic structure; the large-scale hydrodynamic description {\it emerges} after a highly nontrivial averaging over the corresponding quantum degrees of freedom of many molecules.  
Refined methods such as renormalization \cite{wilson1983,cardy-book} were crafted specifically to single out relevant parameters from the rest upon iterative 
\new 
{\it coarse-graining} \cite{Migdal1976,Kadanoff1976}, 
which is a procedure that averages the dynamics over finer-scale degrees of freedom to derive approximate effective dynamics at the coarser scales involving a minimal number of parameters. This way of thinking reveals a fascinating hierarchy of natural phenomena \cite{anderson1972}. 
\keep

For quantifying tissue microstructure by measuring diffusion, transverse relaxation \new or magnetization transfer \keep with MRI, the mesoscopic Bloch-Torrey equation (\ref{BT}) contains all relevant physical processes. This effective theory is as fundamental for the mesoscopic MRI, as the Schr\"odinger equation is for the non-relativistic quantum mechanics, or the Navier-Stokes equation is for the classical hydrodynamics. It is always  the starting point for developing biophysical models relating  the NMR signal to the mesoscopic tissue architecture.


\begin{figure*}[th!]
\includegraphics[width=0.8\textwidth]{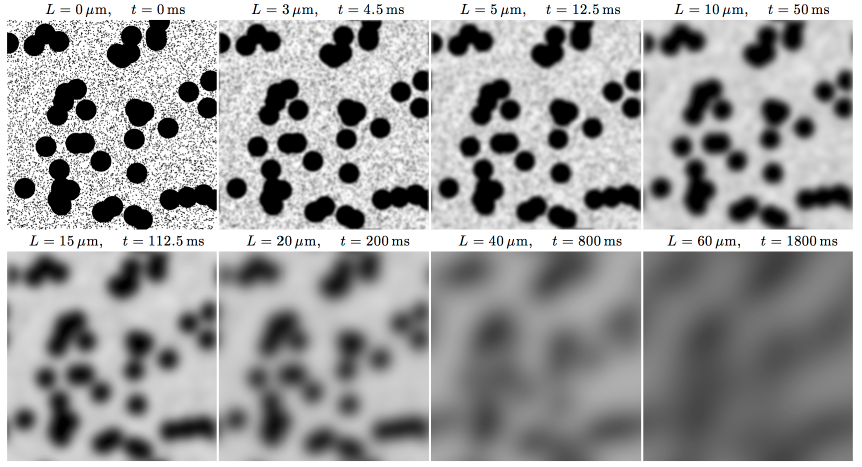}%
\caption{
{\bf Diffusion as coarse-graining.} 
An example of a medium where the mesoscopic structure is created by randomly placing black disks of two different radii, 
$r_{\rm small}=1\,\mu$m and $r_{\rm large}=20\,\mu$m, top left panel. 
To obtain snapshots of the medium as effectively seen by the diffusing molecules at different time scales, we used a Gaussian filter with width $L/2$, where $L(t)=\sqrt{2Dt}$, and ignored the time dependence of $D(t)$ in the definition of diffusion length, using a typical value $D=1\,\mu{\rm m}^{2}$/ms for the illustration purposes (cf.\ Sec.~\ref{sec:long-t} below). 
}
\label{fig:filter}
\end{figure*}

\subsection{Diffusion as coarse-graining}
\label{sec:diff=cg}

\nin
Diffusion in heterogeneous media is a beautiful and simple example of coarse-graining. It can be thought  of as a gradual ``forgetting", or homogenizing over the increasing diffusion length. To illustrate this concept, consider a two-dimensional model example of a two-scale mesoscopic structure, represented by  randomly placed impermeable disks of two different radii, embedded into an NMR-visible space with diffusion coefficient $D_0$, Fig.~\ref{fig:filter}. To be specific, let us assign sizes, typical to  cell dimensions: 
the small disks have radius $r_{\rm small}= 1 \,\mu$m and the large ones are 20 times larger. 
In a (hypothetical) tissue, this could describe diffusion in the extra-axonal space transverse to a fiber tract, hindered by two types of axons. 
Here we consider diffusion as a physical phenomenon; its relation to dMRI is discussed below in Sec.~\ref{sec:qtI}.


At time $t\to 0$, each water molecule only senses its own immediate environment; all molecules see the same ``intrinsic" diffusion coefficient $D|_{t=0} = D_0$, which is of the order $\sim 1\,\units{\mu m^2/ms}$. (For pure water at $37^\circ$C, $D_0=3\,\units{\mu m^2/ms}$.)

As time increases (top row of Fig.~\ref{fig:filter}), molecules get restricted by the walls of both small and large disks. As small disks have much higher net surface area than the large ones, the hindrance occurs mostly due to the small ones. Hence, the decrease of the resulting voxel-averaged diffusion coefficient would happen on the scale of a few ms, mostly dominated by the geometry of the small disks at the scale $\sim 1 \,\mu$m. 

At $t\gtrsim 100\,$ms (bottom row), when the  diffusion length $L(t)$ strongly exceeds the small disk size, the effect of the small disks has become coarse-grained (while the effect of the large disks is not). Now, we can view the medium in-between the large disks as a homogeneous ``effective medium", with some effective diffusion coefficient $D_{\rm small}< D_0$ given by the macroscopic (``tortuosity") limit of a medium with the small disks only.
It is important to note that if we did not have access to shorter times and could only resolve the diffusion times corresponding to the lower row of panels, there would be no way to identify the presence of the small disks --- their effect has been homogenized,%
\footnote{An experienced reader can recall the possibility to apply strong gradients $q\sim 1/r_{\rm small}$ to detect the small disks. 
This, however, practically requires sensitivity to short $t\sim 1/(D_{\rm small}\, q^2)$, cf. Sec.~\ref{sec:high_q} below.} 
and their numerous parameters (e.g., size, coordinates) have become ``irrelevant", with their only role in {\it renormalizing} $D_0$ down to $D_{\rm small}$. 

Hence, from  time $t\gtrsim 100\,$ms on, we can adopt the coarse-grained description which only involves the large disks, immersed in a uniform medium with diffusion constant $D_{\rm small}$. The corresponding Eq.~(\ref{BT}) would have the effective $D(\r)$ varying at the scale associated with large disks, with  $D(\r) \simeq D_{\rm small}$ outside them, and the short-distance spatial harmonics of  $D(\r)$ filtered out as it is obvious from Fig.~\ref{fig:filter}; in Section~\ref{sec:Dt}, we will  rigorously justify and use this intuitive picture. The measured diffusion coefficient would further decrease with $t$ at the scale of a few hundred ms, corresponding to being hindered by the large disks --- the remaining restrictions.  

Eventually, at even longer $t\gtrsim 1000\,$ms, the effect of the large disks also becomes coarse-grained, and the whole sample looks as if it were homogeneous with some macroscopic diffusion coefficient $\Dinf$, such that $0< \Dinf < D_{\rm small} < D_0$. 
From this $t$ onwards, one cannot distinguish this sample from a uniform medium with diffusion constant $\Dinf$. 

Our example shows that the hallmark of  coarse-graining over larger- and larger-scale mesoscopic structure is the time-dependence of the overall diffusion coefficient. In the view of this time dependence, it will be convenient to work with the {\it instantaneous} diffusion coefficient, 
\be \label{Dinst}
\Dinst(t) \equiv  {\partial \over \partial t} {\langle \delta x^{2}(t)\rangle \over 2} \,, 
\quad \delta x(t) = x(t)-x(0) \,, 
\ee 
defined as the rate of change of the mean squared molecular displacement $ \langle \delta x^{2}(t)\rangle$ in a particular direction ${\bf \hat x}$. 
(For simplicity, we assumed that our sample in Fig.~\ref{fig:filter} is statistically isotropic. 
For anisotropic samples, the diffusion tensor components will acquire the time dependence.)

The average $\langle \dots \rangle$ in the definition (\ref{Dinst}) is actually a {\it double average}: 
(i) over the Brownian {\it paths} in the vicinity (of size $\sim L(t)$) of any given initial point $\r_0 \equiv \r(t)|_{t=0}$, 
yielding the local coarse-grained diffusivity value $D(\r_0)$; 
and (ii) over the {\it ensemble} of random walkers (spins) originating from all possible initial points $\r_0$. 
Because of the ensemble average, the measured diffusion characteristics, such as Eq.~(\ref{Dinst}), describe a macroscopic sample \emph{as a whole}. They do not belong to any given Brownian path, but rather emerge as a result of averaging (i) over all possible Brownian paths that could be taken by a given molecule, and (ii) over the initial positions of all molecules in a sample.

Upon taking into account increasing length scales, the effective voxel-averaged $\Dinst(t)$ {\it flows} towards the tortuosity limit 
$\Dinf$, starting, as in our example, from the ``microscopic" $D_0 > \Dinf$.
We used the term ``to flow" because the above picture mimics the {\it renormalization group flow} \cite{wilson1983,cardy-book} according to which the gradual evolution of a physically important parameter, such an elementary particle charge or an effective mass, occurs as a function of the coarse-graining length scale, from the high-energy = short-distance scale, down to the low-energy parameters relevant for the  macroscopic description.%
\footnote{\label{ftn:fixedpoint}
\new 
We also note that the renormalization group flow can have {\it fixed points}, i.e., microscopic parameter sets for which the effective parameters do not change with the increasing coarse-graining scale. 
Approaching the asymptotically normal diffusion with a finite $\Dinf$ is a Gaussian fixed point, which is what happens for most structural arrangements; one can say that diffusion is most tissues is a continuous family of Gaussian fixed points (for each realization of microscopic tissue architecture). 
An example of a non-Gaussian fixed point is the so-called anomalous diffusion, cf. Sec.~\ref{sec:fixedpoint}. 
\keep
}\mpar{new footnote \ref{ftn:fixedpoint}}

Looking back, there was nothing special about requiring the disks to be impermeable (the black regions could have corresponded to some medium with diffusion coefficient $D_1 \neq D_0$); we could have used objects of a non-disk shape, and/or with non-sharp boundaries. Generally, as long as the random walkers can, in the limit $t\to\infty$, reach any point in a given ``compartment", the above coarse-graining picture applies to this compartment. If a voxel contains multiple non-exchanging compartments, it applies to them separately, with the net signal given by a sum of their contributions. 

A similar physical picture qualitatively applies to the effects of spatially varying \new transverse relaxation rate \keep $R_2(\r)$ --- e.g., if the black and white regions in Fig.~\ref{fig:filter} instead represented different local molecular-level $R_2$ values, and spins were able to diffuse everywhere. The above argument would then lead to an effective $R_2(\r)$ entering Eq.~(\ref{BT}) for times $t$ exceeding the corresponding coarse-graining time scale. For instance, for $t\gtrsim 100\,$ms, the effect of small disks would homogenize to produce a uniform $R_{2,{\rm small}}$ rate in-between the large disks, and so on, leading to the time-dependent overall observed $R_2(t)$ with an asymptotic macroscopic rate $R_2|_{t\to\infty}$ observable at very long $t$. 
Likewise, if the mesoscopic structure in Fig.~\ref{fig:filter} represented spatially varying susceptibility $\chi(\r)$, inducing the corresponding $\Omega(\r)$, 
the resulting $R_2^*(t)$ rate would change --- in this case, {\it increase}
with $t$, approaching the $R_2^*(t)|_{t\to\infty}$ macroscopic value as a result of gradual coarse-graining. 

We note that all the above mentioned quantities --- $D_{\rm small}$ and $R_{2,{\rm small}}$;  $\Dinf$ and $R_2|_{t\to\infty}$; $R_2^*(t)|_{t\to\infty}$ --- are {\it nonuniversal}, i.e., they depend on the numerous structural details, such as packing geometry (e.g., periodic versus random arrangement); they would change if the disks were instead squares, etc. Certainly, these quantities are not given by a simple averaging of the microscopic $D(\r)$ or $R_2(\r)$ over the sample. However, the {\it initial} values $\Dinst(0)$ and $R_2(0)$ are  given by the sample-averaged $\langle D(\r) \rangle$ and $\langle R_2(\r) \rangle$, correspondingly, since at $t\to 0$ (practically, at times just exceeding the ps time scale necessary for the local $D(\r)$ and $R_2(\r)$ to emerge), each spin senses only its immediate  environment. 

\begin{SCfigure*}[0.7][th!!]
\includegraphics[width=3.9in]{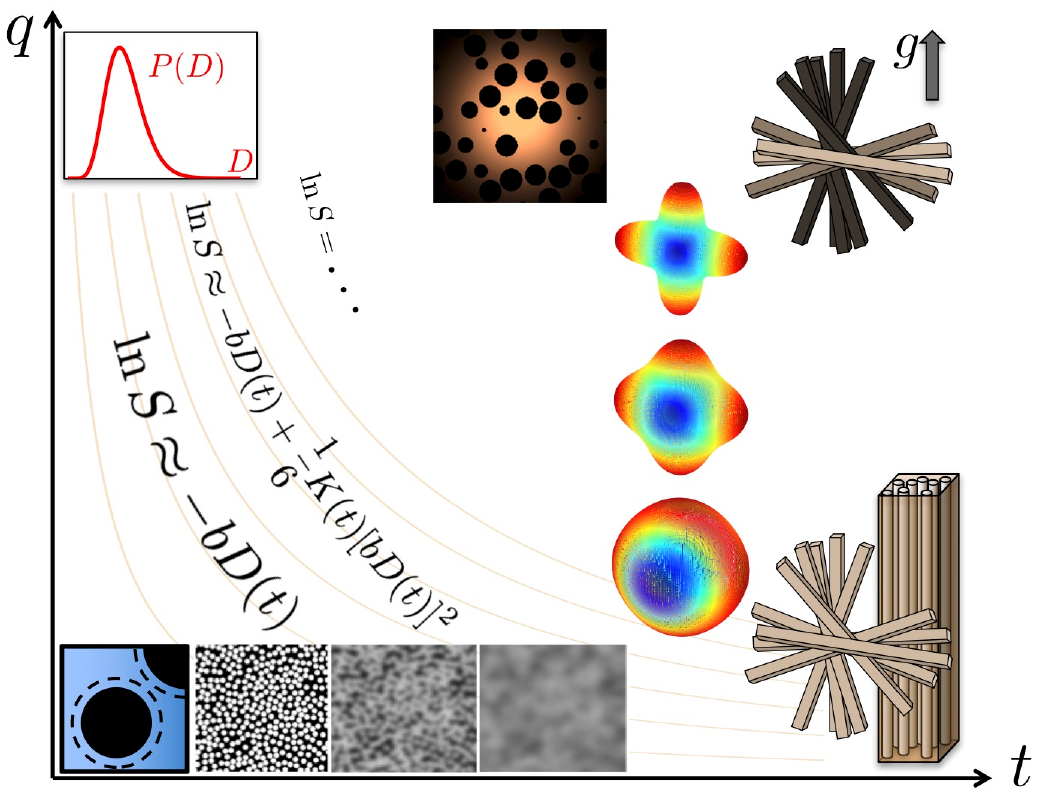}\qquad\qquad
\caption{
The parameter space of dMRI is at least two-dimensional: 
By increasing $q$ one accesses the progressively higher-order diffusion cumulants, Sec.~\ref{sec:cum}, whereas the dependence along the $t$-axis reflects their evolution over an increasing diffusion length scale $L\sim \sqrt{Dt}$, Eq.~(\ref{Dcum}). 
The $b$-value  alone does not uniquely describe the  measurement, unless diffusion in all tissue compartments is Gaussian; contour lines of $b=q^2 t$ are schematically drawn in beige. 
\new
Large-$q$ limits: 
Top-left is high-resolution limit $L(t)\ll l_c$, $ql_c \gg 1$, Sec.~\ref{sec:hierarchy}{\bf (i)}; 
middle is the $L(t)\gtrsim l_c$, $q l_c \gtrsim 1$ limit of probing the pore correlation function, Sec.~\ref{sec:high_q}. 
\keep 
The hierarchy of dMRI models (pictures), cf.\ Fig.~\ref{fig:models}, as well as the cumulant representation with different number of terms, cf. Fig.~\ref{fig:representations}, are superimposed.  
The decrease of the signal from axonal bundles parallel to the increasing gradient is shown by their darkening (top right). 
In Section~\ref{sec:Dt} we move along the 
$t$-axis at low $q$, and in Section~\ref{sec:gauss} we  move along the $q$-axis at asymptotically long $t$. 
Section~\ref{sec:mpfg} is devoted to effects beyond this diagram, contained in voxel-averaged products of  propagators at different $t$ and $q$.
}
\label{fig:qt}
\end{SCfigure*}

The picture of  gradual coarse-graining over an increasing diffusion length has a number of  important consequences: 
\renewcommand{\theenumi}{\bf \arabic{enumi}}
\begin{enumerate}

\item The mesoscopic Bloch-Torrey equation (\ref{BT}) {can be fully determined only after the relevant spatio-temporal scales are specified}, since its parameters $R_2(\r)$, $\Omega(\r)$ and $D(\r)$ are effective and, hence, scale-dependent. 

\item Generally, {the observed voxel-averaged diffusion coefficient and the effective relaxation rate would depend on time} $t$ because of the presence of the mesoscopic structure (such as $\Dinst(t)$ decaying from $D|_{t=0}$ down to $D_\infty$  in our example of Fig.~\ref{fig:filter}). This time $t$ can be set by the measurement sequence, and varying it provides a unique window into the tissue architecture at the scale of the corresponding diffusion length $L(t)$. 

\item This brings us to the fundamental challenge of interpreting such time dependencies in terms of the mesoscopic structural complexity. Practically, we must figure out which features in the effective $R_2(\r)$, $\Omega(\r)$ and $D(\r)$ remain observable after the voxel-wise averaging as a result of a macroscopic acquisition (cf.\ Section \ref{sec:Dt}). 
This is the overarching task --- and justification --- for the theoretical efforts in our community. 

\item If a measurement is  too slow to track the transient processes, we are left (in each non-exchanging  compartment) with the $t\to\infty$ {\it macroscopic} Bloch-Torrey equation, i.e., Eq.~(\ref{BT}) with {\it uniform} effective parameters $D(\r) \to \Dinf$, $R_2(\r) \to R_2(t)|_{t\to \infty}$, $\Omega(\r) \to \langle \Omega(\r) \rangle$. 
Its solution becomes trivial --- mono-exponential relaxation and Gaussian diffusion (cf.\ Section \ref{sec:gauss});
i.e., coarse-graining leads to the {\it universal} $t\to\infty$ dynamics, albeit with nonuniversal macroscopic parameters such as $\Dinf$. 
The mesoscopic information is now lost, as the signal is indistinguishable from that in a uniform medium. Effective macroscopic parameters are in general  different from the intrinsic mesoscopic ones; for instance, $\Dinf$ can be notably lower than the intrinsic water or axoplasmic diffusion coefficient. 

\end{enumerate}


\subsection{dMRI signal as the diffusion propagator;  $qt$ Imaging}
\label{sec:qtI}

\nin
So far we managed to get away with looking at a single equation (\ref{BT}) and wave hands based on drawing parallels with concepts developed in physics. 
It is now time to introduce basic notations; the content of this subsection should be familiar to anyone actively working in dMRI.  

In what follows, for simplicity we will confine ourselves only to the mesoscopic structure as related to diffusion, and will assume the relaxation effects to be trivial (at least in each tissue compartment), setting $R_2(\r) \to R_2$, and a uniform voxel-wise Larmor frequency, $\Omega(\r) \to \langle \Omega \rangle$. (The nontrivial $R_2(\r)$ and $\Omega(\r)$ modify apparent diffusion metrics \cite{zhong1991,kiselev2004}; this is beyond the scope of our review.) 
This allows us to factorize 
the magnetization 
$m(t,\r) \equiv e^{-R_2 t - i\langle\Omega\rangle t} \, \psi(t,\r)$, where $\psi(t,\r)$ is not subjected to the relaxation and frequency shift and obeys the following equation
\be \label{BTD}
\partial_t \psi(t,\r) = \partial_\r \lb D(\r) \partial_\r \psi(t,\r) \rb -  i{\bf g}(t)\r \,\psi(t,\r) .
\ee

We focus here on the most easily interpretable measurement with very narrow (i.e., short) gradient pulses.%
\footnote{%
\new
The focus on narrow pulses helps one to gain physical intuition. 
Finite pulse-width $\delta$ for relatively weak gradients has an effect of a low-pass filter, filtering out the frequencies 
$\gtrsim 1/\delta$, acting on the narrow-pulse solution 
\cite{callaghan-book,Kiselev2010_diff_book, Kiselev2016_NMB_review}, cf. models of restricted \cite{Murday1968,neuman1974,VanGelderen1994muscle} and hindered \cite{burcaw2015,HHL2015,HHL2016-phantom,lee2017} 
diffusion relevant for the brain. 
For strong gradients, long pulses lead to the localization regime, Sec.~\ref{sec:beyond}. 
Arbitrarily-shaped pulses in the Gaussian phase approximation will be considered in Sec.~\ref{sec:2cum}. 
\keep
} 
As we now discuss, serendipitously, this measurement accesses the propagator of the mesoscopic diffusion equation, 
which (cf. Sec.~\ref{sec:diff=cg}) describes evolution of particle density $\rho(t,\r)$
\be \label{DE}
\partial_t \rho(t,\r) = \partial_\r \lb D(\r) \partial_\r \rho(t,\r) \rb . 
\ee

The {\it fundamental solution} of Eq.~(\ref{DE}), or  {\it diffusion propagator} $\G_{t;\r, \r_0}$, satisfies this equation
\be \label{def-Gcal}
\partial_t \G_{t;\r,\r_0} = \partial_\r \lb D(\r) \partial_\r \G_{t;\r,\r_0}  \rb + \delta(t)\delta(\r-\r_0) 
\ee
with the point-like and instant source  at $\r=\r_0$. The source term corresponds to the solution with zero particle density for $t<0$ and with the initial condition $\delta(\r-\r_0)$ instantly appearing at $t=0$. The solution is thus proportional to the unit step function, $\theta|_{t>0} = 1$ and $\theta|_{t<0} = 0$, such that $\partial_t \theta(t) = \delta(t)$. 

The propagator $\G_{t;\r, \r_0}$ is a fundamental quantity describing the diffusion process around the point $\r_0$, with a meaning of the probability distribution function (PDF) of molecular displacements $\r-\r_0$ over time $t$. 
(This PDF can be sampled using Monte Carlo simulations by releasing random walkers all at once from the point $\r_0$.)  
Of course, since the local tissue structure is different around each initial point $\r_0$, the propagator $\G_{t;\r, \r_0}$ depends on the points $\r_0$ and $\r$ separately. 

The fundamental connection between the diffusion process (\ref{DE}) and the NMR measurement stems from the gradient-dependent phase of $\psi(t,\r)$ as described by Eq.~(\ref{BTD}). In the limit of narrow pulses ${\bf g}(\tau) = \q \cdot \lb \delta(\tau-t) - \delta(\tau) \rb$ 
and the initial condition as in Eq.~(\ref{def-Gcal}), the magnetization $\psi(t,\r)$ differs from $\G_{t;\r,\r_0}$ by the position-dependent phase
$e^{-i\q(\r_t-\r_0)}$ acquired during the gradient application. The {\it diffusion-weighted signal}, 
which is a net magnetization $\int\! \d\r\, m(t,\r)$ in a voxel,  
\bea \label{S=G}
{ S(t, \q) \over S(t,\q)|_{{\bf q}=0} }
&=&\int\! {\d\r_0 \d\r_t \over V}\, e^{-i\q(\r_t-\r_0)} \, \G_{t;\r_t,\r_0} \equiv G_{t,\q}  \quad
\eea
becomes equivalent to a spatial Fourier transform of the {\it voxel-averaged propagator} 
\be \label{def-G}
G_{t,\r} \equiv \la \G_{t;\r_0+\r,\r_0} \ra_{\r_0} =   \int\! {\d\r_0 \over V}\,  \G_{t;\r_0+\r,\r_0} \,.
\ee
In Eq.~(\ref{S=G}) we divided by the voxel volume $V$, such that the unweighted signal (the right-hand side) is normalized to unity. 
A thorough discussion can be found e.g.,\ in ref.~\cite{EMT}.   

Note that exact ``local" propagator $\G_{t;\r_t,\r_0}$ is not translation invariant, i.e., it depends on the absolute coordinates $\r_t$, $\r_0$ (and time $t$). The voxel-averaging in Eq.~(\ref{S=G}) automatically restores translation invariance, which means that the measured propagator $G$ is parameterized by the two variables: the spatial displacement $\r \equiv \r_t - \r_0$ and the diffusion time interval $t$ (equivalently, by $\q$ and $t$). 

Hence, the parameter space of dMRI fundamentally consists of $\q$ and $t$, 
Fig.~\ref{fig:qt} (here we dropped the directionality in $\q$ to not overload the picture). 
Literally speaking, mapping the diffusion propagator in the space of $\q$ and $t$ can be referred to as 
$qt$ Imaging.\footnote{{\tt cu-tie imaging, or qtI} {\it (noun)}: A noninvasive medical imaging technique for  spatio-temporal mapping of the diffusion propagator in soft tissues to quantify tissue structure below the nominal MRI resolution. 
Of course, it is nothing but the familiar $q$-space imaging \cite{Callaghan1988,cory1990,callaghan-book} sampled at various $t$, but don't we all need a new acronym once in a while? 
} 
For multiple diffusion encoding, which maps a more complex object than the diffusion propagator (Section~\ref{sec:mpfg}), 
the parameter space in principle depends on the multiple $\q$ and $t$ intervals.

The so-called $b$-value \cite{lebihan1986} has historically become the often single-quoted measurement parameter. 
However, it only defines the measurement {\it if  diffusion is Gaussian in every compartment}, 
in which case the diffusion propagator
\be \label{Gbare}
\Gbare_{t,\q} = \theta(t) \, e^{-Dq^2 t}  \equiv \theta(t)\, e^{-bD} \,, \quad b \equiv q^2 t 
\ee
in each compartment is determined solely by the parameter combination $q^2 t$. Schematically, the contour lines of constant $b$ are outlined in Fig.~\ref{fig:qt}. 
In general, for anisotropic tissues such as brain white matter, Gaussian diffusion in each compartment is described by the  diffusion tensor, $bD \to b_{ij} D_{ij}$, where the $b$-matrix \cite{Basser1994} $b_{ij}  = q_i q_j \, t$.

The Gaussian limit (\ref{Gbare}), and its more general anisotropic Gaussian limit, are hallmarks of ``full" coarse-graining, which occurs in the $t\to\infty$ limit, cf.\ Fig.~\ref{fig:filter}. In this case, no matter how structurally complex the tissue, it can be modeled as a sum of (anisotropic) Gaussian signals.  
Section \ref{sec:gauss} will be devoted to the picture of multiple Gaussian compartments (the Standard Model), cf.\ the column of pictures at long $t$ in Fig.~\ref{fig:qt}.

\subsection{Hierarchy of diffusion models based on  coarse-graining: The three regimes}
\label{sec:hierarchy}

\begin{figure*}[th!!]
\centering
\includegraphics[width=6.1in]{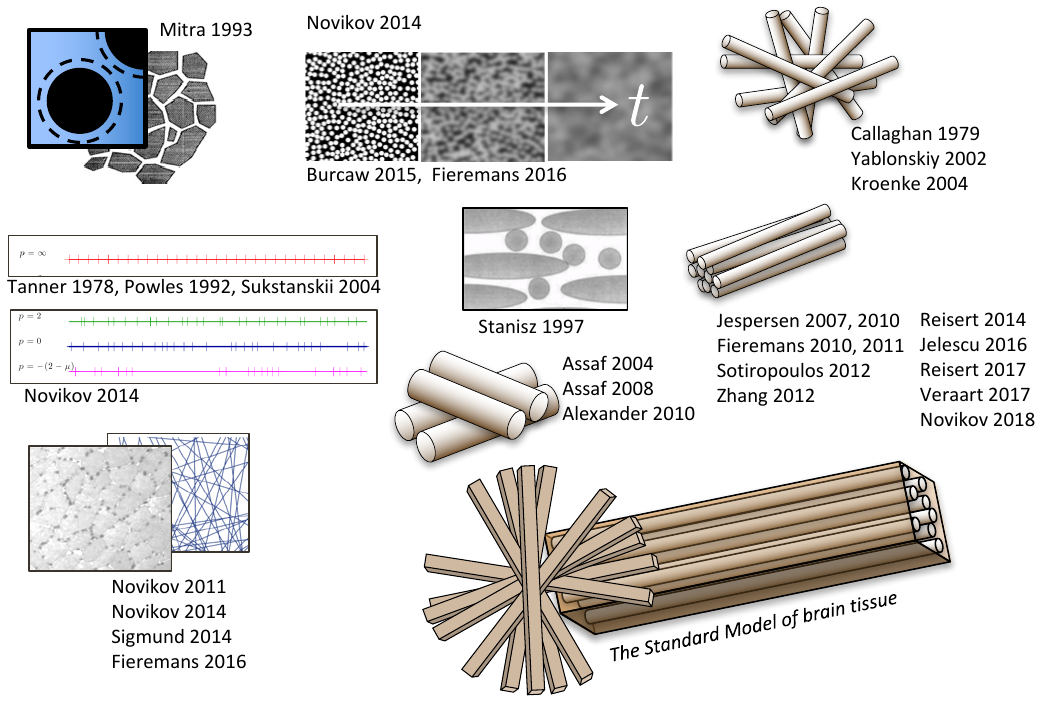}
\caption{
{\bf Models are pictures...} 
Here they are drawn with coarse-graining occurring, roughly, from left to right. 
References: 
Mitra 1993 \cite{Mitra1993}, universal short-$t$ limit; 
Novikov 2014 \cite{mesopnas}, universal long-$t$ behavior; 
Burcaw 2015 \cite{burcaw2015} and Fieremans 2016 \cite{fieremans2016}, long-$t$ behavior transverse and along WM fibers; 
Tanner 1978 \cite{tanner1978}, Powles 1992 \cite{powles1992}, Sukstanskii 2004 \cite{sukstanskii-jmr2004}, periodic 1-dimensional lattice;  
Novikov  2011 \cite{nphys}, random permeable barriers in any dimension, and its application to myofibers
(Sigmund 2014 \cite{sigmund2014} and Fieremans 2016 \cite{fieremans2016rpbm});  
Callaghan 1979 \cite{callaghan1979}, first model of diffusion inside random narrow cylinders;
Yablonskiy 2002 \cite{yablonskiy2002}, diffusion in finite-diameter cylinders modeling lung alveoli; 
Stanisz 1997 \cite{stanisz1997}, first model for WM fiber tracts made of ellipsoids; 
Assaf 2004 \cite{charmed}, CHARMED;
Assaf 2008 \cite{axcaliber}, AxCaliber; 
Alexander 2010 \cite{Alexander2010}, ActiveAx;
Kroenke 2004 \cite{Kroenke2004}, NAA diffusion inside neurites.
The widely adopted $t\to\infty$ picture of narrow ``sticks" for the neurites, embeded in the extra-neurite space (the Standard Model): 
Jespersen 2007 \cite{Jespersen2007},
Jespersen 2010 \cite{Jespersen2010},
Fieremans 2010 \cite{KM},
Fieremans 2011 (WMTI) \cite{wmdki},
Sotiropoulos 2012 (Ball and rackets) \cite{sotiropoulos2012},
Zhang 2012 (NODDI) \cite{noddi},
Reisert 2014 (MesoFT) \cite{reisert2014mesoft},
Jelescu 2016 (NODDIDA) \cite{jelescu2016},
Reisert 2017 \cite{baydiff},
Veraart 2017 (TEdDI) \cite{teddi},
Novikov 2018 (LEMONADE \cite{lemonade-ismrm}, RotInv \cite{rotinv}).
}
\label{fig:models}
\end{figure*}

\nin
From  the unifying coarse-graining point of view,  we  can now categorize biophysical  models  of  diffusion, Fig.~\ref{fig:models},  into the following three regimes. In either of the regimes, the theoretical treatment simplifies.  
The regimes can be arranged according  to the increasing  diffusion length $L(t)$ relative to characteristic mesoscopic tissue length scales:  

\renewcommand{\theenumi}{\bf (\roman{enumi})}
\begin{enumerate}

\item 
\new
No coarse-graining has yet occurred. 
If the local $D(\r)$ varies in space 
\newe over \keep the correlation length scale $l_c$, 
then for $L(t)\ll l_c$ and $q l_c\gg 1$, each molecule senses its own, locally homogeneous $D(\r)$. 
In this {\it high-resolution limit} \cite{EMT}, the signal $S(b) \simeq \int\! \d D \, \P(D)\, e^{-bD}$ 
is a Laplace transform of the histogram $\P(D)$ of all the local values $D(\r)$. 
\keep
A more relevant to biology situation occurs when instead of smooth $D(\r)$ variations, there are sharp barriers. 
The  relevant  parameter is then the net surface-to-volume ratio $S/V$ of all barriers (e.g., cell walls). 
For times such that $L(t)S/V \ll 1$, one observes the $S/V$ universal  short-time  limit  of  the  diffusion  
coefficient \cite{Mitra1993}.


\item Coarse-graining over the structural disorder \cite{mesopnas} results in the power-law approach $t^{-\vartheta}$ of  the instantaneous diffusion coefficient $\Dinst(t)$ towards the $t\to\infty$ limit $\Dinf$. 
 Here,  the  power-law  exponent  $\vartheta$  is  connected  to  the  large-scale  behavior  of  the  density correlation  function  of  the  hindrances  to  diffusion,  and  to  the  spatial  dimensionality,  yielding qualitatively distinct behavior along \cite{mesopnas,fieremans2016} and transverse \cite{fieremans2016,burcaw2015} to the neurites in  the  brain.  In Section \ref{sec:Dt} we argue, following ref.~\cite{mesopnas}, that  the  more  heterogeneous,  or  ``disordered",  the  sample  is,  the  slower the approach (the smaller the exponent $\vartheta$). 
 Conversely, in ordered media, such as in the model of perfectly ordered membranes \cite{mesopnas,sukstanskii-jmr2004}, the approach of $\Dinst(t)$ towards $\Dinf$ is exponentially fast. 

\item Complete  coarse-graining.   Diffusion  in  each  non-confining  tissue  compartment  has  approached  its  $t\to\infty$  Gaussian (tortuosity) limit, as discussed above (cf.\ also a more detailed discussion in Sec.~\ref{sec:fixedpoint} below). If  there  is  no  exchange  between  compartments, we obtain the most common, ``multi-exponential" model. 
For  neuronal tissue, the compartments are anisotropic due to the presence of effectively  one-dimensional neurites. 
In Section \ref{sec:gauss}, we introduce the ``Standard Model" of neuronal tissue that accounts for the neurites with associated extra-neurite space, and with an orientation dispersion (Fig.~\ref{fig:SM}). While known under a plethora of names and acronyms \cite{Kroenke2004,behrens2003,behrens2003nature,Jespersen2007,Jespersen2010,KM,wmdki,sotiropoulos2012,noddi,reisert2014mesoft,jelescu2016,baydiff,lemonade-ismrm,rotinv,teddi},  
from the physics standpoint, this is practically the same model, with differences  in the parameterization of the neurite orientation distribution function and variations in the  descriptions  of  the  extracellular  space,  as  well  as  in  the  model  parameter  estimation  procedures
and employed parameter constraints.  

\end{enumerate}

The crossover between regimes {\bf (i)}  and {\bf (ii)} occurs when the diffusion length, $L(t)$, is commensurate with the characteristic length scale of the structural disorder. The instantaneous diffusion coefficient $\Dinst(t)$ decreases with time within this crossover; 
while no general results are available there, it can be studied using numerical simulations.

\subsection{How to become sensitive to short length scales?}
\label{sec:high_q}

\nin
Working in the $t\to\infty$ limit {\bf (iii)} can only give us compartment volume  fractions and  their  diffusion  coefficients. 
Coarse-graining has already occurred and apparently washed out all traces of other microstructural parameters. 

Determining characteristic $\mu$m-level length scale(s) $l_c$, such as the correlation length of the arrangement of tissue building blocks (e.g., disk radii in Fig.~\ref{fig:filter}), is in principle possible using deviations from the Gaussian signal shape. In the spirit of Fig.~\ref{fig:qt}, varying either $t$ or $q$ can yield the  sensitivity of the diffusion signal (propagator) to the length scale, via the diffusion length $\sqrt{D(t) t}$ [cf. Eq.~(\ref{Dcum}) below], and via $1/q$, respectively. 
However, as we now discuss, these theoretically distinct ways are not that different in practice, because attaining $q\sim 1/l_c$ at times $t \gg t_c$
practically requires sensing the signal contributions that are small at least as some positive power of the small ratio 
$t_c/t \ll 1$, where $t_c \sim l_c^2/D(t_c)$.

\begin{figure*}[t!]
\centering
\vspace{5mm}
\includegraphics[width=5.5in]{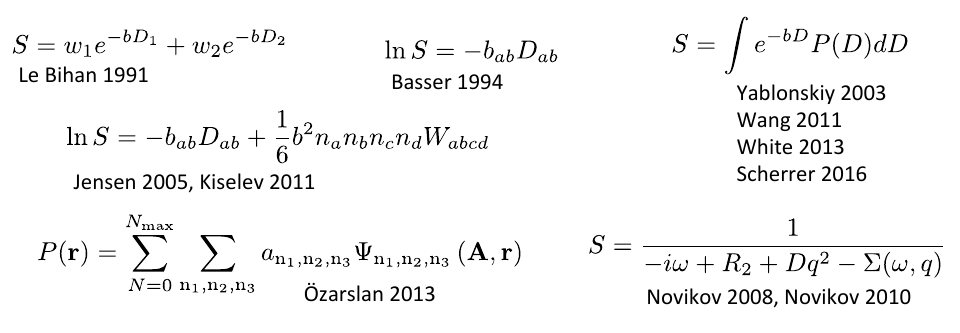}
\caption{
{\bf ...while representations are formulas.}
References: 
Le Bihan 1991 \cite{lebihan1991}, first biexponential representation of dMRI signal from brain;
Basser 1994 \cite{Basser1994}, diffusion tensor imaging (DTI); 
Jensen 2005 \cite{DKI}, diffusion kurtosis imaging (DKI);
Kiselev 2011 \cite{Kiselev2010_diff_book}, cumulant expansion;
Novikov 2008 and 2010 \cite{emt-jmr,EMT}, effective medium theory (transverse relaxation and diffusion, correspondingly);
\"Ozarslan 2013 \cite{ozarslan2013}, expansion in  harmonic oscillator basis; 
Yablonskiy 2003 \cite{yablonskiy2003}, inverse Laplace transform (multi-exponential representation). 
Anisotropic multi-exponential representations:  
Wang 2011 \cite{DBSI}, diffusion basis spectrum imaging (DBSI);
White 2013 \cite{RSI}, restriction spectrum imaging (RSI);   
Scherrer 2016 \cite{DIAMOND}, distribution of anisotropic microstructural environments in diffusion-compartment imaging (DIAMOND).
}
\label{fig:representations}
\end{figure*}
\clearpage

{Varying $t$} amounts to literally observing the diffusive dynamics for short times, when the coarse-graining has not yet fully occurred,  
such as during the regimes  {\bf (i)}  and {\bf (ii)} above. In our example in Sec.~\ref{sec:diff=cg}, to identify the presence of the small disks, one could try, e.g., detecting time dependence in $\Dinst(t)$ or in $D(t)$ at $t\sim r_{\rm small}^2/D_0$.  
The  random  permeable  barrier  model  \cite{nphys,fieremans2016rpbm},  a  candidate  for  diffusion  transverse  to myofibers and for one-dimensional hindrances along the neurites \cite{mesopnas}, allows one to trace the effect of coarse-graining 
in $\Dinst(t)$ or in $D(t)$   across all the regimes {\bf (i) -- (iii)}.

\mpar{par edited, EMT propagator eq. removed, instead ref \cite{Mitra1992} discussed}
\new 
{ Varying $q$}, by employing strong narrow gradients (with width $\delta \ll t_c$ so that $q$ is well-defined \cite{callaghan-book,Kiselev2010_diff_book,neuman1974,burcaw2015,HHL2015,lee2017,HHL2016-phantom,Kiselev2016_NMB_review}), 
can in principle allow one to unravel the coarse-graining, i.e., to observe features at $q\sim 1/l_c$ even when $t\gg t_c$.%
\footnote{In the language of emergent phenomena, Sec.~\ref{sec:emergent}, this would be analogous, e.g., to using neutron scattering (with large wave vectors $q \gtrsim 10\,\units{nm^{-1}}$) to resolve atomic structure of the fluid, for which the coarse-grained (large $t$ and  $r \gg 1/q$) continuous description is  classical hydrodynamics.} 
However, the price to pay for accessing such fine structures is the suppression of the signal. 
To give an example \cite{Mitra1992}, 
consider diffusion in a porous medium with connected pores and isolated grains, cf. Fig.~\ref{fig:qt} (center-top drawing). 
For long diffusion times, the pore structure is effectively coarse-grained (similar to diffusion in-between the small grains in Fig.~\ref{fig:filter}). While having the overall Gaussian-like envelope $\sim \Gbare_{t,\r}$ with $D\approx \Dinf$, the diffusion propagator 
$\sim \Gamma(\r)\Gbare_{t,\r}$ (up to an overall normalization) replicates the pore shape on the fine scale of the order of $l_c$,  Fig.~\ref{fig:qt}, 
with the density correlation function $\Gamma(\r)$ of the pore space arising due to voxel averaging. 
Correspondingly, in $q$-space,
\be \label{ansatz}
G_{t,\q} \approx  \Gbare_{t,\q} + \frac1\phi \int\! {{\rm d}^d {\bf \q'}\over (2\pi)^d}\, \Gbare_{t,\q-\q'}\Gamma(\q')
\ee
where the first term originates from sample-averaged $\langle \Gamma(\r)\rangle \equiv \phi$, the pore water fraction, 
and represents the average spin density spread, 
while the product $[\Gamma(\r)-\phi]\Gbare_{t,\r}$  becomes a convolution (the second term)
of this envelope with the correlation function in  $q$-space (i.e., the pore space power spectrum) $\Gamma(\q)$.  
The longer the time, the sharper is the Gaussian propagator $\Gbare_{t,\q}$ in $q$-space, and the less is the blurring of the pore  correlation function induced by the convolution. 
However, longer times result in a stronger power-law suppression of this nontrivial convolution term, whose magnitude can be estimated as $\sim [l_c/L(t)]^d$ in $d$ dimensions. This estimate comes from noting that we are essentially averaging the pore density fluctuations  over the ``diffusion volume" $L^d(t)$; for the short-range disorder in the grain placement, 
these fluctuations, $\sim 1$ at the scale $\sim l_c$, become uncorrelated at the scales much beyond $l_c$.
\keep

When a tissue  consists  of  non-exchangeable  compartments of different nature, one can tune the experiment to focus on one or the other. 
Consider an example of a tissue consisting of a \new non-confining \keep compartment  (e.g., the extra-cellular  space),  and  a  \new fully  confining, i.e., restricted \keep  one  (e.g., cells  of  size  $a$).  The  diffusional  coarse-graining in the latter stops at the time $t_a \sim a^2/D$. The whole medium possesses  two  relevant  scales,  $L(t)$  and  $a$,  which  are  markedly  different  for  long  times  when $L(t)\gg a$.  An  experimentalist  working  in  this  limit  has  a  choice  of  selecting  the  wave  number  $q$  of  diffusion  measurements  (the  strength  of  diffusion  weighting).  The  choice $q\sim 1/a$ (diffusion  diffraction  \cite{callaghan1991})  enables  measuring  the  size  of  the restricted  compartment,  but  strongly  suppresses  the  signal  from  the  permeable  one. The  choice  $q\sim 1/L(t)$  enables  observation  of  diffusion  dynamics  in  the  permeable  compartment.  The  signal  from  the  restricted  compartment  remains  unsuppressed,  such  that  both  the  signal attenuation $-\ln S \sim (qa)^2 \ll 1$ and the diffusivity $D(t)\sim a^2/t \ll 1$ become negligible; one can  then formally treat such a compartment as Gaussian with its diffusion coefficient $D\to 0$. This leads to the picture of zero-radius  ``sticks" in the brain (neurites with zero diffusivity in the transverse direction), cf. Sec.~\ref{sec:sticks} below.

\subsection{Models versus representations} 
\label{sec:model-represent}

\nin
{\bf Models are pictures} (Fig.~\ref{fig:models}), exemplifying a rough sketch of physical reality, specified  by their {\it assumptions} meant  to  simplify nature's  complexity. This simplification relies on averaging over the irrelevant degrees of freedom, and keeping only a handful of relevant parameters describing the corresponding effective theory. 
Model  assumptions are therefore a claim for the relevant parameters. They are more  important  than  mathematical  expressions,  as  they  prescribe  a  parsimonious way to think about the complexity. Model validation is thereby validation of our frame of  thinking. 

%

A {\it representation}  could  be  defined  as  a model-independent mathematical expression used   to  store,  to  compress, or to compare measurements. It can be realized as a function with a few adjustable parameters or a set of coefficients for a decomposition in a basis (cf.\ Fig.~\ref{fig:representations} for a few most commonly used representations in dMRI). In contrast to models, representations are as general as possible, and have very little assumptions. As there are infinite ways to represent a continuous function, the choice of representation  is often dictated by convenience or tradition. Practically, not all representations are equivalent because one only  uses  a  few  basis  functions  rather  than  an  infinite  set;  from  this  standpoint,  sparser  representations are more favorable.  
By construction, representations do not carry any particular physical meaning and hence do not immediately invoke any picture of physical reality; one can say that  {\bf representations are formulas.}

\new 
A detailed discussion on the choices between modeling and representing can be found in ref.~\cite{manifesto}. In this Review, we mostly focus on  models; however, there exists one fundamentally important representation that we will cover now.
\keep



\subsection{The cumulant expansion as a default representation}
\label{sec:cum}

\nin
\new \mpar{R1.1}
The ubiquitous nature of Gaussian diffusion, at least for sufficiently long times, has prompted 
a Taylor expansion \cite{Kiselev2010_diff_book,Jensen2005}: 
\keep
\be \label{cum}
\ln G(t,\q) \simeq - D_{ij}(t) q_i q_j +  \frac1{6} (\Dbar t)^2 W_{ijkl}(t) q_i q_j q_k q_l  - \dots 
\ee
in the  powers of $\q$, describing the deviation from the Gaussian form (\ref{Gbare}). 
The summation over the repeated coordinate indices $i,j, ... = 1...3$ is implied throughout; 
\new \mpar{R1.2}
$\Dbar = \frac1d  D_{ii}$ is the mean diffusivity  used for normalization. \keep \newv The symmetric tensor $W_{ijkl}$ is called the {\em diffusional kurtosis tensor}, while the {kurtosis} in a given direction, $\n$, is defined as $K(\n) = \Dbar^2 W_{ijkl} n_i n_j n_k n_l / (D_{ij}n_i n_j)^2 $ \cite{Jensen2005}. \keep 

The propagator expansion (\ref{cum}) stems from the corresponding cumulant expansion in  probability theory noticed almost a century ago by Fisher and Wishart \cite{Fisher1932,vanKampen}.
For  diffusion, only even orders in this series are nonzero due to the time-reversal symmetry in the absence of the bulk flow.
Typically, the Taylor series (\ref{cum}) converges within a finite radius in $q$ which is model-dependent  \cite{Froehlich2006,Kiselev2007}.

A general diffusion propagator will have all even cumulant terms $D_{ij}(t)$, $W_{ijkl}(t)$, $\dots$ nonzero and diffusion time-dependent \cite{Kiselev2010_diff_book,EMT}. Experimentally we often access only a few first terms, especially when using low diffusion weighting on clinical systems. \mpar{R1.3}
\new (We assume the narrow-pulse limit throughout. In Sec.~\ref{sec:2cum} we discuss in detail  how the lowest order of the ideal cumulant series (\ref{cum}) is modified by the arbitrary gradient shape).\keep

Upon coarse-graining, for a given tissue compartment the higher-order terms $W_{ijkl}, \dots$ flow to zero, such that the signal approaches the Gaussian form (\ref{Gbare}) as $t\to\infty$. In this limit, the higher-order cumulant terms of the net diffusion propagator can originate only from the partial contributions from different tissue compartments (since a sum of Gaussians is non-Gaussian).  

For any $t$, the series (\ref{cum}) {\it generates} the cumulants $\la x_j x_j \dots \ra_c$ (see e.g., refs.~\cite{Fisher1932,vanKampen,Kiselev2010_diff_book} for definition) of the PDF of molecular displacements\footnote{Since $G_{t,\r}$ is written in terms of the {\it relative} displacements $\r = \r_t - \r_0$, we  re-denote $\delta x_i(t) \to x_i(t)$ in Eq.~(\ref{Dcum}) to simplify the notation, and drop the dependence on the initial position in the view of the translational invariance property (\ref{def-G}).}  
(\ref{S=G}), via taking derivatives at $\q=0$, such as 
\be
\langle x_i x_j \rangle \equiv \int\!\d\r \, x_i x_j \, G_{t, \r} 
= - \left.{\partial^2 \over \partial q_i \partial q_j}\right|_{\q=0}  \int\!\d\r \, e^{-i\q\r} G_{t, \r} \,.
\ee
Based on such averages, it is conventional to define the {\it cumulative} diffusion coefficient 
\be \label{Dcum}
D(t) = \frac{\langle  x^2(t) \rangle}{2t} \,,
\ee 
or, more generally, the cumulative diffusion tensor 
\be
 \label{Dcumtens}
D_{ij}(t) = {\langle  x_i(t) x_j(t) \rangle \over 2 t} 
\ee
(a symmetric $3\times 3$ matrix with 6 independent parameters in 3 dimensions).
These objects are defined  in terms of the {\it average rate of change} of the mean-squared molecular displacement over the whole interval $[0, \ t]$ (in contrast to the instantaneous rate of change (\ref{Dinst}) above). 

The linear estimation problem for $D_{ij}(t)$, referred to as the 
{\it diffusion tensor imaging (DTI)}, has been solved by Basser \etal \cite{Basser1994}. 
It requires%
\footnote{\label{foot:DTI} DTI, contrary to a widespread misconception, does {\it not} assume Gaussian diffusion, as it merely provides the lowest-order cumulant term $D_{ij}$, and tells nothing about the higher-order terms in the series (\ref{cum}).  
DTI applicability is thus dictated by the kurtosis term $W$ to have negligible bias on the estimated $D_{ij}$, and the employed $b$-range
 is practically set by balancing the bias when $b$ is too large and precision loss when $b$ is too small.} 
a diffusion measurement along at least 6 non-collinear gradient directions in addition to at least one more, e.g., the $b=0$ (unweighted) image.

Likewise, the linear estimation problem for both the diffusion and kurtosis tensors, via the expansion up to $\sim q^4 \sim b^2$, 
called {\it diffusion kurtosis imaging (DKI)}, has been introduced by Jensen \etal  \cite{DKI,jensen-helpern-dki}. 
It involves the  4$^{\rm th}$ order cumulant $\la x_i x_j x_k x_l \ra_c$ related to $W_{ijkl}(t)$.
The number of parameters are now $6+15 = 21$, hence one needs at least two $b\neq 0$ shells in the $q$-space, and at least 15 non-collinear directions. 
The weights for unbiased estimation of diffusion and kurtosis tensors for non-Gaussian MRI noise were found recently \cite{veraart2013}. 



A general method to calculate the number of parameters for a given order $l_c$ of the cumulant series (\ref{cum}) in 3 dimensions is 
based on the SO(3) representation theory (known in physics as theory of  angular momentum in quantum mechanics). 
A term $\sim q^{l_c}$ of even rank $l_c$  is a fully symmetric tensor, which can be represented as a sum of the so-called symmetric trace-free (STF) tensors of ranks $l_c$, $l_c-2$, $\dots$, $2$, $0$ \cite{thorne}. Each set of $2l+1$ STF tensors of rank $l$ realizes an irreducible representation of the SO(3) group of rotations, equivalent to a set of $2l+1$ spherical harmonics $Y_{lm}$ \cite{thorne}. Hence, the total number $n_c$ of nonequivalent components in the rank-$l_c$ cumulant tensor is 
\be \label{nc}
n_c(l_c) = \sum_{l=0, 2, \dots}^{l_c} \! (2l+1) = \frac12 (l_c+1)(l_c+2) \,,
\ee
so that $n_c=6$ for DTI ($l_c=2$) and $n_c=15$ for DKI ($l_c=4$). 

Suppose we truncate the cumulant series (\ref{cum}) at an (even) term of rank $l_c = \lmax$. Hence we determine all the parameters of  cumulant tensors (diffusion, kurtosis, ...) of  ranks $2, \ 4, \ \dots,\ \lmax$. The total number of 
\new independent \keep parameters in the truncated series 
\be \label{Nc}
N_c(\lmax) = \sum_{l= 2, 4, \dots}^{\lmax} \!\!\! n_c(l) 
= \frac{1}{12}\, \lmax^3 + \frac{5}8 \, \lmax^2 + \frac{17}{12} \, \lmax
\ee
corresponding to $N_c = 6, \ 21, \ 49, \dots$ for $\lmax=2, \ 4,\ 6, \dots$.
Hence, DTI yields 6 parameters, DKI yields 21, etc.  
(Here we did not include the proton density $S|_{b=0}$ in our counting.)

The cumulants $D_{ij}$, $W_{ijkl}$, $\dots$ of the signal obtained via Taylor-expanding its logarithm in the (even) powers of $q_i$, or equivalently, in the powers of $b$, 
correspond to the cumulants of the genuine PDF of molecular displacements $\r= \r_t - \r_0$ only in the narrow pulse limit, and in the absence of the mesoscopic magnetic structure (uniform $R_2$ and $\Omega$).  
When the finite gradient pulse duration $\delta$ is comparable to the time scale of the transient processes, 
the measurement acts as a low-pass filter with a cutoff frequency $\sim 1/\delta$ \cite{callaghan-book,Kiselev2010_diff_book,Murday1968,neuman1974,VanGelderen1994muscle,burcaw2015,HHL2015,HHL2016-phantom,lee2017,Kiselev2016_NMB_review}.

\subsection{Normal or anomalous diffusion?}
\label{sec:fixedpoint}

\nin
For finite $t$, the diffusion propagator in a heterogeneous medium is never Gaussian. 
The existence of domains with slightly different ``local" $D(\r)$ at a given coarse-graining scale necessarily yields the time-dependent $\Dinst(t)$, 
as well as the higher-order terms in $q$, such as $q^4$, in the Taylor expansion of $\ln \Gbare(t,\q)$ \cite{EMT,mesopnas}.  
Upon coarse-graining, these terms gradually flow to zero, and $\Dinst(t)\to \Dinf$, such that diffusion becomes Gaussian {\it asymptotically} as $t\to\infty$ in each separate \mpar{R1.4}\new non-confining \keep tissue compartment. This was the picture of Sec.~\ref{sec:diff=cg}, cf. Fig.~\ref{fig:filter}. 
In particular, we implied that the diffusion coefficient decreases, as a result of the coarse-graining, towards its {\it finite} {tortuosity asymptote} $\Dinst(t)|_{t\to\infty} \equiv \Dinf > 0$.  How reliable is this picture? What does it take to destroy it?

Existence of finite $\Dinf$ is equivalent to mean squared displacement $\langle  x^2(t) \rangle \simeq 2\Dinf  t$ growing linearly with time for sufficiently long $t$, --- this is a direct consequence of the definition (\ref{Dinst}). One says that  diffusion asymptotically becomes ``normal", i.e., the PDF of molecular displacements over a sufficiently large $t$ approaches normal (Gaussian) distribution, cf.\ Eq.~(\ref{Gbare}) with $D\to\Dinf$. 
Of course, if there are two or more non-exchanging tissue compartments, the total distribution will be non-Gaussian (as a sum of Gaussians with different $\Dinf$), but this non-Gaussianity is in a sense trivial; the total $\Dinf$ would still exist (given by a weighted average for the corresponding compartment values) \cite{KM}, and the scaling $\langle  x^2(t) \rangle \sim  t$ at large $t$ would hold. 

There exists a radical alternative, when $\langle  x^2(t) \rangle \sim t^\alpha$ for $t\to \infty$, with exponent $\alpha \neq 1$ --- the so-called {\it anomalous diffusion} \cite{bouchaud1990}. According to the definition (\ref{Dinst}), $\Dinf = 0 $ for $\alpha < 1$ \new (sub-diffusive behavior), \keep and  $\Dinf = \infty$ for $\alpha > 1$ \new (super-diffusive behavior). \keep In other words, {\it observation of anomalous diffusion is equivalent to stating that the macroscopic diffusion coefficient $\Dinf$ does not exist}. 
(The trivial case $D(t) \sim a^2/t$ for a confining compartment of size $a$ is not considered  anomalous;  
$\langle x^2\rangle \sim a^2$,  $\alpha=0$.)

The absence of $\Dinf$ in a non-confining medium \new is always a drastic claim: it is potentially exciting yet should be thoroughly validated, because the underlying physical assumptions yielding $\alpha\neq 1$ are generally quite peculiar and exceptional, as we discuss below.  \keep
In neuronal tissue, one always observes finite $\Dinf$ in non-confining compartments (e.g., in the extra-cellular space), Section \ref{sec:Dt}, hence diffusion is empirically never anomalous \cite{mesopnas,burcaw2015,fieremans2016} for brain dMRI.%
\footnote{We are not reviewing the MRI literature on anomalous diffusion, since our  goal here is to discuss models which are relevant to observable diffusion effects in  neuronal tissue. A curious reader can find occasional claims of anomalous diffusion, or dMRI signal as a stretched-exponential. We are not aware of examples of a constructive derivation of the non-Gaussian fixed point  \cite{wilson1983,cardy-book} starting from the stationary mesoscopic disorder with properties relevant to the brain. Hence, these claims can merely be viewed as postulates ``proven" by fitting in a finite range of $t$ or $q$. 
\new If the model's functional form contradicts the physics of the signal, the estimated parameters will depend  on the range of $t$ and $q$, thereby characterizing the particular measurement scheme, rather than the tissue \cite{manifesto}.\keep}

From the point of  coarse-graining, anomalous diffusion means that the sample never quite looks homogeneous --- for example, a fractal has a self-similar structure, which implies similar statistics of static structural fluctuations at every length scale. In other words, when the coarse-graining over some scale has taken place, a larger scale looks statistically similar, so that the already averaged structural features are never forgotten, since they are reproduced again and again. In contrast, the  structure in Fig.~\ref{fig:filter} implies that this memory is forgotten for each of the two length scales, correspondingly on the two well-defined time scales. 

Tissues empirically do not look self-similar; usually, when we look at a histological slide without a scale bar, we can still roughly say at which resolution the sample is imaged because usually  cell size is well defined  (for a given tissue type) --- otherwise, medical students would not pass their pathology exams. For instance, when we look at cross-sections of white matter tracts, the majority of axons are of the order of $\sim 1\,\mu$m in diameter \cite{Lamantia1990,aboitiz1992,Tang1997,caminiti2009,Liewald2014},
and the section does not look the same when magnified by factors of 10, 100, or 0.1, 0.01, etc. A more quantitative statement can be made by studying  large-distance scaling behavior of the density-density correlation function of the tissue structure;  recent investigation \cite{burcaw2015} confirms that the structural fluctuations in white matter tracts are short-range (and not diverging at large length scales). 

When {\it can} anomalous diffusion arise? 
In a broader context, this fundamental question has been extensively studied for the Fokker-Planck equation
\be \label{FPE}
\partial_t \psi(t,\r) = \partial_\r \lb D(\r) \partial_\r \psi(t,\r) \rb  - \partial_\r \lb {\bf v}(\r) \psi(t,\r) \rb,
\ee
where in addition to the ``diffusive" flow ${\bf j}(t,\r) = -D(\r) \partial_\r \psi(t,\r)$ (Fick's law), one considers mesoscopic {\it random flow} because of some stationary local ``velocity", or ``force" field ${\bf v}(\r)$ (imagine active streams, such as vortices or currents in an ocean \cite{kravtsov1985}). Equation (\ref{FPE}) arises as a conservation law $\partial_t \psi = -\partial_\r \cdot {\bf j}$, where the total flow  
\[
{\bf j} = -D(\r) \partial_\r \psi(t,\r) + {\bf v}(\r) \psi(t,\r) \,.
\] 

It turns out that the presence of the random flow term ${\bf v}(\r)$ with short-range spatial correlations can drastically change the dynamics in dimensions $d\leq 2$ and drive the system away from the Gaussian diffusion. In dimension $d=1$, random force field causes {sub-diffusive} behavior $\langle \delta x^2\rangle \sim \ln^4 t$, a famous result by Sinai \cite{sinai1982}. 
In $d=2$ dimensions,  super-diffusive behavior occurs when the flow ${\bf v}(\r)$ is solenoidal, ${\rm div}\, {\bf v}(\r) = 0$, 
and sub-diffusive if it is potential, ${\rm curl}\, {\bf v}(\r) = 0$ \cite{fisher1984,aronovitz1984,kravtsov1985,fisher1985}.

In the absence of the random forces, ${\bf v}\equiv 0$, small fluctuations in $D(\r)$ do not destroy the ``trivial" Gaussian fixed point in dimensions $d>0$ \cite{tsai1993,lerner1993} \new (cf. footnote~\ref{ftn:fixedpoint}). \keep In other words, for the spatial short-range disorder in $D(\r)$ to become relevant (i.e., to increase under the renormalization group flow), and for the anomalous diffusion to take over, the spatial dimension should formally be $d=0$. What this tells is that it is very difficult, without the flow term, to break the  Gaussian fixed point of the finite $\Dinf$, at least starting from the weak disorder in $D(\r)$. Extremely strong disorder, which is specially tuned, can induce the {\it percolation transition} \cite{shklovskii-book} $\Dinf\to 0$; another possibility for destroying finite $\Dinf$ is to create the disorder in the mesoscopic $D(\r)$ with anomalously divergent spatial fluctuations \cite{havlin-1989,havlin-benavraham}. To the best of our knowledge, neuronal  microstructure is compatible neither with a  percolation transition, nor with diverging structural fluctuations \cite{burcaw2015}. 

\new \mpar{R1.5}
Another class of phenomena where anomalous diffusion takes place corresponds to systems with  slow dynamics, originating from a broad distribution of time scales, such that the waiting time distribution $p(\tau)\sim 1/\tau^{1+\mu}$ between  ``hops" of random walkers has a power-law tail whose first moment diverges, $0<\mu < 1$. Such broad distributions can emerge, e.g., in highly disordered amorphous solids, where escape times $\tau$ from various ``traps" for electrons are distributed as a power law, first postulated by Scher and Montroll  \cite{scher-montroll}. For the traps, the long tail in $p(\tau)$ can arise due to an exponentially strong dependence of the activation rate $\tau^{-1} \sim e^{-E/k_B T}$ on the energy barrier $E$ at temperature $T$, such that a relatively flat distrubution $\tilde p(E)$ can result in the L\'evy-like $p(\tau) = \tilde p(E(\tau))/\tau$.
Hopping with traps may lead to anomalous transport \cite{Novikov2005} and fluorescence \cite{Wang2011}. 
Anomalously slow dynamics also occurs in viscoelastic systems where elementary components are strongly coupled (Rouse polymer chain \cite{Rouse1953} of monomers tied to each other by elastic springs and undergoing Langevin dynamics). The simulated dynamics of single protein molecules \cite{Hu2016} and of colloidal tracers restricted by crowded dynamical environments such as an F-actin network \cite{Wong2004} can exhibit such a broad distribution of time scales \cite{Metzler2014}. While an active area of investigation, the anomalously slow dynamics is always characterized by strong disorder (e.g., broadly distributed traps) and/or interactions among the random walkers (e.g., parts of a polymer). 
\keep

To recap, coarse-graining over an increasing diffusion length $L(t)$ provides a physical picture for  time-dependent diffusion in mesoscopically disordered samples. 
\new This picture implies gradual ``forgetting" of the memory about the structural heterogeneities. 
In an overwhelming majority of systems, the macroscopic dynamics is characterized by a Gaussian fixed point, the absence of long-term memory, and an asymptotically normal diffusion. 
In short, {\it diffusion is almost always non-Gaussian, but almost never anomalous}. 
\keep
In the brain, it is not anomalous specifically because the density fluctuations of brain structural units do not diverge at large scales, \new traps for water molecules with broad distribution of escape times do not seem to exist, \keep
and  the ``active" flow effects (microstreaming, axonal transport) are negligible \cite{mussel2016}.



\subsection{dMRI methods beyond the scope of this review}
\label{sec:beyond}
\mpar{R1: new sub-sec}

\new
Before proceeding to review brain dMRI models, we would like to mention what we have left out,  because of limited relevance to brain dMRI as of today, and/or due to exhaustive coverage elsewhere.

On the methodological front, the leitmotif of the review is the language of coarse-graining, Fig.~\ref{fig:filter}. It is most intuitive for modeling {\it structurally disordered} systems, typical 
\newe for \keep
\new biology, cf. modeling the time-dependent diffusion in Section~\ref{sec:Dt}, based on including all the restrictions into the spatially varying $D(\r)$ in Eq.~(\ref{BT}).  
This took precedent to approaches to fully confining or periodic geometries, conventionally solved by formulating Eq.~(\ref{BT}) as the Laplace equation with boundary conditions, thoroughly reviewed in ref.~\cite{grebenkov-rmp} in the context of diffusion in porous media. 

We also left out the  physics of the {\it localization regime}, where diffusion in a strong constant gradient suppresses the signal everywhere except next to pore walls, within the gradient-dependent dephasing length $L_g = (D_0/g)^{1/3}$, which 
leads to signal decay  \cite{Stoller1991,DeSwiet1994,Hurlimann1995} $-\ln S \sim L^2(\delta)/L_g^2 \sim  D_0^{1/3}g^{2/3}\delta $.  This is an example where effects of non-narrow pulses lead to decoupling of the gradient magnitude $g$ and the gradient pulse width $\delta$ in the narrow-pulse combination $q=g\delta$. 
The ``edge enhancement" also amplifies the role of the permeability of the walls \cite{Grebenkov2014}. 
Brain structures seem to be too small for the edge effects to be relevant, but such phenomena can become important in body dMRI. 

Playing with $\delta$, e.g., using short-wide pulse combinations, \newe we or one \keep can map the Fourier transform of the  shape of the confining pore \cite{Laun2011,Kuder2013mrm}, which again requires prohibitively strong gradients for the narrow axons and dendrites in the brain, but is applicable in porous media NMR. The relevance of pulse width $\delta$ would add an extra dimension to the phase diagram in Fig.~\ref{fig:qt}. 

\newe 
Detailed review of practical aspects of dMRI measurements and biological applications are beyond our scope here. The reader is referred to the review \cite{Alexander2018} for implementation details of dMRI measurements, recent reviews of dMRI in white matter \cite{Jelescu2017} and in cancer \cite{Reynaud2017}, as well as to other articles in this Special Issue. 
\keep

%% file: Section_II.tex

\section{Time-dependent diffusion in  neuronal tissue}
\label{sec:Dt}

\epigraph{Everything should be made as simple as possible, but not simpler}{Albert Einstein}

\nin
The intuition of Sec.~\ref{sec:diff=cg} suggests that the time-dependence of the diffusion coefficient defined as either Eq.~(\ref{Dinst}) or Eq.~(\ref{Dcum}), is a hallmark of the mesoscopic structure, and the associated  time scale can be translated into the  corresponding mesoscopic length scale. Identifying $\mu$m-level tissue length scales  is the ultimate test of our ability to ``quantify microstructure" --- after all, how else would we know that we are indeed sensitive to the {\it micro}-structure? 
The  focus of this Section is  on determining tissue properties on such length scales. 

Fundamentally, observation of the time dependent overall $D(t)$ is significant because it tells that {\it diffusion is non-Gaussian in at least one of the tissue compartments}. 
\new
Indeed, at the lowest order ${\cal O}(q^2)$  of the cumulant expansion (\ref{cum}) of the signal 
$S = \sum f_i S_i$,  contributions from non-exchanging tissue compartments $S_i$ add up, such that the total diffusion coefficient is a weighted average:
\be \label{D=sumD}
D(t) = \sum f_i D_i(t) \,, \quad \sum f_i = 1\,.
\ee 
\keep
An  overall time-dependent  $D(t)$ necessarily means that at least one of $D_i$ depends on $t$. In  turn,  the time-dependent $D_i(t)$ must necessarily lead to a nonzero kurtosis and higher-order cumulants \cite{EMT} in the $i$th compartment, arising from the same mesoscopic heterogeneity which has not yet been fully coarse-grained --- and, hence, may still be  possible to quantify. 
Conversely, if diffusion is Gaussian in all tissue compartments, all $D_i=\const$, and the overall diffusion coefficient is time-independent. The overall kurtosis is then a nonzero  constant  just because a sum of Gaussians is not a Gaussian.%
\footnote{
\new
This argument can also be extended onto the regime of slow exchange between compartments, 
since Eq.~(\ref{D=sumD}) turns out to be valid in that regime in the long-$t$ limit, 
following the coarse-graining argument \cite{KM} for generalizing the K\"arger model \cite{Karger1985} 
to media with mesoscopic disorder. 
If the overall $D=\const$ (i.e., the full coarse-graining is achieved in each of the compartments), 
and the overall kurtosis $K(t)$ still depends on $t$, 
this $t$-dependence arises due to exchange \cite{DKI,KM}.  
\keep } 
\keep

We begin this Section by reviewing  experimental data on the time-dependent diffusion coefficient and kurtosis in brain, and then discuss the two physically distinct regimes of time-dependence, according to the hierarchy of Sec.~\ref{sec:hierarchy}: the short-time regime {\bf (i)}, and the long-time regime {\bf (ii)} approaching the asymptotically Gaussian diffusion in each non-exchanging  compartment.
 
Certainly, \news  we are almost never in a pure limit experimentally \keep  --- rather, we are typically in some crossover, e.g., in-between the regimes {\bf (i)} and {\bf (ii)}. However, it is still important to understand the behavior of the system in certain limits where it can be modeled with more confidence. Performing experiments in such limits provides a way to validate models through observing {\it definitive functional dependencies} on the measurement parameters \cite{manifesto}; thus-identified relevant degrees of freedom for tissues can then be incorporated into more complex theories of the crossover behavior relevant to a broader range of dMRI studies, and for clinical translation.

\subsection{Time dependent diffusion in the brain: Is there an effect?}
\label{sec:Dt-exper}

\nin
Empirically, observing time-dependence of diffusion in brain tissue has been challenging because this effect occurs at time scales associated with diffusion across length scales featuring neurites (i.e.,\ axons and dendrites). Typically, their diameters as well as the heterogeneities along them (e.g.,\ spines, beads) are of $\sim 1\,\mu$m size, hence, the corresponding diffusion times are expected to be of the order of a few ms. Such short times are quite difficult to access, especially on human systems. Besides, the time-dependence is generally {\it slow} --- which is theoretically expected due to its power-law character \cite{mesopnas}, as discussed below in Sec.~\ref{sec:long-t} --- therefore requiring a sufficiently broad range of times to detect. 

Time-dependence of the {cumulative} $D(t)$, Eq.~(\ref{Dcum}), in brain tissue has been demonstrated using pulse gradient spin echo (PGSE) in several {\it ex vivo studies} for a range of diffusion times encompassing $20 - 250\,$ms \cite{beaulieu1996,stanisz1997,assaf2000,Bar-Shir2008}.  {\it In vivo}, time-dependent diffusion in both longitudinal and transverse directions was also observed in rat corpus callosum at $t$ ranging from 9 to 24 ms \cite{kunz2013}, though another study yielded no change in the mean diffusivity of healthy and ischemic feline brain with respect to $t$ between $20 - 2000\,$ms \cite{vangelderen1994}. 

In the human brain, it has been unclear for quite some time whether {\it in vivo} time-dependent diffusion properties can be observed. While several {\it in vivo} studies report no observable change over a broad range of diffusion times  \cite{clark2001,nilsson2009}, Horsfield \etal \cite{horsfield1994} reported time-dependent diffusion in several white matter regions at times ranging from 40 to 800 ms. Very recently, {\it in vivo} pronounced time-dependence in the longitudinal diffusivity and less pronounced time-dependence in the transverse diffusivity has been reported in several WM tracts of healthy human volunteers for relatively long diffusion times, $t = 45 - 600$\,ms, on a standard clinical scanner using stimulated echo acquisition mode (STEAM)-DTI \cite{fieremans2016}. Subsequently, a similar effect in the transverse direction to WM tracts was observed with STEAM-DTI in the range $t=48-195\,$ms \cite{desantis2016}.  



Oscillating gradient spin echo (OGSE) diffusion-weighted sequences are able to probe shorter diffusion time scales compared to conventional PGSE, and have clearly demonstrated time-dependent diffusion in the brain, including the observation of time-dependent diffusivities {\it in vivo} in normal and ischemic rat brain cortex \cite{Does2003}, as well as  {\it ex vivo} in rat WM tracts \cite{xu2014}. By combining OGSE and PGSE, Pyatigorskaya \etal \cite{pyatigorskaya2014} observed time-dependent diffusion coefficient and a non-monotonic time-dependent kurtosis (with a maximum value $K\approx 0.6$ at $t\approx 10\,$ms) in healthy rat brain cortex at 17.2\,T, and Wu and Zhang \cite{wu2014,wu2016} recently observed time-dependence in mouse cortex and hippocampus. In humans, Baron and Beaulieu \cite{baron2014} found eight major WM tracts and two deep gray matter areas to exhibit time-dependent diffusion using OGSE and PGSE, and Van \etal \cite{van2014} reported a similar effect with OGSE in human corpus callosum. Furthermore, works using double diffusion encoding (cf.\ Section \ref{sec:mpfg}) indirectly point at the time-dependent nature of diffusion in brain tissue. 


Overall, while it is common to assume that diffusivities are approximately diffusion time-independent for $t \gtrsim10\,$ms, the experimental data described above clearly demonstrates  time-dependent diffusion both at short and long times. In what follows, we describe the underlying theory for both limits and discuss the corresponding biophysical interpretation and potential for  applications.

\begin{figure*}[t]
\includegraphics[width=6in]{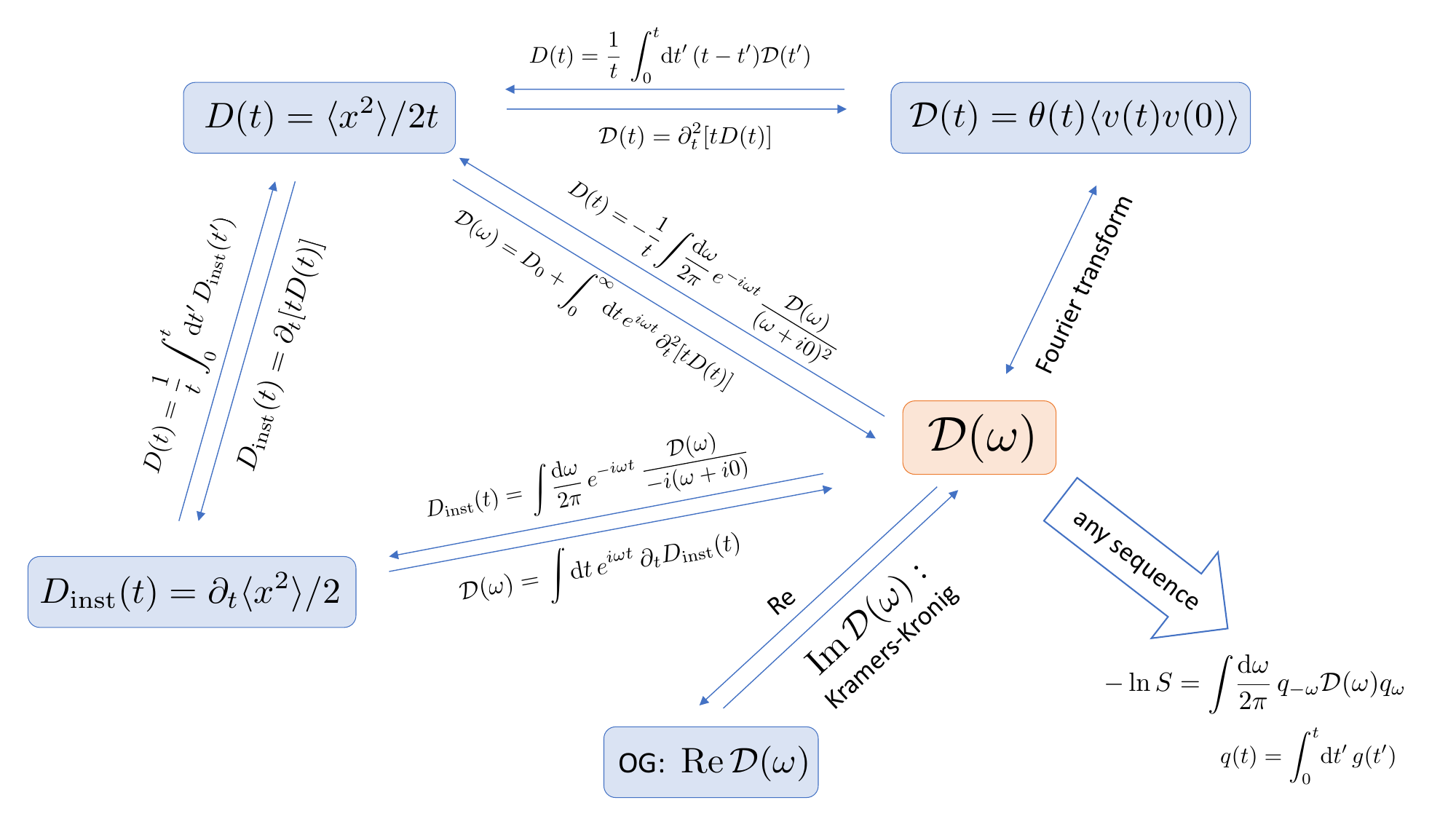}\\
\caption{General relations between the basic diffusion metrics: $\D(\w)$, $\D(t)$, $\Dinst(t)$ and $D(t)$, and the signal attenuation up to ${\cal O}(q^2)$.}
\label{fig:Drelations}
\end{figure*}

\subsection{The second-order cumulant}
\label{sec:2cum}
\mpar{R2: rewritten Sec.~\ref{sec:2cum}, added GPA, Fig.~\ref{fig:Drelations}, etc.}

\subsubsection{Gaussian phase approximation}

\nin
In Sec.~\ref{sec:qtI}, we derived a general relation (\ref{S=G}) between the dMRI signal and the diffusion propagator in the narrow-pulse limit. 
For gradient pulses $g(t)$ of arbitrary shape, there is no such simple relation; 
the signal $S[{\bf g}(t)]$ is a functional of ${\bf g}(t)$ (i.e.,\ a mapping of a function to a number). 
To obtain an explicit dependence of $S$ on ${\bf g}(t)$, one treats the gradient term in Eq.~(\ref{BTD}) perturbatively in ${\bf g}(t)$, generalizing the cumulant series (\ref{cum}). Here, we will stay at the level of  ${\cal O}(g^2)$, the so-called Gaussian phase approximation (GPA), and describe the family of diffusion coefficients which define the second-order cumulant and carry the same information content, yet can be accessible using different techniques, Fig.~\ref{fig:Drelations}. 

The GPA approximates the dMRI signal  \cite{callaghan-book,Kiselev2010_diff_book,Kiselev2016_NMB_review} 
\be \label{S=cum}
S[{\bf g}(t)] = \langle e^{-i\varphi} \rangle \approx e^{-\frac12 \langle \varphi^2\rangle} 
\ee
up to the second cumulant of the accumulated phase 
\be \label{phase}
\varphi(t) = \int_0^t\! \d t'\, {\bf g}(t') \r(t') = - \int_0^t\! \d t' \q(t') \v(t') \,,   
\ee
where we introduced the time-dependent wave vector 
\be \label{q}
\q(t) = \int_0^t\! \d t' {\bf g}(t') 
\ee
such that the Larmor frequency gradient ${\bf g}$ is given by its time derivative, ${\bf g} = \partial_t \q$. 
The first-order cumulant $\langle \varphi\rangle\equiv 0$ in the absence of the net flow. 
The balanced gradient condition sets $\q(t)|_{t>T}\equiv 0$ at the end $t=T$ of the gradient train interval. 
Eq.~(\ref{q}) generalizes the definition of $\q$ for narrow-pulse gradients, where $\q$ remained constant during an interval 
$0<t<T$, cf.  the propagator Eq.~(\ref{S=G}) with $t=T$. 
Writing $\langle \varphi^2\rangle$ as a double integral, and averaging over the Brownian paths, we obtain 
\be \label{cumt}
-\ln S[q(t)] \simeq \frac12 \int_0^T\! \d t_1 \d t_2\, q(t_1) \la v(t_1)v(t_2)\ra q(t_2) \,,
\ee
where we from now on dropped the explicit vector notation of $\q$ and $\v$ 
(the corresponding tensor indices can be easily restored; one can think about isotropic media for simplicity). 

We can see that the diffusion process at the $\O(q^2)$ level is fully characterized by the autocorrelation function 
$\la v(t_1)v(t_2)\ra$ of molecular velocity, an even function of $t_1-t_2$ in stationary media due to time translation invariance and time reversal symmetry of the Brownian motion. 

For uniform media, $\la v(t_1)v(t_2)\ra = 2D_0\delta(t_1-t_2)$, which can be thought of as one of the equivalent definitions of the diffusion constant $D_0$. Technically, there is no such thing in nature as a zero-width $\delta(t_1-t_2)$; we can use this approximation for simple liquids since the correlation time for forgetting the memory about the molecular collisions is of the order $\sim 1-10\,$ps, orders of magnitude below our ms-level time scales. 
We can say that diffusion in simple liquids is thereby Markovian (has no memory) on the relevant NMR time scales. 
This leads to  the standard expression $-\ln S = bD_0$ with $b=\int_0^T\! q^2(t) \d t$, generalizing Eq.~(\ref{Gbare}).

\subsubsection{The dispersive diffusivity}

\nin
For general mesoscopic media, microstructure introduces temporal correlations in  positions and velocities of random walkers. For instance, if a walker just hit a wall, then its velocity will correlate negatively with the velocity just before the hit (since reflection and moving away from the wall is preferred), and this memory will last during the time depending on the wall geometry and the presence of other restrictions. To characterize such correlations,  let us introduce the {\it retarded} velocity autocorrelation function \cite{EMT}
\be \label{Dt=vv}
\D(t) \equiv \theta(t) \, \langle v(t) v(0)\rangle ,
\ee
where $\theta(t)$ is the unit step function, cf.\ Sec.~\ref{sec:qtI}. In terms of $\D(t)$, Eq.~(\ref{cumt}) reads
\be \label{calDqq}
-\ln S[q(t)] \simeq \int\! \d t_1 \d t_2\, q(t_1)\D(t_1-t_2) q(t_2) \,,
\ee
where the double integration can be extended over all real values of $t$, 
since $q(t)$ is nonzero only on a finite interval. 

The time translation invariance of $\D$ allows us to rewrite the double intergral in the $t$-domain as a single integral in the  $\w$-domain, by introducing the Fourier transform of $\D(t)$, the {\it dispersive diffusivity}%
\footnote{
The real part, $\Re\D(\w)$, corresponds to the quantity called ``$D(\w)$" in the NMR literature \cite{Stepisnik1981}. 
} 
\cite{EMT,sv-og}
\be \label{Dw}
\D(\omega) = \int_0^\infty \! \d t\, e^{i\omega t}\, \langle v(t) v(0)\rangle \,.
\ee
Eq.~(\ref{calDqq}) can now be written in terms of the  Fourier-transformed $q_\w=\int\! \d t\, e^{i\w t} q(t)$, as%
\footnote{For anisotropic media, and for arbitrary $q$-space trajectories \cite{RN1646,Topgaard2017}, 
the integrands in Eqs.~(\ref{calDqq})--(\ref{GPA}) are $q_{i}(t_1) \D_{ij}(t_1-t_2) q_j(t_2)$, $v_i(t)v_j(0)$ 
and $q_{-\w, i} \D_{ij}(\w) q_{\w,j}$ respectively (with the sums over repeated indices). 
} 
\be \label{GPA}
-\ln S[q_\w] \simeq \int\! {\d\w\over 2\pi}\, q_{-\w} \D(\w) q_\w \,.
\ee
Here, only $\Re \D(\w)$ contributes, as $\Im \D(\w)$, odd in $\w$, yields zero after being integrated with an even function $|q_\w|^2$. Equivalently, $\Im \D(\w)$ does not contain extra information as it can be restored using the Kramers-Kronig relations \cite{Landau5}. 

The representation (\ref{GPA}) underscores that, knowing the velocity autocorrelator $\D(\w)$, one can evaluate the diffusion-weighted signal  
to  ${\cal O}(g^2)$ for any gradient waveform $g(t)$. Conversely, by selecting a particular form of $q(t)$ according to its Fourier representation $q_\w$, one effectively allocates particular weights to different Fourier harmonics $\D(\w)$ 
contributing to the measured signal (\ref{GPA}). 


The dispersive diffusivity (\ref{Dt=vv}) and (\ref{Dw}), and the cumulative (\ref{Dcum}) and instantaneous (\ref{Dinst}) diffusion coefficients, are related to each other via non-local transformations in the time domain
\bea \label{Dinst=D}
\Dinst(t) &=& {\partial \over \partial t} \lb t D(t) \rb ,
\\ \label{Dw=Dt}
\D(t) &=&  {\partial\over \partial t} \Dinst(t) = {\partial^2\over \partial t^2} \left[ t D(t)  \right]  \,,
\eea
and in the frequency domain%
\footnote{\newv The addition of $i0$ in the denominators preserves  causality (retarded response character) of integrated quantities, see Appendix \ref{sec:app-G} for details.} 
\cite{EMT,sv-og,Kiselev2016_NMB_review}, Fig.~\ref{fig:Drelations}: 
\bea \label{Dinst=Dw}
\Dinst(t) &=& \int\! {\d\omega\over 2\pi}\, e^{-i\omega t} \, {\D(\omega) \over -i(\omega+i0)} \,, 
\\ \label{D=Dw}
D(t) &=& \frac1t\, \int\! {\d\omega\over 2\pi}\, e^{-i\omega t} \, {\D(\omega) \over [-i(\omega+i0)]^2} \,. 
\eea
Conversely, the dispersive diffusivity $\D(\w)$ can be found either by a Fourier transform (\ref{Dw}) of the retarded velocity autocorrelator $\D(t)$, 
Eq.~(\ref{Dt=vv}), or from
the time-dependent diffusion coefficient (\ref{Dcum}), measured by ideal narrow-pulse gradients, via
\be \label{Dw=Dt}
\D(\w) = D_0 + \int_0^\infty\! \d t\, e^{i\w t} \partial_t^2 \lb t D(t) \rb ,
\ee
where $D_0 \equiv D(t)|_{t=0}$ (cf.\ Eq.~(D3) in Appendix D of ref.~\cite{EMT}). 
These relations are summarized in Fig.~\ref{fig:Drelations}.

We underscore that the three diffusion metrics: the dispersive diffusivity $\D(\w)$; the retarded velocity autocorrelator $\D(t)$; and the time-dependent diffusion coefficient $D(t)$ contain the same amount of information about the medium, and thus can be expressed via each other \cite{EMT}.  
However, the practical feasibility of their measurement may differ greatly.  
Generally speaking,  long times are most conveniently accessed using pulse-gradient or stimulated echo-based methods \cite{Merboldt1985}, while short times are best measured in the frequency domain using {\it oscillating gradients}.

\subsubsection{Oscillating  gradients}
\label{sec:og-pg}

\nin
The oscillating gradient (OG) method, typically with a refocussing pulse in the middle of the periodic gradient train (OGSE), was pioneered by Gross and Kosfeld in 1969 \cite{gross-kosfeld}, and was first utilized to measure properties of biological tissue (muscle) by Tanner in 1979 \cite{tanner1979} and applied to porous media later on \cite{Stepisnik1993,Callaghan1995}.   
This sequence is useful for accessing short diffusion times, since the diffusion weighting accumulates over $N$ oscillation periods, $b = N b_1$ \cite{Does2003,sv-og}, cf.\ Eq.~(\ref{Sgen-D}) in Appendix \ref{sec:app-ogse}. 
In this way, the (short) diffusion time $\sim 2\pi/\omega$ 
gets decoupled from the (long) duration $T = N \cdot 2\pi/\omega$ of the total gradient train. 
It can be shown \cite{EMT,sv-og}, that in the limit of large number $N\gg 1$ of oscillations, OGSE measures the real part
$\Re \D(\omega)$ of the dispersive diffusivity (\ref{Dw}). 
\new In Appendix \ref{sec:app-ogse}, we derive the 2nd-order cumulant expression in terms of $\D(\w)$ 
for OGSE  with finite $N$.\keep%

To compare pulse gradient with oscillating gradient methods, a practical question arises: What is the diffusion time in terms of the OGSE frequency (and vice-versa)? How can we plot  results of both types of measurements in the same axes? 

Unfortunately, in the view of relations (\ref{Dinst=D}) -- (\ref{Dw=Dt}), there is no universal answer to the above question. The relation between $D(t)$ and $\Dinst(t)$ on the one hand, and $\D(\omega)$ on the other, is mediated by the Fourier transform, which is {\it non-local} in $\omega$. 
In other words, the conversion between $\omega$ and $t$ depends on the {\it functional form} of either diffusivity --- i.e., on the tissue properties.  
Without understanding the system's physics (embodied by   the functional form of the diffusivity), we are limited to the relations between macroscopic properties:  
\be \label{Dinf}
\D(\omega)|_{\omega\to 0} = D(t)|_{t\to \infty} = \Dinst(t)|_{t\to\infty} \equiv \Dinf \,. 
\ee

Below, as we describe different models, we will demonstrate examples of such conversion for different  functional forms of $\D(\omega)$ and $D(t)$, e.g., Eqs.~(\ref{t-f-Mitra}) and (\ref{t-f-ii}).

\subsection{The short-time limit, regime {\bf (i)}: \\ Net surface area of restrictions}
\label{sec:short-t}

\subsubsection{Theory}

\nin
The qualitative picture of the $S/V$ limit \cite{Mitra1992} was given in Sec.~\ref{sec:hierarchy}{\bf (i)}. 
Quantitatively, the short-$t$ expansion of the diffusion coefficient (\ref{Dcum}) proceeds in  powers of 
$L(t)S/V$, where $L(t)= \sqrt{2D_0t}$ is the diffusion length: 
\be \label{Dt-Mitra}
D(t) = D_0 \lp 1 - {4 \over 3\sqrt{\pi}\, d} {S\over V} \sqrt{D_0 t} + \O(t) \rp ,
\ee
and $S/V$ is the surface-to-volume ratio of the restrictions in $d$ spatial dimensions. 
Identifying the $\sqrt{t}$ term practically involves very short diffusion times. 
Even for a red blood cell suspension, this  limit was barely observable in the time domain \cite{latour-pnas}; for the brain, with structural features even smaller than the red blood cell size, getting to this limit using PGSE  is practically impossible due to very low $b$-values for short $t$.

Hence, regime {\bf (i)} is best accessed using OGSE. 
The corresponding functional form of $\D(\omega)$  for Eq.~(\ref{Dt-Mitra}) was recently derived in the $N\to \infty$ limit
\cite{sv-og}
\be \label{Dw-Mitra}
\Re \D(\omega) \simeq D_0 \lp 1 - {1 \over d\sqrt{2}} {S\over V} \sqrt{D_0\over \omega}  \rp .
\ee
\new\mpar{R2 -- main issue}
For a finite total number $N$ of oscillations, Eq.~(\ref{Dw-Mitra}) is modified, see Appendix~\ref{sec:app-SV}, by a correction factor $c$, 
Eq.~(\ref{Mitra-c}), in front of the $1/\sqrt{\omega}$ term.
This factor approaches its $N\to\infty$ limit $c\to1$ rather fast, $c-1\sim 1/N$, 
such that $c-1<0.05$ as long as $N\geq 4$ for the $\cos$, and $N\geq 10$ for the $\sin$ waveforms. 

From directly comparing Eqs.~(\ref{Dt-Mitra}) and (\ref{Dw-Mitra}), the relation between OGSE frequency $\nu=\omega/2\pi$ and diffusion time $t=\Delta$ (in the narrow-pulse PGSE limit) is as follows \cite{sv-og}: 
\be \label{t-f-Mitra}
\mbox{$S/V$ \ limit \ {\bf (i)}:} \quad t = {9\over 64} \cdot \frac1\nu \,.
\ee

We note that Eq.~(\ref{t-f-Mitra}) differs from the  empirical relation (see, e.g., ref.~\cite{Does2003})
\be \label{t-f-wrong}
\mbox{wrong yet widely used:} \quad t = \frac1{4\nu} \,, \qquad
\ee
which in fact is almost always incorrect [cf.\  Eq.~(\ref{t-f-ii}) below]. 
Relation  (\ref{t-f-wrong}) originates from matching the $b$-value between one OGSE period and PGSE of the same duration. 
Since the whole notion of the $b$-value stems from Gaussian (i.e.,\ time-independent) diffusion, it is not surprising that merely matching the diffusion attenuation between PGSE and OGSE for the constant $D$ falls below the accuracy needed to define the diffusion time for the nontrivial, time-dependent case.%
\footnote{\label{foot:difftime}%
The often quoted relation $t = \Delta - \delta/3$ for the diffusion time of a finite-width PGSE is a myth for the same reasons. 
One can only say that the measurement gives $D(t)$ with $t\approx \Delta$ (with the accuracy of this approximation controlled by $\delta$). More rigorously, the effect of finite pulse width $\delta$ creates a low-pass filter \cite{callaghan-book,Kiselev2016_NMB_review} on $\D(\omega)$, whose effect is again model-dependent, see, e.g., Eq.~(24) in ref.~\cite{burcaw2015}, as well as refs.~\cite{HHL2015,lee2017,HHL2016-phantom,Dhital2017Da}, for the examples of this filter effect on the  models of $\D(\omega)$ relevant for brain. 
}

\subsubsection{Applications}

\nin
Probing the short-time limit either in the time domain (Eq.~(\ref{Dt-Mitra})) or the frequency domain (Eq.~(\ref{Dw-Mitra})) potentially allows for decoupling the geometric effects of the surface-to-volume ratio $S/V$ and the  free diffusivity $D_{0}$. Recently this limit has been demonstrated in phantoms using both PGSE \cite{Papaioannou2017} as well as OGSE \cite{Lemberskiy2017}. For {\it in vivo} brain measurement, OGSE provides the most practically feasible method, with the $1/\sqrt{\omega}$ dependency as the signature functional form (\ref{Dw-Mitra}) of this regime. In the healthy brain, this signature has so far never been observed, since presumably the achievable oscillation frequencies are still too low as compared to those needed to identify the effect of the restrictions from neurite walls with typical radius of curvature $\sim 1\,\mu$m, requiring diffusion times much below 1\,ms (i.e., frequencies $\nu \gg 1\,$kHz).

The search for the $1/\sqrt{\omega}$ regime has prompted using brain tumors with roughly spherical cells of larger size (about $10\,\mu$m), such that the required frequency range can be potentially accessible. Recently, the $1/\sqrt{\omega}$ functional form was observed by Reynaud \etal \cite{Reynaud2016} in a mouse glioma model in the frequency range up to $\omega/2\pi = 225\,$Hz, which for the first time enabled the separation between the geometric ($S/V$) and ``pure" diffusive ($D_0$) tumor features. Further combining the OGSE and PGSE methods has lead to the POMACE \cite{pomace} and IMPULSED \cite{impulsed} methods for quantifying cell size and extra-cellular water fraction, cf.\  topical review \cite{Reynaud2017}. \keep

\subsection{Approaching the long time limit, regime {\bf (ii)}: \\ Structural correlations via gradual coarse-graining}
\label{sec:long-t}

\subsubsection{Theory}

\nin
Over time,  random walkers probe the spatial organization of  the sample's microstructure, which makes the time-dependence of the diffusion metrics intricately tied to an increasingly large number of structural characteristics. Technically,  finding $D(t)$ or $\D(\w)$ analytically in a realistic complex sample is nearly impossible as it amounts to including the contributions from the spatial correlations of the local diffusion coefficient $D(\r)$ and of the positions of all restrictions up to an infinitely high order. 


The intuition based on coarse-graining, Sec.~\ref{sec:diff=cg}, turns out to be helpful in solving this  problem in the long time regime \cite{mesopnas}, 
when the diffusion coefficient (\ref{Dinst}) gradually approaches its macroscopic (tortuosity) value (\ref{Dinf}).
As mentioned in Sec.~\ref{sec:diff=cg},  in the limit  $t\to\infty$, any non-confining tissue compartment effectively looks completely uniform. 

Let us step back just a bit from $t\to \infty$ and consider $t$ long enough (yet finite) for the sample to look  {\it almost} homogeneous from the point of the diffusing molecules, Fig.~\ref{fig:filter}, --- no matter how heterogeneous it is in reality
(e.g., at the cellular scale).  
In this limit, the problem of finding the diffusion propagator maps onto a much simpler problem of finding the diffusion propagator in a {\it weakly heterogeneous} medium (which is the corresponding effective theory), characterized by the diffusion equation (\ref{DE}) with 
\be \label{small}
{ |\delta D(\r)| \over \Dinf }\ll 1 \,, \quad \delta D(\r) \equiv D(\r) -\Dinf \,.
\ee 
This problem admits a perturbative solution \cite{EMT,mesopnas}, with Eq.~(\ref{small}) defining a small parameter, as long as the macroscopic (tortuosity) limit (\ref{Dinf}) exists,  $0<\Dinf<\infty$  (i.e., diffusion is not anomalous, which is practically always the case for dMRI in tissues, cf.\ Sec.~\ref{sec:fixedpoint}). The lowest order in $\delta D(\r)$ vanishes, and the second order in the parameter (\ref{small}) yields  
\be \label{Dinst(t)}
D_{\rm inst}(t) \simeq \Dinf + \frac1d \, {\la (\delta D(\r))^{2}\ra|_{L(t)} \over \Dinf} 
\ee
in $d$ spatial dimensions.

The last term in Eq.~(\ref{Dinst(t)}) involves the variance of the slowly-varying $D(\r)$ {\it at a given coarse-graining length scale} defined \news by \keep  the diffusion length $L(t)$. This variance {\it decreases} as a result of {\it self-averaging}, i.e., when different diffusing molecules on average begin to experience more and more similar mesoscopic structure with an increasing $L(t)$, such that 
any sample begins to approximate the  ensemble of different disorder realizations of $D(\r)$ more and more precisely. 
The always positive ``fluctuation correction" to $\Dinf$ (the last term) elucidates why the diffusion coefficient can only decrease with $t$; observation of its increase with diffusion time is a red flag for imaging artifacts. 

To be more rigorous, Eq.~(\ref{Dinst(t)}) can be expressed as \cite{mesopnas}
\be \label{Dinst(t)-Gamma}
D_{\rm inst}(t)|_{t\gtrsim t_0} \simeq \Dinf + \frac1{\Dinf \, d} \, \int\! {\d^d\k \over (2\pi)^d}\, \Gamma_D(\k) \, e^{-\Dinf k^2 t}  \,, 
\ee
in terms of the power spectrum $\Gamma_D(\k) = \int\! \d\r\, e^{-i\k\r}\, \Gamma_D(\r)$ of the underlying effective $D(\r)|_{L(t_0)}$ 
coarse-grained over the diffusion length $L(t_0)$ corresponding to some sufficiently long time scale $t_0$, for which the relative deviation (\ref{small}) from $\Dinf$ is sufficiently small. 
The correlation function
\be \label{Gamma-D}
\Gamma_D(\r) = \la \delta D(\r+\r_{0}) \delta D(\r_{0})\ra_{\r_{0}}  
\ee
embodies the fluctuation correction in Eq.~(\ref{Dinst(t)}). 
We can see that diffusion  indeed acts as a Gaussian filter (cf.\ Fig.~\ref{fig:filter} in Sec.~\ref{sec:diff=cg}), with a filter width 
$\sim L(t) \sim \sqrt{\Dinf t}$, over the effective medium defined via the  correlation function of the weakly heterogeneous $D(\r)$. 

Hence, for sufficiently long $t$, Eqs.~(\ref{Dinst(t)}) and (\ref{Dinst(t)-Gamma}) become {\it asymptotically exact} with $L(t)\to\infty$, 
no matter how strongly heterogeneous the ``true" (microscopic) $D(\r)$ is. 
From the renormalization group flow standpoint, we can say that the time-dependent corrections  (last terms of Eqs.~(\ref{Dinst(t)}) and (\ref{Dinst(t)-Gamma})) to the asymptotically Gaussian propagator become {\it irrelevant} as a result of  integrating out the fluctuations of the locally varying $D(\r)$ over larger and larger scales. Likewise, the kurtosis and higher-order cumulants in this compartment will decay to zero, as governed by  similar fluctuation terms.

How to relate the time-dependence (\ref{Dinst(t)}) and (\ref{Dinst(t)-Gamma}) to the mesoscopic structure? 
Here, one realizes \cite{mesopnas} that the coarse-grained $D(\r)|_{L(t)}$ depends on the similarly coarse-grained local density $n(\r)|_{L(t)}$ of mesoscopic restrictions to diffusion (e.g., the disks in Fig.~\ref{fig:filter}). 
Hence, the variance of $D(\r)|_{L(t)}$ entering Eq.~(\ref{Dinst(t)})  is proportional to a typical  density fluctuation $\la (\delta n)^{2}\ra|_{L(t)}$ of the restrictions in a volume of size $L^d(t)$ in $d$ dimensions (this becomes valid when the deviations 
$\delta n(\r)|_{L(t)} = n(\r)|_{L(t)}-\langle n \rangle$ from the mean sample density $\langle n \rangle$ become small). 
This proportionality, asymptotically exact at small $k$ (i.e., after coarse-graining over large distances, 
cf. Eq.~(\ref{Dinst(t)-Gamma}) for long $t$), leads to the proportionality 
\be \label{Gamma-Gamma}
\Gamma_D(k) \propto \Gamma(k)  \,, \quad k\to 0
\ee
between the correlation functions (power spectra) of $D(\r)$  
and of the underlying structure $n(\r)$, 
\be \label{Gamma}
\Gamma(\r) = \la n(\r+\r_{0})n(\r_{0})\ra_{\r_{0}} \,.
\ee 

The structural correlation function can behave qualitatively distinctly at large distances, i.e., small $k$:
\be \label{Gamma=kp}
\Gamma(k) \sim k^p\,, \quad k\to 0 \,.
\ee
The {\it structural exponent} $p$ in Eq.~(\ref{Gamma=kp}) defines the {\it structural universality class} to which a sample belongs, according to its large-scale structural fluctuations embodied by its correlation function (\ref{Gamma}). 
The greater the exponent $p$, the more suppressed are the structural fluctuations at large distances (low $k$); conversely, negative $p$ signify strong disorder, where the fluctuations are stronger than Poissonian (for which $p=0$). 
Hence, $p$ characterizes global structural complexity, taking discrete values robust to local perturbations. 
This enables the classification  of mesoscopic disorder \cite{mesopnas}, and its relation to the Brownian dynamics, as we now explain.  

From Eq.~(\ref{Dinst(t)-Gamma}) it directly follows that the time-dependent { instantaneous} diffusion coefficient 
(\ref{Dinst}) approaches the finite bulk diffusion constant $\Dinf$ as a {\it power law} 
\be \label{Dinst-approach}
D_{\rm inst}(t)
 \simeq \Dinf + \mbox{const}\cdot t^{-\vartheta} \,, \quad \vartheta>0 \,,
\ee
with the {\it dynamical exponent} \cite{mesopnas}: 
\be \label{theta=p+d}
\vartheta= (p + d)/2  
\ee
related to the statistics of large scale structural fluctuations via the structural exponent $p$, and to the spatial dimensionality $d$. 

To illustrate the above general relations, consider Poissonian disorder (uncorrelated restrictions, e.g., completely randomly placed disks in Fig.~\ref{fig:filter}). Their density fluctuation within the ``diffusion volume" $L^d$ scales as the inverse volume, 
$\langle (\delta n(\r))^2 \rangle \sim 1/L^d(t) \sim t^{-d/2}$ according to the central limit theorem. Equivalently, $\Gamma(k) \to \mathrm{const} \sim k^0$ as $k\to 0$, i.e., \news the \keep  exponent $p=0$. 
As a result,  when restrictions are uncorrelated (or, more generally, {\it short-range disordered}, i.e., have finite correlation length in their placement), the instantaneous diffusion coefficient approaches its macroscopic limit as
\be \label{Dinst-uncorr}
D_{\rm inst}(t) \simeq \Dinf  + \mbox{const} \cdot t^{-d/2} 
\ee
in $d$ dimensions, i.e., $\vartheta=(0+d)/2$. This is the intuitive picture behind the formal results \cite{Ernst-I,Visscher}. 

The approach described in ref.~\cite{mesopnas} generalizes this picture onto any universality class of structural disorder and enables identifying relevant structural fluctuations by measuring the dynamical exponent (\ref{theta=p+d}). This exponent manifests itself in the power law tail of the molecular velocity autocorrelation function (\ref{Dt=vv})
\be \label{vaf}
\D(t)  \sim t^{-(1+\vartheta )} 
\ee
and in the dispersive diffusivity,%
\footnote{The dispersive terms reads $i\omega \ln (-i\omega)$ for the special case of $\vartheta=1$, 
hence $\Re \D(\omega)$ will depend on $\omega$ as $|\omega|$, 
cf.\ ref.~\cite{burcaw2015}.}
Eq.~(\ref{Dw}), 
\be \label{Dw-approach}
\D(\omega) 
 \simeq \Dinf + \mbox{const}\cdot (i\omega)^{\vartheta} 
\ee
whose real part is accessible with OGSE, Sec.~\ref{sec:og-pg}. 

Relation \eqref{theta=p+d} provides a way to determine the exponent $p$ (or the effective dimensionality $d$) and, thereby, the  
structural universality class, 
using any type of macroscopic time-dependent diffusion measurement.
Local properties, contributing to biological variability, affect the non-universal coefficients, e.g., the values of $\Dinf$ and  the prefactor of $t^{-\vartheta}$ in 
Eq.~(\ref{Dinst-approach}), but not the exponent $\vartheta$. The latter exponent is {\it universal}, i.e.,\ is independent of microscopic details, and is robust with respect to variations between samples of a similar \news nature\keep . 
From the point of dMRI in biological tissues, 
the {\it exponent (\ref{theta=p+d}) is robust with respect to biological variability}.  

We can also see that the stronger the fluctuations (the smaller the exponent $p$), the smaller is the dynamical exponent $\vartheta$, i.e., the  slower is the approach to $\Dinf$. Physically, this happens because it takes longer for the coarse-graining to self-average the sample's structural fluctuations. Conversely, if a sample is regular (a periodic lattice, formally equivalent to $p\to \infty$), the approach of $\Dinf$ will happen exponentially fast (i.e., faster than any finite inverse power law) \cite{mesopnas}. 

The above approach exemplifies the power of an effective theory way of thinking, where, to make fairly general statements about the relation between the diffusive dynamics and the structural disorder, we did not have to solve the full nonperturbative problem (starting from the microscopic restrictions $n(\r)$), but instead ended up solving a relatively simple problem of finding lowest-order corrections \cite{EMT,mesopnas} to Gaussian diffusion in a weakly heterogeneous medium. 

\newv
Note that the undefined constants in Eqs.~(\ref{Dinst-approach}) and (\ref{Dw-approach}) are different. It is possible to find a more precise correspondence between the time-dependent terms in $D_{\rm inst}(t)$ and $\D(\omega)$ by using Eq.~(\ref{Dinst=Dw}) \news followed by contour integration in the complex plane of $\omega$, yielding \keep
\be \label{t_to_omega}
t^{-\vartheta} \longleftrightarrow 
 {\pi/2 \over \Gamma(\vartheta) \, \sin {\pi \vartheta \over 2}} \cdot \omega^\vartheta \,, \quad \vartheta < 1 
\ee
(here $\Gamma(\vartheta)$ is Euler's $\Gamma$-function; cf.\ also ref.~\cite{mesopnas}, compare Supplementary Eqs.~[S17] and [S18].)

The above ``conversion" between PGSE and OGSE works for $\Dinst(t)$. However, the cumulative $D(t)$ follows the behavior (\ref{Dinst-approach}) only for $\vartheta < 1$;   for $\vartheta \geq 1$, the PGSE $D(t)$ expansion at long $t$ will begin with the $1/t$ term, due to the integral in inverting the relation (\ref{Dinst=D}), 
\be \label{D-Dinst}
D(t) = \frac1t\, \int_0^t \! \d\tau\, \Dinst(\tau)
\ee
converging at short $t$ for the $t^{-\vartheta}$ term with $\vartheta>1$ in $\Dinst(t)$. 
Therefore, the structure-specific dynamical exponent (\ref{theta=p+d}) is masked in PGSE if it exceeds unity; to reveal it, one has to use $\Dinst(t)$, which amounts to differentiating noisy experimental data \cite{mesopnas,Papaioannou2017}. The borderline case $\vartheta = 1$ has been considered in detail in ref.~\cite{burcaw2015}; 
the PGSE diffusion coefficient has a $(\ln t) /t$ tail due to the logarithmically divergent integral in Eq.~(\ref{D-Dinst}), 
\be \label{Dt-theta=1}
D(t) \simeq \Dinf + A \cdot {\ln (t/\tilde t_c) \over t} \,, \quad t \gg \tilde t_c \sim {\rm max}\, \{ t_c , \ \delta \} \,, 
\ee
whereas the OGSE counterpart is given by
\be \label{Dw-theta=1}
\Re \D(\omega) \simeq \Dinf + A \cdot {\pi \over 2} |\omega| \,, \quad |\omega| t_c \ll 1\,.
\ee
Here $t_c\sim l_c^2/\Dinf$ is the time to diffuse across the correlation length of the corresponding disordered environment (e.g., correlation length of the disordered axonal packing \cite{burcaw2015} in the case $p=0$ and $d=2$ considered below in Sec.~\ref{sec:long-t-app}). When  the  pulse duration $\delta$ exceeds $t_c$, it starts to play the role of a  cutoff time for the power-law tail 
\cite{burcaw2015,HHL2015,lee2017,HHL2016-phantom}.

\begin{figure}[t!]
\centering
\includegraphics[width=3.2in]{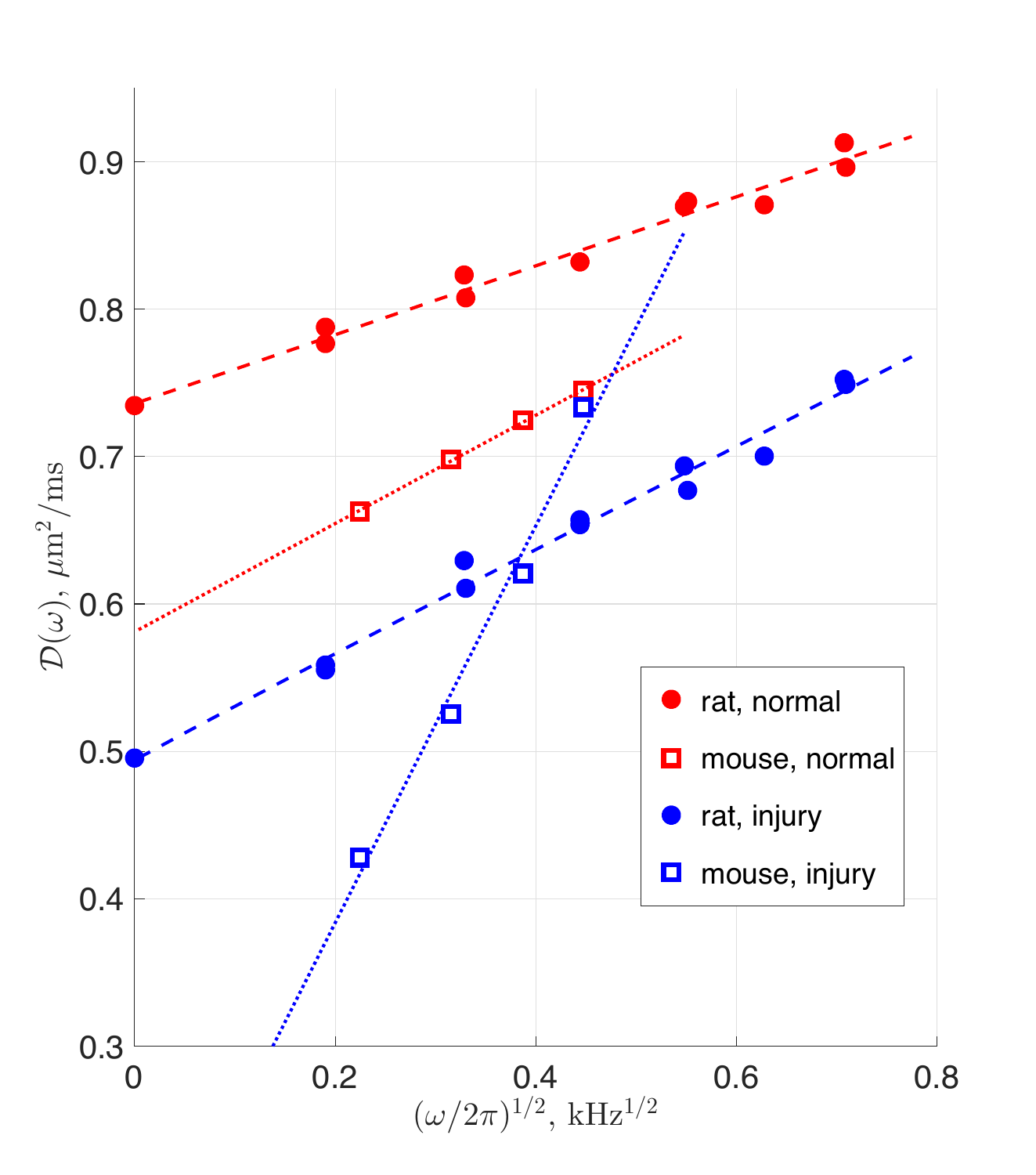}
\caption{
OGSE measurements in cortical GM: circles are data from average of 5 rats \cite{Does2003} 
and squares from 6 neonatal mice at 24 hours after unilateral hypoxic ischemic injury \cite{wu2014}. 
Red: normal rat brain and contralateral side of mouse brain. 
Blue: globally ischemic rat and ipsilateral side of hypoxia-ischemia injured mouse brain.  
PGSE data not shown. Dashed lines are fits from Fig.~4 of ref.~\cite{mesopnas}, dotted lines are $\omega^{1/2}$ fits (shown as guide to the eye; power-law exponent fit for mouse data was not robust due to narrow frequency range). 
Note that while the absolute $\D(\omega)$ values differ between rat and mouse, the general features are similar: data is well described with $\omega^{1/2}$ behavior for normal and ischemic GM (except, possibly, the ischemic mouse, where major structural changes may have occurred in 24h); 
and the coefficient in front of $\sqrt{\omega}$ (the slope) increases in ischemia, consistent with short-range structural disorder increase along the neurites (e.g., due to beading). 
}
\label{fig:wu-does}
\end{figure}

Finally, we give one more  illustration  of the absence of any universal relation between PGSE diffusion time and OGSE frequency $\omega = 2\pi \nu$ addressed in Sec.~\ref{sec:og-pg}. The PGSE-OGSE correspondence, empirically, means that the constants in
the tail $t^{-\vartheta}$ in $D(t)$,  and in the $\w^\vartheta$ tail of Eq.~(\ref{Dw-approach}) are equal. 
According to Eqs.~(\ref{t_to_omega}) and (\ref{D-Dinst}), 
\be \label{t-f-ii}
\mbox{regime {\bf (ii)}:} \quad t = \lb \frac{2\Gamma(\vartheta) \sin{\pi \vartheta\over 2}}{\pi(1-\vartheta)} \rb^{1/\vartheta} \cdot \frac1{2\pi \nu} \,, \quad \vartheta < 1 \,.
\ee
This relation is neither obvious, nor has  it   anything to do with the empirical Eq.~(\ref{t-f-wrong}).  
For example, for $\vartheta = 1/2$ (random permeable barrier model \cite{nphys} or short-range disorder in $d=1$ relevant for the neurites, Sec.~\ref{sec:long-t-app} below), we obtain $t = (4/\pi^2)/\nu$.
And besides, Eq.~(\ref{t-f-ii}) only applies for 
$\vartheta < 1$; for media characterized by larger $\vartheta$, the power-law PGSE and OGSE tails will not match. 
\keep 

\subsubsection{Validation and applications }
\label{sec:long-t-app}

\nin
Probing the diffusivity time-dependence at long times potentially allows for identifying the disorder universality class  
of the mesoscopic structure, which then helps to parsimoniously model the relevant features of tissue architecture and extract the corresponding tissue length scale(s) and other parameters.

\new
{\bf Validation with Monte Carlo simulations:}  Equation (\ref{theta=p+d}) was verified in $d=1$ MC simulations using different placements of identical permeable barriers, according to periodic, short-range, hyperuniform, and strong disorder classes \cite{mesopnas}. The same relation for the $p=-1$ universality class of random permeable membranes in $d=2$ was verified in ref.~\cite{nphys}. The borderline ``$\log$" case of $\vartheta=1$ for $p=0$ in $d=2$ dimensions, Eq.~(\ref{Dt-theta=1}), was verified in ref.~\cite{burcaw2015} in the $t$-domain. In ref.~\cite{Ginsburger2018}, the same universality class was considered in the $\w$-domain, verifying Eq.~(\ref{Dw-theta=1}) and  revealing  the dependence of the prefactor $A$ on the degree of  disorder in fiber packing.  
Subsequently, the scaling (\ref{theta=p+d}) was verified with MC along synthetic model neurites, dimension $d=1$, featuring realistic spines, leaflets and beads placed randomly according to the Poissonian statistics, $p=0$ \cite{Palombo2017}. 
The  slight deviation from the $\vartheta=1/2$ power law at the longest $t$ 
is attributed to the periodic boundary conditions for a relatively short sample. 
In the same setting,  Eq.~(\ref{Dw-Mitra}) transverse to the neurites was verified for the sub-ms times, regime {\bf (i)}.  

A more empirical approach \cite{Palombo2016pnas} was employed for diffusion of cell-specific metabolites up to $t=2\,$s that was measured by diffusion-weighted MR spectroscopy in vivo. Due to the broad time range, the $D(t)$-dependence was more pronounced, in comparison to the earlier measurements \cite{ackerman-neil-nbm2010} for the narrower $t$-ranges. 
Distinct tissue morphologies were recognized by comparing with large-scale MC simulations for particles diffusing in many synthetic cells generated as tree-like structures, by varying statistics of the number of processes, branches, and segment lengths. While the simulations matched the measurements, the relevant structural degrees of freedom and the associated functional forms of the $t$-dependence were not unequivocally identified. 

{\bf Validation in phantoms:}
The exponent (\ref{theta=p+d}) corresponding to the short-time disorder, $p=0$,  has been demonstrated in $d = 2$ dimensions in  an anisotropic fiber phantom mimicking the extra-axonal space \cite{burcaw2015} ($\vartheta = 1$, leading to  the $\ln t$ singularity in the PGSE $D(t)$, Eq.~(\ref{Dt-theta=1})), as well as more recently in a $d = 1$-dimensional phantom \cite{Papaioannou2017} for which $\vartheta = 1/2$ for $p=0$ and $\vartheta=3/2$ for hyperuniform placement of permeable membranes ($p=2$). 
\keep

{\bf Cortical GM} was probed with OGSE in rat \cite{Does2003} and mouse \cite{wu2014,wu2016}, Fig.~\ref{fig:wu-does}. 
It was suggested \cite{mesopnas}, that the apparent $\omega^\vartheta$ behavior with $\vartheta=1/2$ can be  explained by the dominance of the effectively $d=1$-dimensional diffusion along the narrow neurites  with the short-range disorder (e.g., spines, beads, varicosities) along them  \cite{Shepherd30042002, Shepherd2003,Debanne555}. The varicosities are known \cite{zhang-murphy-2005,li-murphy-2008} to become more pronounced in ischemia, which is consistent with the increase in the structural disorder-induced coefficient in front of the $\omega$-dependent term in Eq.~(\ref{Dw-approach}). 
Conversely, assuming  short-range disorder (e.g., in the varicosity placement), 
the power law exponent $\vartheta=1/2$ then validates the effectively $d=1$-dimensional diffusion along the so-called ``sticks" (narrow channels used to model the intra-neurite space), cf.\ Sec.~\ref{sec:sticks} below.

{\bf Human WM} was probed {\it in vivo} in both the longitudinal \cite{fieremans2016} and transverse directions \cite{fieremans2016,desantis2016}. 
The time-dependence of longitudinal diffusivity suggests restrictions are present along axons, which, similar to the GM case above, augments the commonly used ``hollow tube" model for diffusion inside and outside neurites (cf.\ Sec.~\ref{sec:sticks}). 
The ``hollow tube" filled with some effective Gaussian medium with diffusion coefficient $\Dinf$ 
becomes an effective theory technically valid only in the $t\to\infty$ regime {\bf (iii)}; 
for finite $t$,  non-Gaussian effects (time-dependent $D(t)$ and higher-order cumulants) will be present.  
 

Recent quantitative analysis \cite{fieremans2016} based on Eq. (\ref{Dinst-approach}) for $d= 1$, revealed that this time-dependence is compatible with short-range disorder in  the placement of restrictions along axons. Intriguingly, the corresponding correlation lengths of about $3-7\,\mu$m are similar to those reported in the literature for varicosities along axons \cite{Shepherd30042002, Shepherd2003,Debanne555}, suggesting them as potential sources for the reduction of the longitudinal diffusivity with time. Varicosities are often found to be  rich in mitochondria and 	 could therefore    form obstacles for the diffusion along the fibers, or they could act as temporary traps for the longitudinal diffusion. Additional potential sources  of   the short range disorder could be axonal undulations \cite{nilsson2012,Dhital2017Da}, or  functional gap junctions unevenly spaced between 20 and 60\,$\mu$m along the myelin sheath in sciatic nerve \cite{Schain2014}. 

\newv 
Note an interesting observation that the reduction in the diffusivity in acute stroke patients occurs predominantly along the axons when measured at the frequency $\omega/2\pi=50\,\units{Hz}$, while the decrease in both the longitudinal and transverse directions directions is observable for the diffusion time $40\,\units{ms}$ \cite{Baron2015}. Originally explained by axonal beading, this effect  has   to be taken into account in more general models of the diffusion response  to   tissue damage. \keep  

\subsubsection{Axonal diameter mapping}
\label{sec:adm}

\nin
Quantifying $\mu$m-level structure of neuronal tracts in vivo has been brought to the forefront of neuroscience research primarily due to the axonal diameter mapping (ADM) concept, developed within the  CHARMED and AxCaliber frameworks  (\cite{charmed,axcaliber,Barazany2009}) and their extensions \cite{Alexander2010,Zhang2011}. 
Their common theme is the focus on the intra-axonal compartment (assuming no exchange)  as the source of the diameter sensitivity, while typically  approximating axons as impermeable cylinders, and building on exact solutions \cite{neuman1974,Stepisnik1993,callaghan-book,grebenkov-rmp}. Water diffusion in the extra-axonal space in all of the above approaches is  approximated as Gaussian (time-independent).

Large overestimation of axonal diameters, by factors 3-15 in humans, cf., e.g., refs.~\cite{Alexander2010,huang2015}, provoked a debate \cite{Horowitz2015,Innocenti2015,Horowitz2015response} about the feasibility of the method. It has been since recognized that ADM may be confounded by two issues. 

{\it The first ADM issue} is the smallness of the signal attenuation for typical thin axons. 
The attenuation inside a cylinder up to $\O(g^2)$ (GPA) is given by van Gelderen's formula \cite{vangelderen1994} that depends on the PGSE sequence timings $\Delta$ and $\delta$. However, 
in the practically relevant case  $\delta \gg r^2/D_0$, the dependence on $\Delta$ drops out, hence the diffusion time is not actually being used to ``probe" the cylinder diameter. This is the Neuman's limit \cite{neuman1974}, in which attenuation
\be \label{logS-neuman}
-\ln {S\over S_0} \simeq {7\over 96}\, {g^2 r^4 \cdot 2\delta \over D_0} \approx 4.3 \cdot 10^{-6}
\ee
 is proportional to the total time $2\delta$ the gradients are on. 
 The proportionality to the time $2\delta$ can be  understood in terms of the mapping onto the transverse relaxation in the diffusion-narrowing regime, 
 where $\sim (gr)^2 \cdot r^2/D_0$ is the effective $R_2^*$ rate \cite{lee2017}.  
The above attenuation was evaluated for typical values of the   Larmor frequency gradient $g = 0.0107\, (\units{\mu m \cdot ms})^{-1}$ corresponding to 40 mT/m; free axoplasmic\footnote{\label{foot:D0}
This value is based on the observation \cite{Beaulieu2002} that axoplasmic diffusion coefficient in squid giant axon is 20\% below the water diffusion coefficient at the same temperature, and is consistent with the recent estimate of $\Da\approx 1.9-2.4\,\units{\mu m^2/ms}$ {\it along} axons in human WM at $t= 50\,$ms   \news obtained  \keep 
by suppressing extra-axonal compartment  using either high $b$ \cite{highb} or planar diffusion encoding \cite{Dhital2017Da}, 
which sets \news a \keep  lower bound for $D_0$. Another large axon study in excised lamprey spinal cord \cite{Takahaski_PNAS2003} reported a similar deviation of about 25\% for the longitudinal diffusion coefficient from the free water diffusion coefficient.   A somewhat larger value was reported in excised pig spinal cord \cite{Jespersen2017}. Alternatively, using NAA as an intracellular reporter molecule, the ratio for the {\it in vivo} measured parallel  diffusion coefficient in the corpus callosum relative to its diffusion coefficient in dilute aqeous solutions, ranges from 0.4 up to 0.46  \cite{Kroenke2004}, corresponding to estimates of water 
$\Da\sim 1.2-1.8\,\units{\mu m^2/ms}$. }
$D_0 = 2.4\,\units{\mu m^2/ms}$; pulse duration $\delta = 10\,$ms, and a typical inner diameter $2r = 1\,\mu$m \cite{Lamantia1990,aboitiz1992,Tang1997,caminiti2009,Liewald2014}.\footnote{Here we consider brain; axons are about factor of 5 thicker in the spinal cord, and the ADM prospects are much better there \cite{Komlosh2013,xu2014,Xu2016,Duval2015,Benjamini2016}, due to the $r^4$ scaling in Eq.~(\ref{logS-neuman}); see, however, the need for beyond-GPA corrections discussed below.} 

\new
Likewise, the OGSE attenuation $- \ln {S\over S_0} \simeq  b\cdot \Re \D(\w)$ in the relevant $\w r^2/D_0 \ll 1$ limit 
[cf.\ Eqs.~(\ref{Sgen-D}) and (\ref{Dw-beta}) in Appendix \ref{sec:app-ogse}]
\mpar{paragraph added about OGSE}
\be \label{Dw-cyl}
- \ln {S\over S_0} =  \frac7{96} {(g_0^2/2)\, r^4 \cdot T \over D_0}  
\ee
becomes {\it independent of the OGSE frequency} $\w$ and of the OGSE initial phase $\phi$, since $b\sim T/\w^2$ and 
$\Re \D(\w) \sim \w^2$.  
The analogy with Eq.~(\ref{logS-neuman}) becomes obvious 
if we realize that, following the mapping onto the transverse relaxation in the presence of an (oscillating) gradient $g=g_0 \cos(\w t - \phi)$, what matters is the total time $2\delta \to T$ the gradients are on (here 
$T = N \cdot 2\pi/\w$ is the total OGSE train duration), and the time-averaged gradient power 
$g^2\to \la g^2(t)\ra =  g_0^2/2$.  In other words, in the low-frequency limit, OGSE is just the Neuman's limit (\ref{logS-neuman}) albeit with the reduced average gradient amplitude, since the gradients are not at their peak value all the time. 
This yields that OGSE is not beneficial to map small compartment sizes, and does not provide any independent parameter combination, in the limit $\w r^2/D_0 \ll 1$ (cf.  Appendix \ref{sec:app-ogse}).
Obviously, the most optimal setting is to keep the diffusion gradient at its maximum all the time $2\delta \lesssim T_2$, cf.\ Eq.~(\ref{logS-neuman}). 
Practical resolution limits for axonal radii depending on the SNR and fiber geometry were considered for both pulse-gradient and OGSE sequences \cite{drobnjak2010,Drobnjak2016,nilsson2017limit}.  
\keep

{\it The second ADM issue} 
is potentially more significant. Had the above smallness been the only problem, we would just not see any  
dependencies of intra-axonal signal on experimental parameters. 
Yet the fits of ADM model to data do show definitive trends in the estimated ``apparent diameters" --- for instance, with the gradient strength \cite{huang2015} --- suggesting some unaccounted physical phenomenon.  
This has prompted taking into account the coarse-graining {\it outside} 
\cite{burcaw2015} randomly packed axons. The $(\ln t)/t$ term (\ref{Dt-theta=1}) from the extra-axonal space appears to completely overwhelm the weak attenuation (\ref{logS-neuman}) in simulations \cite{burcaw2015} and in the recent $D(t)$ measurements transverse to  human WM fiber tracts \cite{fieremans2016,desantis2016}. This reveals an exciting unexpected {\it mesoscopic effect:} The structural disorder in axonal packing within a WM fiber bundle completely changes the interpretation of ADM at low to moderate diffusion weightings.  Furthermore, recently observed logarithmic dependence on the pulse duration $\delta$ in human WM \cite{lee2017}, cf.\ Eq.~(\ref{Dt-theta=1}), instead of the linear one, Eq.~(\ref{logS-neuman}), and validated in a fiber phantom \cite{HHL2016-phantom}, confirms the mesoscopic extra-axonal origin of the ``apparent" ADM effects. 

The decreasing apparent diameter trend with increasing gradient strength \cite{huang2015} is consistent with eventual suppression, 
as $\sim e^{-b(\Dinf + A \ln t /t)}$, of the  extra-axonal contribution; 
however, since the transverse $\Dinf \lesssim 0.5\,\units{\mu m^2/ms}$ (cf.\ Section \ref{sec:gauss}), 
very large $b$-values are needed to fully suppress this effect \cite{sepehrband2016}.
\new 
However, when sufficiently strong gradients are used in animal settings, the GPA results of van Gelderen \cite{vangelderen1994}
 and Neuman \cite{neuman1974} should be corrected. Unfortunately, no analytical solution exists beyond GPA for finite $\delta$. 
 Lee \etal \cite{lee2017} estimated that the GPA will break down when $g \gtrsim g^* = D_0/r^3$ for axons of radius $r$.
 This may become relevant for large axons: e.g., for $r=3\,\units{\mu m}$ and $D_0=2\,\units{\mu m^2/ms}$, the critical gradient $g^*$ corresponds to $277\,\units{mT/m}$. 
 \mpar{large-$g$ discussion added}
Furthermore,  the next-order $\O(g^4)$ correction to the right-hand side of Eq.~(\ref{logS-neuman}) will be of the same sign as the main effect, scaling as $\sim g^4 r^{10}\delta/D_0^3$, implying that strong gradients cause extra attenuation relative to what GPA predicts. If the GPA is used instead of the exact solution, the GPA-derived radii will be overestimated, which may explain some residual overestimation of axonal radii in a recent animal study with gradients as large as $1.3\,\units{T/m}$ \cite{sepehrband2016}.
\keep
 
We note that an even stronger dominance of the extra-axonal contribution occurs in the OGSE domain, since the fully confined water (within an impermeable cylinder transverse to its axis) 
yields a regular, $\omega^2$ contribution  to $\Re \D(\omega)$ (Appendix \ref{sec:app-ogse}),
whereas the extra-axonal water would contribute {\it linearly}, as $|\omega|$, cf.\ Eq.~(\ref{Dw-approach}) with $\vartheta=1$. The linear term will dominate at low $\omega$, in agreement with the linear dispersion observed transverse to fibers with OGSE by Portnoy \etal \cite{Portnoy2013} and analyzed in ref.~\cite{burcaw2015}. 
Such linearity has been also observed in rat spinal cord by Xu \etal \cite{xu2014,Xu2016}. 

The predominance of the $|\w|$ scaling means that using OGSE is not optimal for probing inner axonal diameters --- not just numerically as discussed after Eq.~(\ref{Dw-cyl}), but parametrically! 
However, the $|\w|$ scaling makes OGSE parametrically better for probing the extra-axonal space geometry. 
We warn  that the gradient waveform optimization which does account for the mesoscopic effects in the extra-axonal space may sometimes give unfair preference to OGSE \cite{drobnjak2010,Drobnjak2016,nilsson2017limit}.

Overall, the above mesoscopic effects, measured {\it in vivo} on both animal and human scanners, may enable a novel kind of structural contrast at the micrometer scale (e.g.,\ axonal loss and demyelination), and open up exciting possibilities
of monitoring subtle changes of structural arrangements within GM and neuronal tracts in disease, aging, and development.

\subsection{Mesoscopic fluctuations}
\label{sec:meso-fluct}

\nin
When ADM is feasible (e.g., spinal cord, due to much larger $r$), Neuman's $r^4$ scaling (\ref{logS-neuman}), together with 
 volume-weighting $\sim r^2$, 
gives a large weight to a  small number of axons with largest diameters,  effectively measuring \cite{burcaw2015}
\be \label{r-neu}
r_{\rm Neuman} = \lp {\overline{r^6} /\overline{r^2}} \rp^{1/4} , 
\ee
where the averages are taken over the voxel-wise axonal distribution. 
Hence,  the metric (\ref{r-neu}) may become susceptible to the {\it mesoscopic fluctuations} governed by the tail of the distribution (practically, for sufficiently small voxels in which such fluctuations can be pronounced). Additionally,  sampling fluctuations confound  the comparison of dMRI measurements with histology, --- where the metric derived based on, e.g., Eq.~(\ref{r-neu}) can be strongly sample-dependent, especially if small fields of view are utilized. 
 
This general phenomenon of rare structural configurations determining the measurement outcome has parallels with similar effects found in hopping conduction in disordered semiconductors, kinetics of reaction-diffusion systems, and other phenomena in disordered media 
\cite{lifshitz1964,mott-book,shklovskii-book}. 
In our case, an incidentally large number of thick axons may significantly skew the 
intra-axonal attenuation for a particular voxel. This could lead to strongly enhanced variations (relative to those expected based on the measurement noise alone) in the corresponding parametric maps. 

The issue of the mesoscopic flucutations is fundamental, and the separation of the effects of biological variabiilty from the   randomness in  measurement outcomes due to the thermal noise requires model-independent ways of estimating local noise level \cite{MPnoise,veraart2016}, as well as precisely quantifying  the tails of the corresponding distributions of biophysical tissue parameters (e.g., of the axonal diameter distribution \cite{Lamantia1990,aboitiz1992,Tang1997,caminiti2009,Liewald2014}).


%% file: Section_III.tex
\section{The $t\to\infty$ limit, regime {\bf (iii)}: \\ Multiple Gaussian compartments}
\label{sec:gauss}

\epigraph{All science is either physics or stamp collecting}{Ernest Rutherford}



\nin
The  flamboyant century-old quote of a founder of the atomic age could be excusable, 
as scientific disciplines other than physics in his days were mostly collecting empirical information. 
Today, with so much more knowledge about the world and the associated abundance of data, 
Rutherford's quote could as well sound ``All science is either physics or fitting". While the purpose of physics remains to identify relevant parameters and to produce an explanation (an effective theory), \new and its instance for a particular measurement --- {\it a model} \cite{manifesto}, \keep the complexity of models and the amount of data  have turned {\it parameter estimation} into a field on its own, if not into a multitude of fields, employing a wealth of approaches, 
\new 
known under different names and incorporating advanced  tools of statistics, machine learning and artificial intelligence. 
Within MRI,  modern parameter estimation approaches are tied to the idea of undersampling, typically of the $k$-space data, which spurred the applications of compressed sensing \cite{candes2006,lustig2007} and  MR fingerprinting \cite{ma2013}.
\keep 

Tissue microstructure mapping presents its own set of parameter estimation challenges. As we will illustrate in this Section, while from the  physics standpoint, the dMRI models in the $t\to\infty$ regime become trivial (a sum of Gaussians = exponentials in $b$), 
their  number of parameters, \new and the inherent degeneracy of the fitting landscape in face of the typically low SNR of dMRI acquisitions, have turned parameter estimation into an active area of investigation. In other words, the problem remains largely unsolved --- even with a densely sampled $q$-space, and fully sampled $k$-space. \keep
So far, arguably, most intellectual efforts in the regime {\bf (iii)} (as defined in Sec.~\ref{sec:hierarchy}) have been spent on the  ``fitting" rather than on the ``physics". This Section is hence primarily about the parameter estimation aspect of modeling (cf.  Sec.~\ref{sec:BT}).

Below, after introducing the {\it stick} compartment in Sec.~\ref{sec:sticks}, we formulate the overarching Standard Model of diffusion in neuronal tissue as a sum of anisotropic Gaussian compartments (Sec.~\ref{sec:SM}, Figs.~\ref{fig:models} and \ref{fig:SM}), and then discuss challenges of its parameter estimation, Sec,~\ref{sec:paramest}, focussing on its degeneracies. 
We subsequently review  works involving constraints on the Standard Model parameters (Sec.~\ref{sec:constr}), followed by the unconstrained, rotationally-invariant methods (Sec.~\ref{sec:rotinv}), and conclude this Section with a summary of unresolved issues (Sec.~\ref{sec:branch}).

\subsection{Neurites as ``sticks"} 
\label{sec:sticks}

\subsubsection{Theory and assumptions}
\nin
In this Section, we assume that the $t\to\infty$ regime {\bf (iii)} has been practically achieved,
and neglect the time-dependent power-law ``tails" describing the approach of the diffusion coefficient to its tortuosity limit,
discussed in Section~\ref{sec:Dt}.%
\footnote{\label{foot:long-t}%
Mathematically speaking, a power-law approach, being scale-invariant, means that the $t\to\infty$ regime is never fully achieved --- there is no time scale that tells us where we can neglect the residual non-Gaussian effects in each compartment. 
However, practically, their detection limit is set by a finite SNR.} 

Full coarse-graining in the intra-neurite space then leads to the most anisotropic Gaussian compartment possible 
--- the so-called ``stick" compartment ---  
first introduced by 
Kroenke \etal \cite{Kroenke2004} and Jespersen \etal \cite{Jespersen2007} in 2004--2007. 
Its main features are: 
\renewcommand{\theenumi}{\bf \arabic{enumi}}
\begin{enumerate}
\item A stick is a cylinder whose radius $r$ is negligible compared with 
the ``free" diffusion length $\sim \sqrt{D_0 t}$ at given $t$.  
Equivalently, the transverse diffusion coefficient \newe inside neurites \keep $D_a^\perp \simeq r^2/4t \sim 0.01\, \units{\mu m^2/ms}$ for typical  $t\sim 100\,$ms is negligible   compared to $D_0 \approx 2.4\,\units{\mu m^2/ms}$,%
\footnote{See footnote \ref{foot:D0} in Section~\ref{sec:Dt}}
and hence can be set to zero, $D_a^\perp \to 0$. 
In other words,  the measurement is insensitive to neurite radii (cf.\ discussion in Sec.~\ref{sec:long-t-app}). 
\item The longitudinal diffusion inside a neurite becomes Gaussian, with the macroscopic (tortuosity) asymptote $\Da$. Of course, $\Da$, being the effective coarse-grained parameter, can be notably reduced relatively to the intrinsic axoplasmic diffusion coefficient $D_0$, cf.\ Section~\ref{sec:Dt}. The parameter $\Da$ takes into account all restrictions, such as varicosities (beads) and undulations, along the (average) neurite direction. 
Hence, it can have important biophysical and diagnostic value in the cases when the structure along neurites changes, e.g., in acute stroke \cite{zhang-murphy-2005,li-murphy-2008,Baron2015} and in Alzheimer's disease \cite{beads-alz}.
\item Exchange between intra- and extra-neurite water can be neglected, at least  at the  time scales $t$ used in clinical dMRI. 
 \mpar{edited this paragraph; red text added; R1.6} 
Measuring exchange times {\it in vivo} is very difficult, making this assumption hard to validate. The consensus so far has been that this assumption holds for WM tracts, where sticks represent (myelinated) axons (and possibly some glial processes). 
\news The filter-exchange study in \keep \cite{Lampinen2016} supports the presence of two compartments in healthy brain WM with the exchange time of about $1\,$s, consistent with 
assuming negligible exchange on the clinical $t\sim 100\,$ms time scale. 
At which $t$ this assumption might break for glial cells, dendrites in GM, or for unmyelinated axons, is a subject of investigation. 
\new 
It was hypothesized \cite{Badaut2011} that transcytolemmal water exchange in astrocytes is fast, 
since inhibition of aquaporin-4  significantly reduced the diffusion coefficient already for $t < 25\,$ms, 
without modifying tissue histology.  
However, recent work \cite{Yang2017} of measuring $T_1$  in the presence of fast extra-cellular flow for cultures of astrocytes and neurons grown on beads puts the intracellular residence time around $570$ and $750\,$ms, correspondingly.  
Likewise, MR relaxation measurement \cite{Bai2018} in the live rat brain organotypic cortical cultures yields the net cellular water efflux rate $2.02\,\units{s^{-1}}$, with a significant fraction ($\sim 34-45$\%) of this exchange rate attributed to {\it active} transcytolemmal exchange related to the Na$^+$-K$^+$-ATPase activity.  
\keep
\end{enumerate}

From the modeling standpoint, the stick compartment is the defining feature of dMRI inherent to the neuronal tissue, as compared to all other kinds of soft tissues. It is chiefly responsible for the anisotropy of the diffusion propagator in the brain (at least, in the white matter), and in spinal cord (where finite axonal radii can be detected, see footnote~\ref{ftn:cord}). 

The diffusion propagator for a stick pointing in the unit direction $\n$, measured in the unit gradient direction $\g$,
\be \label{stick}
G_{\n}\left(\g,b\right) = e^{-b\Da(\g\cdot\n)^2} 
\ee
is determined by $\cos\theta \equiv \g\cdot\n$, where $\theta$ is the angle between $\g$ and $\n$. The signal is not suppressed for $\g\perp\n$ and decays fast with $b$ when $\g\parallel \n$. 
Hence, when $b\Da \gg 1$, the stick dMRI response (\ref{stick}) becomes a thin ``pancake",
non-negligible when 
 $|\n\cdot\g|\lesssim (b\Da)^{-1/2}$ {\it nearly transverse to} $\g$, whose angular thickness scales as 
$\delta\theta \sim 1/\sqrt{b\Da}$ \cite{jensen2016,Mckinnon2017,highb}. 
 Both estimates follow  from setting the argument of the exponential to unity.

\begin{SCfigure*}[0.65][th!!]
\centering
\includegraphics[width=4.6in]{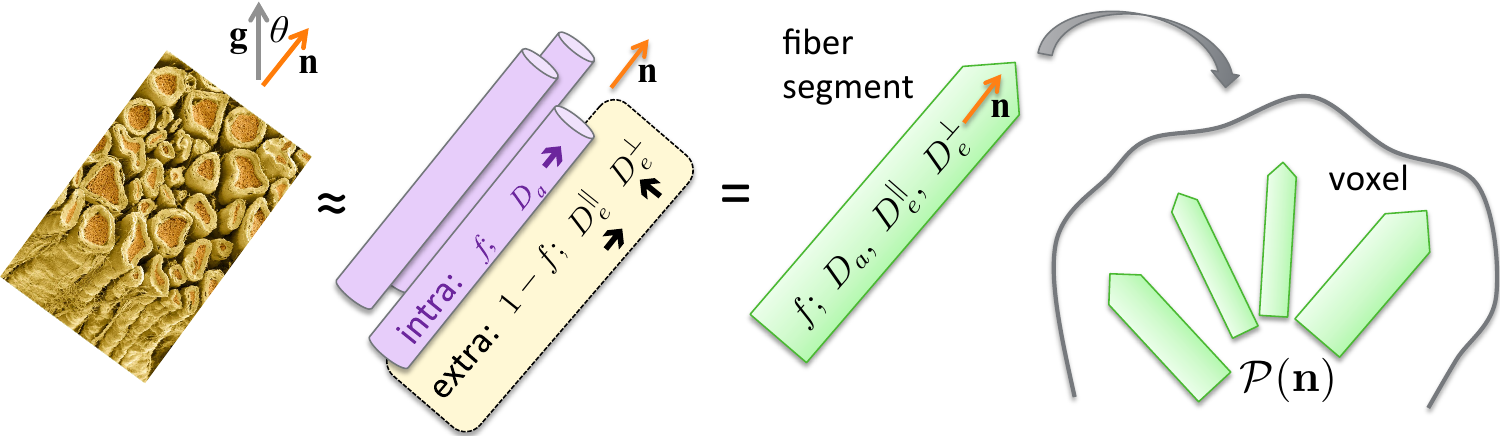}\qquad
\caption{
{\bf The Standard Model of diffusion in neuronal tissue, Eq.~(\ref{S}).} In the $t\to\infty$ regime {\bf (iii)}, elementary fiber segments (fascicles), 
consisting of intra- and extra-neurite compartments, 
are described by at least 4 independent parameters: $f$, $\Da$, $\Depar$ and $\Deperp$. 
CSF can be further added as the third compartment, cf. Eq.~(\ref{K}). 
Within a macroscopic  voxel, such segments contribute to the directional dMRI signal according to their ODF $\P(\n)$. 
}
\label{fig:SM}
\end{SCfigure*}

\subsubsection{Validation of the picture of sticks} 
\label{sec:sticks-val}

\nin
%
An extensive review of dMRI validation studies is beyond the scope of this article. However, given the essential role sticks play in dMRI models, we mention the following two kinds of  results. 

First, metabolites, such as N-Acetylaspartic acid (NAA),  intrinsic to the intra-neurite space, can be used to identify the stick compartment, as reviewed  by Ackerman and Neil \cite{ackerman-neil-nbm2010}. 
A seminal NAA study was performed by Kroenke, Ackerman and Yablonskiy \cite{Kroenke2004}, who demonstrated a very good agreement between the dMRI signal from large voxels in rat brain averaged over three gradient orientations, 
at diffusion times $t=50 - 100\,$ms, and the isotropically averaged stick signal,
\be \label{stick-iso}
\int\! \d\n \, G_{\n}\left(\g,b\right) = \sqrt{\pi \over 4 b\Da} \, \erf \!\lp \sqrt{b\Da} \rp ,
\ee 
where $\erf$ is the error function. 
Taking a large voxel, which presumably has all neurite orientations, and subsequently averaging over 3 directions, maps the signal to that from a completely random stick arrangement, first considered by Callaghan \cite{callaghan1979}  in 1979, and subsequently by Yablonskiy \etal \cite{yablonskiy2002}  in 2002 for $^3$He diffusion in the lung, resulting in Eq.~(\ref{stick-iso}) (Fig.~\ref{fig:models}). 
The agreement with the theory (\ref{stick-iso}) was very good in the whole range $0 < b  \lesssim 20\,\units{ms/\mu m^2}$ \cite{Kroenke2004}. 
Recent directional NAA imaging by Ronen \etal \cite{ronen2014}, quantifying the dMRI signal anisotropy, agrees well with the structure tensor \cite{Budde2012} derived from the axonal histology in the corpus callosum.  

Second, for the water dMRI, identifying a distinct {\it functional form} inherent to the stick compartment can also validate the pictures of sticks at sufficiently large $b$, when the extra-neurite signal becomes exponentially suppressed (because the extra-neurite diffusion coefficient is nonzero in any direction), while the intra-neurite signal is only suppressed as a very slowly decaying power-law $\sim b^{-1/2}$, scaling as the width of the pancake-shaped stick response function (\ref{stick}), cf. Eq.~(\ref{stick-iso}) and refs.~\cite{jensen2016,Mckinnon2017,highb}. 
The recently observed \cite{highb,Mckinnon2017}  
 power-law water signal attenuation in human WM {\it in vivo},  
isotropically averaged over multiple gradient directions $\g$,
\be \label{b^-1/2}
\overline{S}|_{b\Da \gg 1} \simeq \beta \cdot b^{-\alpha} + \gamma \,,
\ee
with exponent $\alpha = 1/2$, provides a unique signature of axons as sticks for water dMRI. 
The isotropic average (cf. Sec.~\ref{sec:isoavg} below)
of the signal makes it equivalent to that from  isotropic set of sticks, cf. Eq.~(\ref{stick-iso}) above, with $\erf$ approaching $1$ at large $b$.

Either detectable axonal diameter values, or a notable exchange rate between intra- and extra-axonal water, would destroy the very particular stick-related $b^{-1/2}$ scaling (\ref{b^-1/2}).  In particular, the analysis in ref.~\cite{highb} shows that, 
had the axonal diameters been notably higher than their histological estimates of $\sim 1\,\mu$m \cite{aboitiz1992,Lamantia1990,caminiti2009}, as the ADM results often show (cf. Sec.~\ref{sec:long-t-app}), the power law exponent $\alpha$ would differ from $1/2$ for $b\leq 10\,\units{ms/\mu m^2}$. 
Hence, human dMRI measurements are practically insensitive to axonal diameters even with  gradients of $80\,\units{mT/m}$ employed, confirming the first stick assumption in Sec.~\ref{sec:sticks}. 
The human measurement \cite{highb} also revealed that the immobile water fraction, not decaying with $b$ for any direction, is below  detection limit.  The same conclusion was made independently 
using isotropic diffusion weighting \cite{Dhital2017}. 
\new Small and slightly {\it negative} $\gamma$, of about $\sim 1\%$, is a signature of the breakdown of zero-radius stick picture relevant at very high $b$ \cite{highb}. \keep 

\newv 
A low, but measurable transverse intra-axonal diffusivity, even in \news vanishingly \keep thin axons, can emerge from deviations of their form from \news perfectly \keep straight sticks \cite{nilsson2012,Dhital2017Da}. Measurement with suppression of extra-axonal signal in the human brain suggests $2D_a^\perp = 0.13\pm 0.04\, \units{\mu m^2/ms}$ for the diffusion time about $120\,\units{ms}$, which was obtained as the difference between the trace of intra-axonal diffusion tensor and the longitudinal diffusivity, $\Da$ \cite{Dhital2017Da}.

\keep


\subsection{The Standard Model of diffusion in neuronal tissue} 
\label{sec:SM}

\subsubsection{Theory}

\nin
With the intra-neurite diffusion modeled as a collection of sticks, and the isotropically fully restricted water out of the picture, 
how should we model the remaining water? 
The answer depends on the coarse-graining length scale. 

If the diffusion time is as large as needed for water molecules to sample a statistically representative part of the extra-neurite space (ENS) within a voxel, then diffusion in this space should become Gaussian and be described by the overall ENS diffusion tensor, $S_{\rm ENS} \sim \exp(- b_{ij} D^{\rm ENS}_{ij})$, 
where the $b$-matrix 
$b_{ij} = q_i q_j t \equiv b\,  g_i g_j$ depends on the components of the unit diffusion gradient $\g$, cf. Eqs.~(\ref{Gbare}) and (\ref{cum}). 
The ENS tensor $D^{\rm ENS}$ would then by definition describe all ENS water, including cerebrospinal fluid (CSF) that could, e.g., contribute if a voxel contains part of a ventricle. 

Practically, such very long-time limit is never realized for a macroscopic voxel. For typical diffusion times $t=50-100\,$ms, the corresponding diffusion length $L(t) \sim 10\,\mu$m defines the coarse-graining window, where the diffusion properties locally become (almost) Gaussian. 
At this scale, the neuronal tissue, at least in the WM, looks as a highly aligned fiber segment (fascicle)%
\footnote{cf. footnote \ref{foot:long-t}; effectively, we neglect $D(t)-\Dinf$ within each fascicle.}%
, Fig.~\ref{fig:SM}, the leftmost image. Because the fibers are locally coherent at the scale $L(t)$, the local ENS tensor $D_{ij}^{\rm ENS}(\r)$ can differ at the different positions $\r$ within a voxel if they are separated by distance much exceeding $L(t)$ (for instance, if a voxel contains fiber crossings).\footnote{This picture might be not as definite in GM, where the dendrites are more intertwined, and the concept of the overall ENS tensor may be better justified \cite{Jespersen2007,Jespersen2010}. But then, much less is known about diffusion in GM overall, and the precise knowledge of in vivo neurite residence times is still lacking. 
Therefore, here we mostly discuss SM in the context of WM, leaving GM  modeling  as one of the unresolved problems in Sec.~\ref{sec:unresolved}.} 
This local microscopic anisotropy of ENS suggests that, strictly speaking, the ENS diffusion is never Gaussian --- but is rather a sum of local anisotropic Gaussian contributions, which are highly aligned with the corresponding local stick arrangements. 

These coarse-graining considerations lead to the general picture of anisotropic  compartments, Fig.~\ref{fig:SM}. 
The signal measured in the unit direction $\g$, 
is a convolution\footnote{\label{foot:dn}
\new The convolution is on a unit sphere $|\n|=1$ \cite{Healy1998}. 
We normalize \cite{rotinv} $\d\n \equiv {\sin\theta \d\theta\, \d\phi \over 4\pi}$ 
such that $\int\!\d\n \cdot 1  \equiv 1$, while $\K|_{b=0} = S_0 \equiv S|_{b=0}$. \keep
}  
\be \label{S}
S_{\g}(b) = \int_{|\n|=1}\!\! \d\n\, \P(\n) \, \K(b, \g\cdot\n)
\ee
between the fiber orientation distribution function (ODF) $\P(\n)$ 
normalized to $\int\d\n\, \P(\n)\equiv 1$, and the response kernel $\K$
from a perfectly aligned fiber segment (fascicle) pointing in the direction $\n$. 
The kernel $\K(b, \g\cdot\n)$ depends on the relative angle $\theta$, $\cos\theta \equiv \g\cdot\n$ (cf. Eq.~(\ref{stick}).
\new \mpar{refs added}
The general representation (\ref{S}) gave rise to a number of methods for deconvolving  
the fiber ODF from the dMRI signal for a given $|\q|=q$ shell in $q$-space, using different empirical forms of the kernel  
\cite{Tournier2004,Anderson2005,Tournier2007,DellAcqua2007,Jian2007,Kaden2007,White2009}. 
\keep

Following the coarse-graining arguments above, the kernel's functional form  
\bea \non
\K(b,\xi) &=& S_0  \lb f e^{-b\Da \xi^2} + (1-f - f_{\rm CSF}) e^{-b\Deperp-b(\Depar-\Deperp)\xi^{2}} 
\right. \\ 
&& \left. +\  f_{\rm CSF} \, e^{-b D_{\rm CSF}} \vphantom{e^{-xi^2}}\rb , \quad \xi = \g\!\cdot\!\n \,, 
\label{K}
\eea
is a sum of exponential (in $b$) contributions from the aligned intra-\newe neurite and extra-neurite spaces, respectively modeled by  a stick compartment,  by the axially symmetric Gaussian compartment with transverse and longitudinal diffusivities $\Deperp$ and $\Depar$ and the principal direction along the stick, \keep and \del{by} the CSF compartment, cf. refs.~\cite{lemonade-ismrm,rotinv,baydiff} and Fig.~\ref{fig:SM}.


\new 
The myelin water compartment is typically neglected due to its short $T_2$ time \cite{Mackay1994} as compared to \newe the echo times $T_E$ employed in clinical dMRI\keep. We emphasize that the fractions $f$ and $1-f$ are the relative {\it signal fractions}, and not the absolute \newe water \keep volume fractions, due to neglecting myelin water, as well as
due to generally different $T_2$ values for the intra- and extra-neurite compartments \cite{Dortch2013,teddi}. 
\keep
The isotropic CSF compartment has $D_{\rm CSF} \approx 3\,\units{\mu m^2/ms}$. 
Because of the ODF normalization $\int\! \d\n \, \P(\n) \equiv 1$, the CSF term can be included in the kernel or added separately to signal (\ref{S}); 
we choose to include it in the kernel (\ref{K}), as it makes the formulation (\ref{S}) more elegant. 
In what follows, we will describe in depth the two-compartment kernel \eq{K} with $f_{\rm CSF}\equiv 0$. 

The overarching model (\ref{S}) and (\ref{K}) includes \newe nearly \keep all previously used models 
\cite{Kroenke2004,Jespersen2007,Jespersen2010,KM,wmdki,noddi,sotiropoulos2012,reisert2014mesoft,lemonade-ismrm,jelescu2016,rotinv,baydiff,teddi}
(at least in the WM) as particular cases, also described in taxonomy  studies in refs.~\cite{panagiotaki2012,ferizi2015} where some of the ``models" are actually representations, in the sense of Sec.~\ref{sec:model-represent}. 
In other words, the numerous models (and acronyms), corresponding to the right part of Fig.~\ref{fig:models} in Section~\ref{sec:intro}, 
{\it describe the same physics, and hence they are really the same model}. 

Because of the overall popularity and inclusiveness of the above picture, here we suggest to call the model (\ref{S})--(\ref{K})  
the Standard Model (SM)%
\footnote{The name is suggested by the tongue-in-cheek analogies with the Standard Model in particle physics. In both communities, SM represents the consensus knowledge about the subject, satisfactorily describes (almost) everything, has been out there for a while, and yet one really hopes that there is exciting physics beyond it --- which is far more difficult to access. Our community has a doubtless advantage in that investigations beyond SM are much cheaper (cf. Section \ref{sec:Dt}) than building particle accelerators.}.

We  note  that a major limitation of the SM kernel (\ref{K}) is \news using the same scalar parameter values for \keep  different fiber tracts passing through a voxel (noted, e.g., in refs.~\cite{Assaf2005,desantis2016mrm}), which prompted assigning different (albeit constant) fiber responses to different 
tracts  to deconvolve the ODF \cite{Sherbondy2010,Tournier2011,Girard2017}, as an alternative.

\subsubsection{Specificity and relevance of SM parameters}

Over the past decade, it has become clear that the {\it scalar parameters} $f$, $\Da$, $\Depar$ and $\Deperp$, and the {\it spherical tensor parameters} 
(the spherical harmonics coefficients of the ODF $\P(\n)$), carry distinct biophysical significance. Deconvolving the voxel-wise fiber ODF, instead of relying on the empirical directions obtained by Fourier-transforming the dMRI signal from the $q$ to $r$ space, in principle provides a more adequate starting point for  fiber tractography, 
\new \mpar{refs added}
an essential tool for mapping structural brain connectivity and for presurgical planning \cite{Behrens2007,Descoteaux2009,Farquharson2013,Sotiropoulos2013a,Wilkins2015}.
\keep

Furthermore, as illustrated further in Sec. \ref{sec:wmti}, the ability to estimate scalar parameters of the kernel (\ref{K}) would make dMRI measurements {\it specific} --- rather than just sensitive ---  to $\mu$m-level manifestations of disease processes, such as demyelination  \cite{axlossdemyel,shrink-remove,jelescu-cpz-2016} ($\Deperp$), axonal/dendritic loss \cite{jelescu-cpz-2016,Khan2016NI,Khan2016DiB,Vestergaard2011} ($f$),  beading \cite{budde-frank2010}, inflammation and oedema   
($f_{\rm CSF}$, as well as, potentially, mostly $\Da$ for cytotoxic and mostly $\Depar, \Deperp$ for vasogenic oedema \cite{unterberg2004}). 
Combining $f$ with the extra-axonal volume fraction derived either from tortuosity modeling based on $\Deperp$  \cite{shrink-remove,axlossdemyel} or from the myelin volume fraction via relaxometry, would ultimately allow one to determine axonal $g$-ratio
\cite{stikov2015}. 
Since the precise nature and pathological changes in microarchitecture of restrictions leading to the scalar parameter values are unknown, ideally, to become specific to pathology, one needs to estimate $f$, $f_{\rm CSF}$, $\Da$, $\Depar$ and $\Deperp$ separately. 


The first attempt to \newe estimate the parameters and to \keep validate the 2-compartment SM was performed by Jespersen \etal \cite{Jespersen2010} 
using direct fitting to an extensive {\it ex vivo} data set covering both WM and GM, while parametrizing the ODF using spherical harmonics $Y_{lm}$ up to $l \leq 4$. (The ENS diffusion in this work was described by the overall tensor, whose 6 components were estimated.)  
Data was acquired on a 16.4\,T magnet using 16 shells with $0 \leq b \leq 15\,\units{ms/\mu m^2}$ and 144 directions, with  acquisition taking over 15 hours. 
The fraction $f$ correlated well with AMG staining for the neurites, and the ODF directionality agreed with the histology. Further quantitative comparison of the predictions to histology was carried out in \cite{Jespersen2012}, where the neurite ODF was determined from Golgi stained cortical neurons in immature ferret brains.

While ``brute-force"  fitting of $\sim 20$ parameters could work  for  an extensive data set \cite{Jespersen2010}, clinical dMRI data is far noisier, with much less $q$-space coverage. \mpar{edited slightly} Hence, because of the high dimensionality of parameter space and the unfavorable fitting landscape \cite{jelescu2016}, SM parameter estimation for Eqs.~\eq{S}--\eq{K} from realistic noisy clinical dMRI data has emerged as an overarching challenge (Sec.~\ref{sec:paramest}), \news which \keep has until recently 
\news been addressed by introducing parameter constraints, \keep as discussed further in Sec.~\ref{sec:constr}\news.
The open challenge of parameter estimation also means that the literature \cite{Kroenke2004,Jespersen2007,Jespersen2010,KM,wmdki,sotiropoulos2012,noddi,reisert2014mesoft,lemonade-ismrm,jelescu2016,rotinv,baydiff,teddi} 
differs largely in the ways SM parameters have been constrained.  \keep

\subsection{The challenge: SM parameter estimation}
\label{sec:paramest}

\subsubsection{SM parameter count}

\nin
To quantify the problem's complexity, we find here how many parameters $N_p$ the model \eq{S}  should have --- and, hence, how many we have to estimate from (noisy) dMRI data. We follow here the treatment in ref.~\cite{rotinv}.
 
The answer depends on the maximal power $\lmax$ 
of the diffusion weighting $b^{\lmax/2} \sim q^{\lmax}$ to which an acquisition is sensitive, at a given SNR (by the time-reversal symmetry of diffusion, only even $\lmax$ are considered). 
This can be seen  either from the cumulant expansion (\ref{cum}) of $\ln S_\g(b)$, or, equivalently \cite{Kiselev2010_diff_book}, from the 
Taylor expansion of the signal \eq{S} 
\be\label{taylor}
{S_\g(b) \over S(0)}  = 1 - b M^{(2)}_{i_1 i_2} \, g_{i_1} g_{i_2} + \frac{b^{2}}{2!} \, M^{(4)}_{i_1 \dots i_4} \, g_{i_1} \dots g_{i_4} - \dots
\ee
in the fully symmetric moments $M^{(l)}_{i_1\dots i_l}$. 
These moments are proportional  to angular averages $\langle n_{i_1}\dots n_{i_l} \rangle$ over the ODF $\P(\n)$, 
as it is evident from expanding the exponential terms containing $\xi = n_i g_i$ in  kernel (\ref{K}), such that subsequent terms have the form $b \langle n_i n_j \rangle g_i g_j$, 
$b^2 \langle n_{i_1}\dots n_{i_4}\rangle g_{i_1}\dots g_{i_4}$, etc. The maximal (even) order $l$ of the product  $\langle n_{i_1}\dots n_{i_l} \rangle$ always appears with
the corresponding power $b^{l/2}$ of the diffusion weighting. 

The symmetric tensors  $\langle n_{i_1}\dots n_{i_l} \rangle$, after subtracting all possible traces, can be turned into the corresponding symmetric trace-free tensors (STF) of rank $l$ \cite{thorne}, which are  equivalent to the set $Y_{lm}$ of the spherical harmonics (SH) discussed in Sec.~\ref{sec:cum} above.
In other words, the ODF averages $\langle n_{i_1}\dots n_{i_l} \rangle$ correspond to the SH coefficients $p_{lm}$ of the ODF, 
\be \label{P=Y}
\P(\n) \approx 1 + \sum_{l = 2, 4, \dots}^{\lmax} \sum_{m=-l}^l p_{lm} Y_{lm}(\n) \,. 
\ee
In particular, the highest-order cumulant $C^{(\lmax)}$ or the moment 
$M^{(\lmax)}$ still practically resolvable from the signal, sets the maximal order $\lmax$ 
for the even-order SH expansion (\ref{P=Y}). 
The correspondence between $\lmax$ in Eq.~(\ref{P=Y}) and the maximal order $b^{\lmax/2}$ in expansion (\ref{taylor}) embodies 
the perturbative radial-angular connection in the $q$-space \cite{rotinv}.

We note here an obvious corollary from the radial-angular connection in $q$-space: Oversampling the directions within the low-$b$ shells does not improve angular resolution in estimating $\P(\n)$ --- in other words, optimal $q$-space coverage should match the sensitivity to the power 
$b^{\lmax/2} \sim q^{\lmax}$ of the shell radius with the minimal number $n_c(\lmax)$ of directions per shell. 
Naive sampling, say, $256$ directions at $b\approx 1\,\units{ms/\mu m^2}$ would not in principle yield better angular ODF resolution than, say, $\sim 10$ averages of $25$ directions. Indeed, the clinical dMRI signal at this $b$-value can be fully described using
$\O(b)$ (DTI, $\lmax=2$), or, at best, $\O(b^2)$ (DKI, $\lmax=4$) cumulant terms, corresponding to being sensitive to the ODF expansion coefficients $p_{lm}$ up to $l=2$ or $l=4$ (containing 5 or 14 parameters). There is no way to determine, say, $p_{6m}$ and beyond, if the diffusion weighting is too weak for the $b^3$ terms to be discernible at a given SNR.

Coming back to counting the SM parameters,  the (minimum) $N_s = 4$ {\it scalar} parameters from the kernel \eq{K} in the absence of CSF (or $N_s = 5$ if the CSF compartment is added), 
are complemented by the $n_c(\lmax)-1 = \lmax(\lmax+3)/2$ {\it tensor} parameters $p_{lm}$, where $n_c(l)$ is the number of the even-order spherical harmonics coefficients up to the order $l$ given by Eq.~(\ref{nc}) in Sec.~\ref{sec:cum}, and 
we subtracted one parameter because $p_{00}\equiv \sqrt{4\pi}$ is set by the ODF normalization.  
This yields the total SM parameter count 
\be \label{Np}
N_p(\lmax) = N_s + \lmax(\lmax+3)/2
\ee
such that $N_p = 9, \ 18, \ 31, \ 48, \ \dots$ for  $\lmax = 2, \ 4, \ 6, \ 8, \ \dots$ already for the two-compartment kernel \eq{K}, without including $S(0)$ and $f_{\rm CSF}$ in the count. 

Equation (\ref{Np}) reveals that the model complexity grows fast, as $\lmax^2$, if we are to account for the rich orientational content of realistic fiber ODFs in the brain. For an achievable  $\lmax \sim 4-8$, the dMRI signal in principle ``contains'' a few dozen parameters, none of which are known  {\it a priori}.

\subsubsection{How many parameters are necessary?}
\label{sec:SMchallenge}

\nin
We can now contrast the SM parameter count (\ref{Np}) with the 
number $N_c(\lmax)$ [Eq.~(\ref{Nc}) in Sec.~\ref{sec:cum}] 
of independent parameters ``contained" in the cumulant or moment series truncated at $l=\lmax$. 
Since $N_c(l) \sim l^3$ grows faster with $l$ than $N_p(l) \sim l^2$, the moments/cumulants estimated from the signal should, beyond some $l$, contain more than enough information to determine all the corresponding SM parameters. Direct comparison of 
Eqs.~(\ref{Nc}) and (\ref{Np}) yields that 
\be \label{Nc>Np}
N_c(\lmax) > N_p(\lmax) \quad \mbox{for} \quad \lmax \geq 4 
\ee
for both 2- and 3-compartmental SM. This naive counting would let one believe that, having mastered sufficiently precise DKI parameter estimation ($\lmax=4$), we would be able to find all the scalar SM parameters, 
as well as estimate arbitrary fiber ODF up to $p_{4m}$, Eq.~(\ref{P=Y}). 

This intuition, however, is deceptive, as {\it the information content is not evenly distributed among all the $N_c(\lmax)$ components}. 
It turns out that the minimal order $\lmax$ for which the moments/cumulants contain enough information to determine all SM parameters is $\lmax=6$, while at the $\lmax=4$ level, there exists a one-dimensional manifold (for $N_s=4$), which can look as a single curve, or as two disjoint continuous ``branches", or families, of scalar parameters, which {\it exactly match} the signal's moment tensors $M^{(2)}_{i_1i_2}$ and $M^{(4)}_{i_1\dots i_4}$ (equivalently, the diffusion and kurtosis tensors) \cite{lemonade-ismrm,rotinv}. The two families of solutions, or the two parts of the above one-dimensional curve (``bi-modality") technically emerge as the two branches of a square root in a solution for a quadratic equation. In what follows, we illustrate this effect with a toy model of parallel fibers \cite{KM,wmdki} and then show that increasing the model complexity does not cure the problem.

\nin
\subsubsection{A toy model of bi-modality: Parallel fibers}
\label{sec:toy}

\nin
Let us now see how two branches appear as solutions of a quadratic equation involving directional diffusion and kurtosis values for a very simple ODF of perfectly aligned fibers, for the 2-compartment SM case. 
Here we follow the cumulant-series DKI approach as in ref.~\cite{KM}; an equivalent formulation in terms of the moments, cf. Eq.~(\ref{taylor}), can be found in ref.~\cite{rotinv}. A similar approach was used to estimate white matter tract integrity (WMTI) metrics from DKI \cite{wmdki} and subsequently adapted 
 \cite{Hansen2016AxWMTI} to enable their estimation with reduced data requirements using axially symmetric DKI \cite{Hansen2016}. 

Staying at the $\O(b^2)$ level (DKI),  the overall radial and axial components of the diffusion tensor, estimated from an ideally measured signal (the left-hand side), correspond to the following combinations of the scalar parameters (the right-hand side):
\bea \label{Dperp}
D^{\perp}&=&(1-f)D_{e}^\perp \, \\ 
D^{\parallel}&=& f D_{a} + (1-f) D_e^\parallel \,,
\label{Dpar}
\eea
and for kurtosis components \cite{KM,wmdki}
\bea \label{Kperp}
K^{\perp} &=& {3f \over 1-f} \,, \\
K^{\parallel} &=& 
{3 f(1-f)(\Da - \Depar)^2 \over {D^{\parallel}}^{2}}\,.
\label{Kpar}
\eea
``Transverse" parameters $f$ and $D_e^\perp$ are uniquely determined from $K^\perp$ and $D^\perp$:
\be
f = {K^\perp \over K^\perp +3}\,, \qquad D_e^\perp = \left( 1 +  {K^\perp \over 3} \right) D^\perp \,.
\ee
However, there are {\it two} possible solutions in the parallel direction. 
The duality arises from choosing \cite{wmdki,rotinv} the $\zeta = \pm$ branch of the square root in Eq.~(\ref{Kpar}),
\be \label{Depar}
\Da - \Depar = \zeta \cdot \sqrt{K^\parallel \over 3f(1-f)} \cdot D^\parallel \,, \quad \zeta = \pm 1  \,. 
\ee
Here $\sqrt{K^\parallel}  \propto | \Da - \Depar | \equiv  \zeta (\Da - \Depar)$, where $\zeta = \sgn (\Da - \Depar)$. 
Note that, since the ground truth is unknown, our choice for the branch  $\zeta$ may differ from the correct one. 

From Eqs.~(\ref{Dpar}) and (\ref{Depar}), we find that, not surprisingly, the correct choice of $\zeta$ yields the true values $\Da$ and $\Depar$. 
However, if the sign choice is wrong,  then the ``apparent" diffusivities do not agree with the true ones:
\bea
D_a^{\rm app} &=& (2f-1) D_a + 2(1-f) D_e^\parallel \,, \\
{D_e^\parallel}^{\rm app} &=& 2f D_a + (1-2f) D_e^\parallel  \,. 
\eea
Note, that in this case, as expected, ${D_e^\parallel}^{\rm app} - {D_a}^{\rm app} = -({D_e^\parallel} - D_a)$, i.e., the difference has the same absolute value and the wrong sign. In particular, for $f=1/2$, the diffusivities are swapped, --- i.e {\it we mistake $\Depar$ for $\Da$ and vice-versa}. Yet the above ``apparent" values can seem completely biophysically plausible, especially if $f\approx 0.5$. From the above derivation it is evident that the bi-modality of the parameter estimation originates from having two tissue compartments, and that the branch choice is certainly not obvious  based on the parameter values estimated at low $b$.

\subsubsection{Bi-modality beyond parallel fibers. \\ Flat directions in the fitting landscape}
\label{sec:bimodality}
\nin
The simplest nontrivial model revealing general degeneracies in parameter estimation, NODDIDA \newe (Neurite Orientation Dispersion and Density Imaging with Diffusivities Assessment) \keep \cite{jelescu2016}, is a two-compartment SM variant ($N_s=4$) that assumes a 1-parameter Watson ODF shape \cite{noddi} and sets $f_{\rm CSF}\equiv 0$. In this model, unconstrained nonlinear fitting has revealed two families (trenches) of biophysically plausible solutions to fit optimization already in the relatively small, (4+1)-dimensional, parameter space, and {\it flat directions} along them \cite{jelescu2016}.


Hence, NODDIDA exemplifies the two-fold nature of the parameter estimation challenge. Beyond the existence of multiple parameter branches (a ``discrete" degeneracy, as in the toy model above), each of them represents a shallow ``trench"  (a ``continuous" degeneracy) in the parameter landscape of nonlinear fitting. 

The flatness, or continuous degeneracy, can be formulated as having the number of estimated parameters exceeding the number of relations between the parameters obtainable from the data. In its simplest form, this problem exists already for the simplest single-directional fitting  with a biexponential function \cite{Kiselev2007}.  A normalized biexponential $S = f e^{-bD_1} + (1-f) e^{-bD_2}$ has 3 parameters; however, if $b$ is low enough so that we are practically only sensitive to the terms $\sim b$ and $\sim b^2$ --- i.e., when the DKI representation works well --- we can only estimate 2 combinations of 3 model parameters, and will have a flat direction in the corresponding 3-dimensional fitting parameter landscape. 
\mpar{R1.7}\new (This problem persists in the presence of exchange, as discussed in ref.~\cite{Nguyen2015}.) \keep

The expansion (\ref{taylor}) of the 2-compartment SM with {\it any} ODF into moments has been analytically and numerically shown to possess a similar kind of degeneracy. This framework, called LEMONADE (Linearly Estimated Moments provide Orientations of Neurites And their Diffusivities Exactly) \cite{lemonade-ismrm,rotinv}, exactly relates the moment tensors $M^{(l)}$ to SM parameters. It turns out that, at the $\O(b^2)$ level, there are only 4 independent equations, which relate rotationally invariant combinations of moments $M^{(2)}$ and $M^{(4)}$ to  5 SM parameters --- the 4 scalar ones: $f$, $\Da$, $\Depar$, $\Deperp$, and the ODF invariant $p_2$ (that characterizes \news the \keep ODF anisotropy, defined in Sec.~\ref{sec:rotinv} below).  Hence, the existence of the flat trenches in nonlinear fitting of NODDIDA is actually completely general; both the discrete bi-modality and the continuous trenches follow from the exact relations \cite{lemonade-ismrm,rotinv} between the moments and the SM parameters, and will be present for any fiber ODF. Hence, it is only capturing the moment $M^{(6)}$ that can lift both kinds of degeneracies --- as we mentioned briefly after Eq.~(\ref{Nc>Np}) above --- which is practically quite difficult to become sensitive to.   


As for the ``discrete" degeneracy, the works \cite{KM,wmdki,jelescu2016,lemonade-ismrm,rotinv} have collectively raised the fundamental question:
{\it Which ``branch" of parameters should be chosen, out of at least two biophysically plausible ones?}
\new
In ref.~\cite{rotinv}, the discrete branch choice was formulated in terms of the the ratio $\beta$ between ground truth compartment diffusivities falling or not falling within the interval 
\be \label{branch}
4 - \sqrt{\ts{40\over 3}}  < \beta  < 4 + \sqrt{\ts{40\over 3}} \,, \quad \beta =  \frac{\Da - \Depar}{\Deperp} 
\ee
determined by the discriminant of a quadratic equation. 
Note that this condition involves all three compartment diffusivities, rather than \news just \keep the two axial ones as in Sec.~\ref{sec:toy}. 
Of course, only one branch corresponds to the truth;  other(s) should be discarded.  \newe Obviously, selecting the wrong branch can radically change biophysical and diagnostic implications of the estimated parameters. Yet, \keep branch choice is nontrivial, since often times, both parameter sets look equally biophysically plausible \cite{wmdki,rotinv}, and it is very difficult to have a precise enough idea about the ground truth values, especially of $\Deperp$ whose precision crucially affects $\beta$.  
\keep
In Sec.~\ref{sec:branch}, we will comment on the branch selection.


\begin{SCfigure*}[0.5][]
\centering
\includegraphics[width=4.4in]{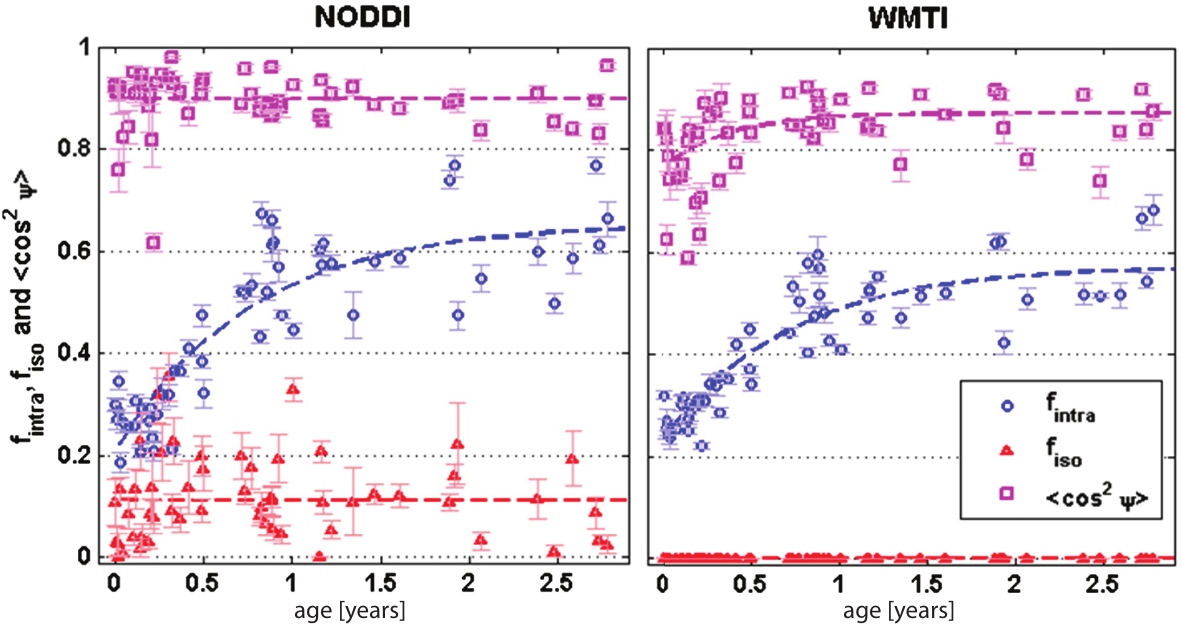}\qquad
\caption{ 
Comparison of NODDI (Sec.~\ref{sec:NODDI}) and WMTI (Sec.~\ref{sec:wmti}) 
parameter evolution with age in human corpus callosum splenium \cite{JelescuNI2015}. 
A qualitatively (but not quantitatively) similar trend of continued increase in the intra-axonal water fraction $f_{\rm intra} \equiv f$ was observed for both models, consistent with on-going myelination. 
WMTI displays a trend of increased fiber alignment (expressed by the orientation dispersion $\langle \cos^2\psi \rangle$, \newe derived from the intra-axonal diffusion tensor \keep), which could be a manifestation of continued pruning in the first year of life, while NODDI does not. 
The CSF fraction is set, $f_{\rm iso}\equiv f_{\rm CSF} = 0$, in WMTI. 
}
\label{fig:NODDIvsWMTI}
\end{SCfigure*}

\subsection{SM parameter estimation using constrain\news t\keep s}
\label{sec:constr}

\nin
Due to the challenge explained above (Sec. \ref{sec:paramest}) of fitting the SM to noisy dMRI data, especially those acquired in clinical setting, most attempts of SM parameter estimation so far are based on superimposing additional constraints on both the scalar SM parameters, as well as the fiber ODF, to improve robustness of the fitted SM parameters.

An overview of employed models used so far is given in Figure \ref{fig:models}. In what follows we consider two representative modeling approaches that have been popular because of their robustness, which potentially allows for clinical translation, and try to explain the quantitative differences in parameter estimates between them in the light of each model's assumptions and consequent biases.





\subsubsection{White Matter Tract Integrity metrics (WMTI)} 
\label{sec:wmti}
\nin
WMTI, as proposed by Fieremans \etal \cite{KM,wmdki}, extracts the 2-compartment SM  parameters by relating the DKI components to scalar parameters in the aligned-fiber framework \cite{KM}, already explained above (Sec.~\ref{sec:toy}). Subsequently, the perfectly aligned approximation was somewhat relaxed by allowing for some dispersion within the fiber bundle, \newe as described by an intra-axonal diffusion tensor, while the diffusion in the extra-axonal space is still modeled as an overall Gaussian compartment \keep \cite{wmdki}. While the advantage of using DKI eliminated the need for direct nonlinear fitting to the diffusion signal, two different biophysically plausible solutions still exist similar to the two branches as described above (Sec.~\ref{sec:toy}). In WMTI, the branch was chosen as $\Da < \Depar$ based on the available data \cite{wmdki}. Parameter histograms corresponding to this choice, yielded $f \approx 0.5\,$, $\Da \approx 1.2\,\units{\mu m^2/ms}$ and $\Depar \approx 2.5\,\units{\mu m^2/ms}$ in human corpus callosum. 

Since no specific model is assumed for the tortuosity $\Depar/\Deperp$, as $\Depar$ and $\Deperp$ are fitted separately, along with $\Da$ and $f$, it was suggested the WMTI parameters could be used to disentangle between acute damage such as neurite beading, as reflected in $\Da$ \cite{hui-stroke}, and chronic damage including different types of demyelination and axonal loss, reflected in changes in the tortuosity, $\Deperp$ and $f$ \cite{axlossdemyel,shrink-remove}.
As an {\it in vivo} validation, the age-related changes in the WMTI metrics were studied during the first two years of healthy brain development \cite{JelescuNI2015} (Figure \ref{fig:NODDIvsWMTI}), showing significant nonlinear increases in $f$, and $\Deperp$, related to increased myelination and axonal density, while no changes in the longitudinal compartment diffusivities, $\Da$ and $\Depar$, as expected.  
{\it Ex vivo} animal validation studies provided reasonably accurate estimates of $f$ in a mouse model of hypomyelination \cite{KelmNI2016} and de-, and remyelination \cite{jelescu-cpz-2016,Falangola_NBM2014}. Furthermore, mouse validation studies 
demonstrated that $\Da$ decreased during acute inflammation, while the axonal water fraction $f$ decreased during the chronic phase of cuprizone intoxication \cite{GuglielmettiNI2016}, whereby $\Deperp$ and $f$ were found to be respectively more sensitive to global and patchy demyelination \cite{jelescu-cpz-2016}. These validation studies suggest increased {\it specificity} of the WMTI parameters to microstructural changes as compared to empirical diffusion metrics. 

However, while the WMTI metrics correlate as expected with the concentration of (purely intra-axonal) NAA under the assumption $\Da < \Depar$ \cite{GrossmanNI2015}, \newe it should be noted that the measured values of $\Da$ with low $b$ dMRI protocols in the range of $1.0 - 1.2 \,\units{\mu m^2/ms}$ are significantly lower compared to recently reported values for $\Da$ measured in the range of $1.9 - 2.4 \,\units{\mu m^2/ms}$ \mpar{value 2.2 corrected} using advanced diffusion protocols providing additional information by varying TE \cite{teddi}, \news diffusion time \cite{Jespersen2017}, or using double diffusion encoding \cite{coelho2017}, cf. Sec.~4\keep, planar \cite{Dhital2017Da}, isotropic \cite{ISO_ismrm2018}, or very strong multi-directional (linear) \cite{highb} 
diffusion encoding. Further research is warranted to understand whether the discrepancy is due to a wrong choice of branch ($\Da > \Depar$, cf. Sec.~\ref{sec:branch} below, which would affect primarily estimates of $\Da$ and  $\Depar$, but not so much of $f$ or $\Deperp$), or, alternatively, due to a potential bias when estimating cumulants dMRI data over a to large $b$-range \cite{Chuhutin2017}
(affecting estimates for all model parameters). \keep  

We also note that branch selection (\ref{branch}) for the unconstrained problem (\ref{S})--(\ref{K}) is qualitatively similar but quantitatively different from that in the WMTI highly-aligned tracts case \cite{KM,wmdki}, cf.\ the toy model of Sec.~\ref{sec:toy}. 
While qualitatively, the ``wrong branch" in both the full model  (\ref{S})--(\ref{K}) and WMTI \cite{wmdki} corresponds, roughly, to swapping of intra- and extra-neurite parameters, there is no exact correspondence \newe with the full model that includes dispersion; \keep for instance,  $f$ and $\Deperp$ are also different between the branches.   
The difference between WMTI and the full model comes from the fact that in the toy model (WMTI prototype), the perfectly-aligned fiber constraint $p_2 = p_4 = 1$ has been implemented, together with effectively mixing the LEMONADE equations with moments $M^{(4),2m}$ and $M^{(4),4m}$. Therefore, the branch choice based on $\sgn(\Da-\Depar)$ is sufficiently different from that of Eq.~(\ref{branch})\news. An intermediate case between the two including dispersion was obtained in \cite{jelescu2016,Jespersen2017}, by constraining the ODF to the Watson distribution, effectively mitigating the degeneracy by parameterizing all $p_l$ in terms of the Watson distribution concentration parameter $\kappa$.\keep

\subsubsection{Neurite Orientation Dispersion and Density Imaging (NODDI)} 
\label{sec:NODDI}

\nin
NODDI, proposed by Zhang \etal \cite{noddi}, is a 3-compartment SM that assumes a Gaussian-like (Watson) ODF shape characterized by one \cite{noddi} or two \cite{noddi-bingham} parameters. In addition, the three diffusivities are effectively fixed, in the following way:  
\begin{enumerate}

\item $\Depar = \Da$   

\item $\Da = 1.7\,\units{\mu m^2/ms}$

\item Mean-field tortuosity model \cite{szafer1995}, $\Deperp /  \Depar= 1-f$. 

\end{enumerate}     
The estimated parameters are $f$ and $f_{\rm CSF}$, as well as the ODF parameters (one or two parameters, depending on the Watson  \cite{noddi} or Bingham  \cite{noddi-bingham} distribution used). 

Using high resolution {\it ex vivo} imaging, it was recently showed that Bingham-NODDI is able to capture the cortical fibers known to exhibit fanning/bending in human neocortex \cite{Tariq-ismrm2015}. \newe It was also shown that NODDI-derived dispersion agrees with histology measures in post-mortem normal and demyelinating lesions in spinal cord samples \cite{Grussu2017}. Furthermore, recent work from Schilling \etal \cite{Schilling2018} shows a strong overall correlation between the fiber orientation dispersion index (ODI) derived from NODDI versus derived from histology based on 3D confocal z-stacks in areas to the size of an MRI voxel in adult squirrel monkey brains. 

However, the same study \cite{Schilling2018} also showed a small, but systematic overestimation of the true histology-based ODI, as well as a correlation of NODDI-derived ODI with the estimates $f$ and $f_{\rm CSF}$. 
Furthermore, \keep recent extensive human measurements up to $b \leq 10 \, \units{ms/\mu m^2}$ \cite{rotinv} also suggest that the above three parameter constraints generally do not hold, and therefore may bias the estimates of the fractions and fiber dispersion.

While both NODDI and WMTI rely on the same overarching SM, they \newe have different constraints, particularly in terms of the \keep compartment diffusivities (fixed in NODDI, fitted in WMTI), the ODF (Watson in NODDI, single bundle in WMTI), and number of compartments (3 in NODDI, 2 in WMTI). The effect of these different constraints have been evaluated by studying changes in the model parameters through normal human early development \cite{JelescuNI2015} (Figure \ref{fig:NODDIvsWMTI}). In this work, qualitatively similar trends were observed in $f$, 
in full agreement with expected on-going myelination, fiber classification and asynchrony of development. The quantitative estimates, however, are model-dependent, exhibiting biases and limitations related to the models' assumptions. Similarly, changes during the first two years in fiber dispersion in the splenium corpus callosum were qualitatively different between NODDI and WMTl. 
This illustration clearly calls for extreme caution when interpreting modeling studies based on limited clinical dMRI data, where accuracy is typically sacrificed in favor of precision. \newe Indeed, both WMTI and NODDI have made assumptions that allowed for a robust, rather than accurate estimation of the SM model parameters that are not fixed according to each model. This prompts both for improved SM parameter estimation methods (discussed next in Sec.~\ref{sec:rotinv}), as well as for ``orthogonal" and more comprehensive validation methods to gain better understanding of the relevant tissue features of modeling (discussed in Sec.~\ref{sec:branch}), prior to applying them to clinical dMRI data. \keep

\subsection{ODF factorization and rotational invariants }
\label{sec:rotinv}

\nin
Let us now introduce the recently proposed family of approaches to SM parameter estimation that do not rely on specifying the ODF shape,
by factoring it out in a rotationally invariant way. This will enable  separation of estimating the scalar and the tensor (ODF) parameters. Of course, all the degeneracies of the parameter estimation will persist --- and in fact, factorization has been used as a tool to prove that the above discussed degeneracies are completely general \cite{rotinv}. 

Much like convolutions become products in the Fourier domain, the convolution (\ref{S}) between the individual fiber response $\K$ and the ODF $\P$ becomes a product in the ``spherical Fourier" domain (i.e., the SH basis) \cite{Healy1998}:
\be \label{fact}
S_{lm}(b) = p_{lm} \, K_l(b)
\ee
where $K_l(b)$ is the projection of the kernel $\K(b,\xi)$ onto the Legendre polynomial $(-1)^{l/2}P_l(\xi)$ \cite{Jespersen2007,rotinv,baydiff}.

Since any rotation transforms SH components $S_{lm}$ and $p_{lm}$ according to a unitary transformation belonging to the $(2l+1)-$dimensional irreducible representation of SO(3) group labeled by ``angular momentum" $l$, the 2-norms 
$\| p_{lm} \| \equiv \sqrt{\sum_{m = -l }^l | p_{lm} |^2}$ and $\| S_{lm}\|$ (defined likewise) are conserved under rotations, i.e., are rotational invariants. 
It is thus convenient to introduce%
\footnote{The idea to operate with a single ``energy" $L_2$ norm per each ``frequency" band $l$ of SH  has been previously applied, e.g., to the problem of shape matching in computer graphics \cite{kazhdan2003} and recently for dMRI data harmonization \cite{Mirzaalian2016}.} 
{\it rotational invariants}  
$p_l \equiv \| p_{lm}\| /{\cal N}_l$ and $S_l \equiv \| S_{lm}\| /{\cal N}_l$, where normalization ${\cal N}_l = \sqrt{4\pi(2l+1)}$ 
is chosen so that $0 \leq p_l \leq 1$. 
Hence, equations (\ref{fact}) for the $(l,m)$ SH components give rise to the corresponding equations for the rotational invariants 
\cite{rotinv,baydiff}, 
\be \label{S=pK}
S_l (b,x) = p_l \, K_l(b,x) \,, \quad l = 0, \ 2, \ \dots \,,
\ee
where we denoted by $x$ the dependence on the kernel's scalar parameters $x = \{ f, \Da, \Depar, \dots \}$ to be estimated. 
The invariant $p_0\equiv 1$ is trivial (ODF normalization); the remaining ODF invariants $p_l$, one for each $l$, characterize its anisotropy irrespective of the chosen basis.

\subsubsection{ Isotropic $l=0$ invariant $K_0(b)$}
\label{sec:isoavg}

\nin
The $l=0$ invariant for Eq.~(\ref{S=pK}) has been independently 
introduced as ``powder averaging" and ``spherical mean" 
\cite{jespersen2013,lasic2014,kaden2015,Hansen2015,smt,DIVIDE}. 
The ODF factorization in this case simply follows 
from swapping the order of integrations over $\g$ and $\n$: 
\bea \non
S_0 &\propto& \int\!  \d\g \, \int\! \d\n\, \P(\n) \, \K(b,\g\cdot\n)  
\\ 
&=& 
\int\!  \d\n \,\P(\n) \, \int\! \d\g \, \K(b,\g\cdot\n) 
\equiv \int_0^1\! \d\cos\theta \, \K(b,\cos\theta) \,,  
\non
\eea 
since $\int\! \d\g \, \K(b,\g\cdot\n)$ is independent of fiber direction $\n$ due to the ``translational invariance" on a unit sphere, 
and the ODF is normalized to $\int\! \d\n \, \P(\n) \equiv 1$. The last identity above gives the projection of kernel (\ref{K}) onto the $l=0$ Legendre polynomial $P_0(\xi) \equiv 1$, where $\xi = \cos\theta$ in our case; 
for a stick compartment, this projection yields Eq.~(\ref{stick-iso}) above.

\subsubsection{ Rotational invariants $K_l(b)$ for  $l=2,4,\dots$}

\nin
Equation (\ref{S=pK}) formally yields an infinite family of rotational invariants $K_l(b)$ \cite{baydiff,rotinv}, one for every  $l=2,4,\dots$. However, it turns out that by far the most useful is the next-order, $l=2$ invariant, since the projections of $e^{-bD \xi^2}$ onto the Legendre polynomials with $l>2$, giving the compartment contributions to $K_l(b,x)$, are too slowly varying \cite{Jespersen2007} \newe and thereby \keep adversely affecting the sensitivity to the estimated parameters $x$. 

We also note that including the $l>0$ invariants in system \eq{S=pK} is only possible for anisotropic ODFs, with $p_l > 0$. 
Physically, it is expected since the less symmetric the system, the more inequivalent ways it enables for probing it.%
\footnote{This intuition underlies theory of quantum-mechanical excitations of non-spherical nuclei \cite{bohr-mottelson}, where analogs of our rotational invariants are the corresponding irreducible tensor operators underpinning the Wigner-Eckart theorem.} 
In the brain, the ODF is at least somewhat anisotropic; its lowest-order invariant $p_2$ 
is generally nonzero even in GM.

Parameter estimation based on the ODF factorization via the rotational invariants amounts to inverting the nonlinear relations
 (\ref{S=pK}) with respect to model parameters $x$ and $p_l$. 
 Such inversion has so far been technically implemented in four distinct ways:  
\renewcommand{\theenumi}{\alph{enumi}}
\begin{enumerate}

\item Analytically inverting  relations between their Taylor expansions --- i.e., expressing model parameters in terms of the moments of the signal (LEMONADE)  \cite{lemonade-ismrm}.  
\new
At typical $bD\sim 1$, the biases in estimating the moments cause notable bias \news in \keep the model parameters. 
 
 \item Using the LEMONADE output as initialization for the RotInv solution of Eqs.~(\ref{S=pK}) via nonlinear fitting using the gradient-descent optimization of the corresponding objective function \cite{rotinv}. 
 \new This notably increases  the accuracy of LEMONADE. \keep

\new
\item The prevalence method \cite{rotinv}: To avoid the branch selection issue, initialize the fit objective function for Eq.~(\ref{S=pK}) with a large number ($\sim 20-100$) of \newe {\it random} \keep \new starting points within the  plausible parameter range (e.g., $0<f, p_2<1$, and $0<D<3$ for all diffusivities),  observe that the fit outcomes cluster around a few sets in the parameter space, and select 
the mean of the largest cluster (after excluding outcomes outside the bounds). 
The method  works best for large $b$, say, $b\gtrsim 5\,\units{ms/\mu m^2}$, 
since increasing $b$ broadens the basin of attraction of the true minimum \cite{jelescu2016}.  
\keep

 \item Machine learning framework: Generate  distributions of the invariants based on the prior distributions of $x$, and numerically invert these relations based on the training set \cite{baydiff}. \newv The invertibility of these relations requires the resolution of the bi-modality problem (section~\ref{sec:paramest}). In particular, the constraint of close traces of intra- and extra-axonal tensors, $|\Depar+2\Deperp-\Da| < 1.5\,\units{\mu m^2/ms}$, was applied according to results obtained using isotropic diffusion weighting \cite{Dhital2017}. While it is the fastest data processing method, 
 \new it is sensitive to the way the training data are \newv generated. \keep
 
\end{enumerate}

Overall, our current experience tells that, no matter the implementation, the sensitivity to different scalar parameters varies dramatically (e.g., $f$ is obtained reasonably well while the sensitivity to $\Depar$ and $\Deperp$ is much worse) \cite{rotinv,baydiff}; 
with decreasing SNR, methods (a)--(c) yield noisier parameter maps while method (d) yields ``too clean" maps completely dominated by the mean values of parameter priors;  branch multi-modality manifests itself in the  need for the branch selection (\ref{branch}) in all these approaches.

It is yet difficult to evaluate {\it accuracy} of these methods in vivo because of lack of understanding of what the ground truth is, and because all these methods are strongly dependent on the branch selection/initialization/priors.  

The lack of precision (due to the ``continuous" degeneracy of shallow trenches) generally exists due to the multi-compartmental nature of the kernel (\ref{K}) \cite{rotinv}. One can say that any standard (directional) dMRI measurement  effectively under-samples the scalar part of the model (\ref{S}), not providing enough relations between the scalar parameters (cf. Sec.~\ref{sec:bimodality} above), and over-samples the tensor (ODF) part. In other words, the system's true complexity lies within the kernel's parameters hidden in functions $K_l(b,x)$, Eq.~(\ref{S=pK}) --- while the ODF is in some sense  ``on the surface". 

This prompts the need for ``orthogonal" measurement schemes \cite{Dhital2017,szczepankiewicz2015,martins2016,DIVIDE,Skinner2017,CODIVIDE,teddi,Dhital2017Da}   
which probe the scalar parameters in different combinations than entering the kernel projections $K_l(b,x)$, as we are now going to discuss. 
 

\subsection{Open questions: Precision and branch selection}
\label{sec:branch}

\nin
\mpar{sec rewritten}
Estimating precise maps of ground truth values, as well as the branch selection (\ref{branch}), remains an essential problem for quantifying neuronal microstructure, and is currently an active topic of research.  Recent experiments using advanced dMRI protocols have been either employing  very strong diffusion gradients (e.g., on unique Connectome scanners with gradients up to 300\,mT/m) \cite{highb}, or adding ``orthogonal'' acquisitions such as extra-neurite water suppression by strong  unidirectional gradients \cite{Skinner2017} or planar diffusion weighting \cite{Dhital2017Da}, isotropic diffusion weighting \cite{dhital-isodw,szczepankiewicz2015,DIVIDE,martins2016,CODIVIDE,ISO_ismrm2018},  and varying other parameters, such as the echo time \cite{teddi} and the diffusion time \cite{Jespersen2017,lemo-t-ismrm}. 

The choice of the branch, and an independent estimation of the compartment diffusivities $\Da$ and $\Depar$ is of particular interest. \keep   
Isotropic weighting (spherical tensor encoding) yields
\be \label{Siso}
S(b)/S_0 = f e^{-b\Da} + (1-f) e^{-b(\Depar+2\Deperp)} \,, 
\ee
which seems to produce relations $\Da \approx \Depar + 2\Deperp$ due to an empirically small iso-weighted kurtosis of signal (\ref{Siso}) \cite{CODIVIDE,Dhital2017}.  While this can be interpreted as favoring one of the branches, this relation cannot be used as a global constraint: \citet{szczepankiewicz2015} show it failing in thalamus (note however that thalamus is a GM/WM mixture).
 Another possibility \news for using orthogonal measurements to resolve the parameter estimation degeneracy is the application of double diffusion encoding (DDE), see Sec.~4, with promising preliminary results \cite{coelho2017}. In rat spinal cord, DDE seems to indicate the branch-merging case  $\Da \approx \Depar$ \cite{Skinner2017}.\keep\ Note that such assumption is made in NODDI (section \ref{sec:NODDI}), albeit this model fixes (rather than fits) the compartment diffusivities to equal values.  This assumption does not seem to universally hold in the human brain \cite{rotinv}. 
 As the ultimate goal of biophysical modeling is to study pathological and other changes (e.g., aging and development), it is imperative to estimate the compartment diffusivities independently, because changes in one of them may indicate the earliest sign of a pathological or other process of interest.

Overall, the Standard Model presents a microcosm of parameter estimation challenges: a relatively low SNR in clinical dMRI coupled with both discrete and continuous degeneracies, require careful validation and prompt employing the widest possible arsenal of measurements, to probe  parameters from as many vantage points as possible. Achieving compartmental specificity, crucial in studying pathological and other processes, 
remains a difficult but worthy goal.



%% file: Section_IV.tex

\section{Multiple diffusion encodings} 
\label{sec:mpfg}

\epigraph{The whole is greater than the sum of its parts}{Aristotle}

\subsection{MDE basics}

\begin{figure}[b!]
\centering
\vspace{-5mm}
\includegraphics[width=0.9\columnwidth]{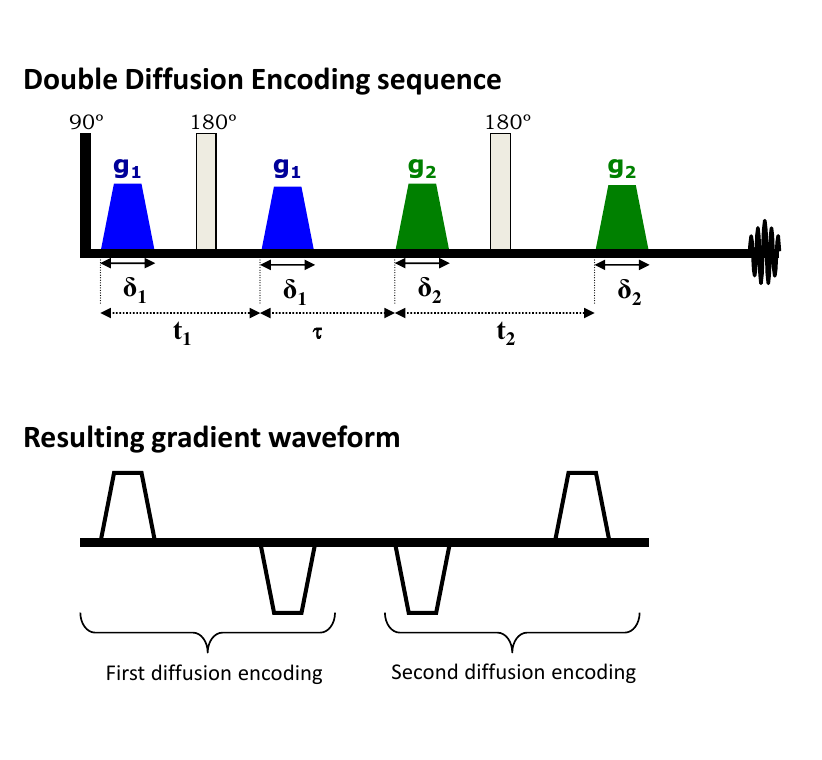}
\vspace{-10mm}
\caption{
(a) Example of a DDE sequence within the framework of a double spin echo. (b) The resulting gradient waveform {\news obtained by multiplying each gradient by $(-1)^{n_\pi}$, where $n_\pi$ is the number of $\pi$ pulses following the given gradient. }
In {\news the} text, we assume narrow-pulse approximation, such that $\delta_i \to 0$, with the Larmor frequency gradients $g_i$
sufficiently large to yield finite $q_i = g_i \delta_i$ (no summation over $i$). 
}
\label{fig:dpfg}
\end{figure}

\nin
Multiple diffusion encoding (MDE) generalizes the Stejskal-Tanner (Sec.~\ref{sec:qtI}) pulse sequence design by adding one or more extra diffusion weighting blocks, as illustrated in Fig.~\ref{fig:dpfg} for the case of double diffusion encoding (DDE) \cite{RN200,RN226,RN47,RN1646}. Figure~\ref{fig:dpfg} also defines the main pulse sequence parameters for DDE, which in addition to the familiar pulse-gradient parameters of each block, includes a mixing time $\tau$. 

In the following, we will restrict our attention to the narrow pulse limit.
Generally, each diffusion weighting block is characterized by an independent diffusion wave vector ${\q_n}$ and diffusion time%
\footnote{Set by the corresponding interval $\Delta_n$ between the fronts of the gradient pulses, Fig.~\ref{fig:dpfg}.  
For finite pulse width $\delta_n$, see footnote \ref{foot:difftime} in Section \ref{sec:Dt}.}
 $t_n$, and the mixing times define delays between blocks. 
 Thus, a rich set of experimentally controllable parameters can in principle enable qualitatively different ways of probing the microstructure,
 as compared with the conventional, single diffusion encoded (SDE) sequences. 

The fundamental question of the information content of MDE signal $S_N$ relative to a set of independently acquired SDEs $S_1$ can be 
formulated using an example of the DDE signal $S_2$ (here we ignore the trivial $e^{-R_2 t}$ factors, assuming the unweighted signal to be normalized to unity): 
\begin{widetext}
\bea \non
S_2(\q_1,\q_2, t_1,t_2,\tau) 
&\equiv &
\left\langle e^{ i\q_1 \cdot [\r(0) - \r({t_1})] + i\q_2 \cdot [\r({t_2} + \tau ) - \r({t_1} + {t_2} + \tau )]} \right\rangle 
\\ &=&  
\int\! \frac{\d{\r_{1a}}\d{\r_{1b}}\d{\r_{2a}}\d{\r_{2b}}}V 
e^{ i{\q_1} \cdot ({\r_{1a}} - {\r_{1b}}) + i{\q_2} \cdot ({\r_{2a}} - {\r_{2b}})}  \, 
{\G_{{t_2};{\r_{2b}},{\r_{2a}}}} 
{\G_{\tau ;{\r_{2a}},{\r_{1b}}}}
{\G_{{t_1};{\r_{1b}},{\r_{1a}}}}
\stackrel{?}{=} 
{G_{{t_2};{\q_2}}} {G_{{t_1};{\q_1}}} \,. \qquad
\label{SneqGG}
\eea   
\end{widetext}
Technically, the above question is as follows: When does the convolution of the local propagators (\ref{def-Gcal}) defined in Sec.~\ref{sec:qtI}
contain more information than the product of the voxel-averaged translation-invariant  SDE propagators (\ref{def-G})?

Let us first outline the three cases when Eq.~(\ref{SneqGG}) holds, i.e., there is {\it no} extra information in MDE relative to SDE. 
\begin{enumerate}
\item {\it Microscopic translation invariance:} If $\G_{t; \r_b,\r_a} \equiv G_{t,\r_b - \r_a}$ depends only on the relative displacement, the above equality holds for {\it any} $G_{t,\r}$. 
Of course, this is true for the Gaussian diffusion, when $G_{t,\q}$ is described by Eq.~(\ref{Gbare}), but the statement is much broader, since its proof (by change of integration variables $\r_1 = \r_{1b}-\r_{1a}$ and $\r_2 = \r_{2b}-\r_{2a}$) involves only the translation invariance requirement. 
Practically, this means that the time scales involved in 
Eq.~(\ref{SneqGG}) exceed the time needed for the coarse-graining to restore sample's  translation invariance, whether this implies the Gaussian fixed point (\ref{Gbare}) or its anomalous counterpart, cf.\ Sec.~\ref{sec:fixedpoint}. 

\item {\it Long mixing time limit of a single pore:} If all spins are confined in the same pore of volume $V$, and $\tau$ exceeds the time to diffuse across the pore size, the ``mixing" propagator ${\G_{\tau ;{\r_{2a}},{\r_{1b}}}} \to 1/V$ approaches a constant, 
and Eq.~(\ref{SneqGG}) again factorizes, irrespective of the (non-translation-invariant) functional form of $\G_{t; \r_b,\r_a}$. 

\item {\it Weak diffusion weighting:} Equality (\ref{SneqGG}) holds for any $\G$ at the level of $\O(q^2)$ \cite{jespersen-equiv}, cf.\ Sec.~\ref{sec:mde-q2} below. For this statement to hold, it is only required that the 
\news cumulant expansion (\ref{cum}) has a nonzero convergence radius. (This common property breaks down for a diffusion propagator of a stretched-exponential form, whose assumptions contradict experimental evidence  \cite{manifesto}.) 
\keep
\mpar{R1.8} 

\end{enumerate}

Generally, the above requirements do not hold --- tissues are microscopically {\it not} translation-invariant, 
a voxel can contain multiple pores of various shapes (cf.\ Sec.~\ref{sec:mde-q4} below), and diffusion weighting can be strong, 
so that the $\O(q^4)$ terms are relevant. This justifies using MDE to obtain extra information. 
 
To get a feel for the difference between DDE and SDE, it is instructive to consider the long mixing time limit $\tau\to\infty$
in a system of disconnected pores. The average over Brownian paths splits into two parts: an average  $\langle \dots \rangle$ over paths within a single pore, followed by an average (denoted by an overbar) over pores $\alpha=1,\dots$, with volume fractions 
$w_\alpha = V_\alpha/V$ adding to unity. 
For example, for the SDE signal (\ref{S=G}):
\begin{widetext}
\bea \non
S_1(\q_1,t) &=&
\overline{\left\langle e^{ i\q_1 \cdot (\r(0) - \r(t)) }\right\rangle}_{\textrm{paths in pores} } 
\equiv 
\sum\limits_{{\rm{pores:\ }}\alpha}  w_\alpha \, \langle e^{  i\q_1 \cdot (\r(0) - \r(t)) } \rangle_{\textrm{paths in }\alpha } \,. \quad
\eea

For DDE, spin displacements in each of the diffusion-weighting blocks become independent of one another within each pore in the limit  $\tau\to\infty$, 
${\G_{\tau ;{\r_{2a}},{\r_{1b}}}} \to 1/V_\alpha$ if  $\r_{2a}$ and $\r_{1b}$ in Eq.~(\ref{SneqGG}) are from the same pore $\alpha$
with volume $V_\alpha$,
while the probability to hop between pores is zero: ${\G_{\tau ;{\r_{2a}},{\r_{1b}}}} \equiv 0$ if $\r_{2a}$ and $\r_{1b}$ belong to different pores
$\alpha \neq \beta$. 
This effective Kroeneker $\delta_{\alpha\beta}$ 
eliminates the cross-terms between different pores that are present in the right-hand side of Eq.~(\ref{SneqGG}): 
\bea 
S_2(\q_1, \q_2, t_1, t_2, \tau) 
&=& \overline{\left\langle e^{ i\q_1 \cdot (\r(0) - \r(t_1)) + i\q_2 \cdot [\r({t_2} + \tau) - \r(t_1 + t_2 + \tau)] } \right\rangle}
= \sum_\alpha w_\alpha  \left\langle e^{  i{\q_1} \cdot (\r(0) - \r({t_1}))} \right\rangle_\alpha \left\langle e^{ i{\q_2} \cdot (\r({t_2} + \tau ) - \r({t_1} + {t_2} + \tau ))} \right\rangle_\alpha . \qquad
\label{SneqSS1}
\eea
Here the subscript ``paths in $\alpha$'' was replaced with $\alpha$ for brevity. On the other hand, for the product of two SDE's we have 
\bea \non
S_1({\q_1},{t_1})\, S_1({\q_2},{t_2}) &=& \overline {\left\langle e^{  i{\q_1} \cdot (\r(0) - \r({t_1}))} \right\rangle } \,\cdot \,{\overline {\left\langle e^{ i{\q_2} \cdot (\r({t_2} + \tau ) - \r({t_1} + {t_2} + \tau ))} \right\rangle }} \\
&=&
\sum_\alpha w_\alpha w_\beta {\left\langle e^{  i{\q_1} \cdot (\r(0) - \r({t_1}))} \right\rangle_\alpha }   { {\left\langle e^{ i{\q_2} \cdot (\r({t_2} + \tau ) - \r({t_1} + {t_2} + \tau ))} \right\rangle_\beta }} \neq  S_2(\q_1, \q_2, t_1, t_2, \tau)  \,.
\label{SneqSS2}
\eea   
\end{widetext}
The physical meaning of the above equations is as follows: 
it is not possible in general to split the coherent averaging of the product (\ref{SneqSS1}) over pores 
into the product of the averages \eq{SneqSS2}. 

The coherent disorder averaging of the propagators in equation (\ref{SneqGG}) is also the reason that the effective medium theory \cite{EMT,mesopnas} for the disorder-averaged SDE propagator (\ref{S=G}) has to be further augmented to incorporate the coarse-graining effects for MDE, relevant at finite $\tau$.

\subsection{Equivalence between MDE and SDE at $\O(q^2)$}
\label{sec:mde-q2}

\nin
Historically, DDE was noted to provide a method for determination of compartment dimensions \cite{RN47} 
at low diffusion weighting in the limit of zero mixing time and long diffusion times. 
Taylor-expanding Eq.~(\ref{SneqSS1}) in this limit, Mitra \etal      
 \cite{RN47} showed that in a system of identical pores
\be \label{S-Mitra}
S_2(q\n_1, q\n_2, t, t, 0) \ \mathop{\longrightarrow}_{t\to\infty} \ 1 - \frac13 q^2 \langle r^2 \rangle \lb 1 + 2\cos^2{\theta\over 2} \rb ,
\ee
where $\theta $ is the angle between the directions $\n_1$ and $\n_2$ of the diffusion wave vectors, and 
$\langle {r^2}\rangle = \int\! \d\r\d\r' \,  (\r-\r')^2 /2V^2 \equiv  \int\! \d\r \,  (\r-\r_{\rm cm})^2/V$
is the pore mean squared radius of gyration ($\r_{\rm cm}$ is pore center-of-mass), a measure of pore size. 

Hence, a measure of the pore size can be determined from the signal dependence on diffusion wave vector angle in isotropic systems, or more simply from the signal difference between parallel and antiparallel diffusion wave vectors. 

Equation (\ref{S-Mitra}) has since been generalized to take into account, e.g., partial volume, multiple concatenations, pulse sequence timings (e.g., finite gradient width) for various geometries \cite{RN214,RN1374,RN501,RN216,RN1051,RN1053,RN1376}. This has later been demonstrated by several groups in model systems and biological samples {\it ex vivo} \cite{RN206,RN428,RN1341,RN492,RN213,RN204,RN215}, and {\it in vivo} in humans \cite{RN809,RN493,RN204,RN810}. 

However, it was recently realized  \cite{jespersen-equiv} that 
this property, \mpar{R2.1} \news  i.e., 
the sensitivity (\ref{S-Mitra}) to pore gyration radius, \keep is a general feature of any diffusion-weighted signal at the $\O(q^2)$ level, 
and hence it does not rely on information beyond that already contained in the SDE signal, which in the same regime behaves as \cite{RN808}:
\bea \label{S-Sune}
\non
S_1(\q, t) \mathop{\longrightarrow}_{t\to\infty}  & & 1  - \frac1{2}q_iq_j \langle (x_i(t)-x_i(0))(x_j(t)-x_j(0))\rangle \\ 
=  & & \ 1 - q^2 \langle r^2 \rangle.
\eea
More generally, it was shown \cite{jespersen-equiv} that up to order $\O(q^2)$, 
\begin{widetext}
\bea \non
\ln {S_2}(\q_1, \q_2, t_1, t_2, \tau) &=&  - {q_{1i}}{q_{1j}}{D_{ij}}({t_1}){t_1} + {q_{2i}}{q_{2j}}{D_{ij}}({t_2}){t_2} 
\\ \non
 &+& {q_{1i}}{q_{2j}}
 \lb {{D_{ij}}({t_1} + {t_2} + \tau )({t_1} + {t_2} + \tau ) + {D_{ij}}(\tau )\tau  - {D_{ij}}({t_1} + \tau )({t_1} + \tau ) - {D_{ij}}({t_2} + \tau )({t_2} + \tau )} \rb
 \\ 
 &+& \O(q^4) \,, 
 \qquad
\label{S-Dt}
\eea   \end{widetext}
where $D_{ij}(t)$ is the cumulative diffusion tensor, Eq.~(\ref{Dcumtens}). This explicitly demonstrates that the signal is fully characterized by the time-dependent diffusion tensor, a quantity which is obtainable from the  SDE acquired at a few diffusion times. This statement is valid for {\it any} diffusion sequence up to second order in the diffusion wave vector \cite{RN1641,jespersen-equiv}, and is a consequence of the existence of the cumulant series, whose lowest order can be completely reproduced by knowing the diffusion tensor for all $t$ or, equivalently, for all $\omega$  \cite{Kiselev2010_diff_book,sv-og}, 
\begin{equation}\label{cum2}
\ln S = -\frac12 \int \la v_i(t_1) v_j(t_2) \ra q_i(t_1)q_j(t_2) \,\d t_1 \d t_2 ,
\end{equation}
where $\q(t)$ is the time integral of the arbitrary-shaped applied gradient, $\v$ is the molecular velocity, $\v=\partial_t \r$, and no bulk flow is assumed as usual. The (symmetric) autocorrelation function 
$\la v_i(t_1) v_j(t_2) \ra \equiv \D_{ij}(|t_1-t_2|)$ is constructed out of its retarded counterpart
\be \label{defD}
\D_{ij}(t) \equiv \theta(t) \la v_i(t) v_j(0) \ra  = {\partial^2 \over \partial t^2} \lb t D_{ij}(t) \rb 
\ee
defined by generalizing Eqs.~(\ref{Dt=vv}) and (\ref{Dw=Dt}) to the anisotropic case. 

\begin{figure}[t]
\centering
\includegraphics[width=0.7\columnwidth]{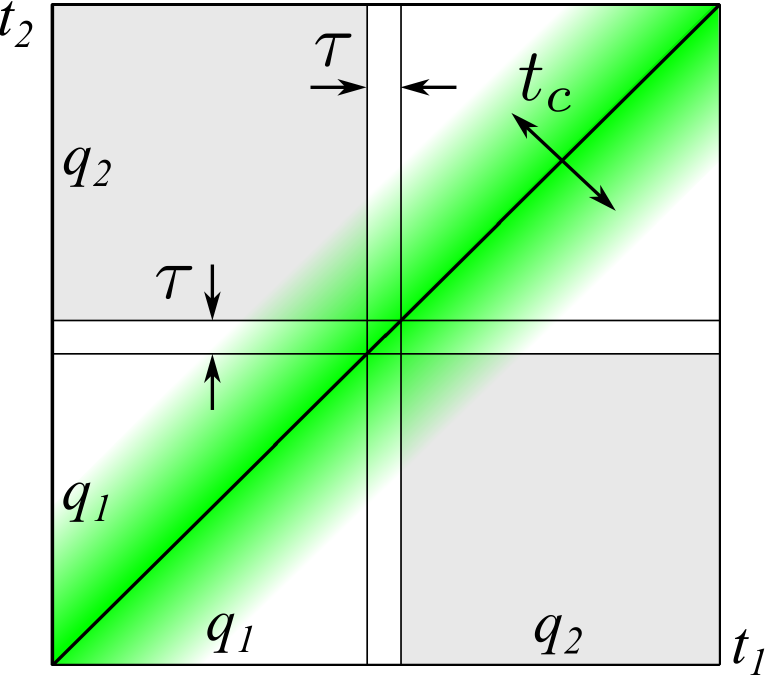}
\caption{
Two-dimensional temporal integration involved in the second-order cumulant, Eq.\,(\ref{cum2}) leading to Eq.~(\ref{S-Dt}) for the DDE measurement. Labels $q_1$ and $q_2$ indicate the time interval in which $\q(t)$ equals to $\q_1$ and $\q_2$, respectively; {\news for simplicity, the vector indices are not shown.} The green shaded area along the diagonal symbolizes $\D_{ij}(|t_1-t_2|)$, Eq.~(\ref{defD}), where it significantly deviates from $0$, 
with the width of this region set by the correlation time $t_c$. The nontrivial cross-term $q_{1i}q_{2j}$ in Eq.~(\ref{S-Dt}) arises from the off-diagonal quadrants. As this contribution is weighted with the velocity autocorrelation function, it tends to zero when the mixing time, $\tau$ (indicated by the thin lines along each dimension) becomes larger than the correlation time, $\tau\gg t_c$. 
In particular, no non-trivial cross-term is present for Gaussian diffusion, for which $t_c\to 0$. 
}
\label{fig:dpfg-corr}
\end{figure}

The function (\ref{defD}) is generally {\it nonlocal} in $t$ 
\cite{Does2003,Froehlich2005,Froehlich2008,Kiselev2010_diff_book,mesopnas,EMT,sv-og,Kiselev2016_NMB_review}.
Figure \ref{fig:dpfg-corr} illustrates how this nonlocality, integrated in Eq.~(\ref{cum2}), 
 gives rise to the cross-term $\sim q_{1i} q_{2j}$ (second line of Eq.~(\ref{S-Dt})); 
this term disappears in the Gaussian diffusion limit, when $\la v_i(t_1) v_j(t_2)\ra  = 2D_{ij}\delta(t_1-t_2)$ is infinitely narrow, 
and the function (\ref{defD}) is concentrated along the diagonal.  
Hence, the cross-term $\sim q_{1i} q_{2j}$ directly probes the time-dependence of the diffusion coefficient in Eq.~(\ref{S-Dt}), 
cf.\ Section \ref{sec:Dt}.     


\twocolumngrid

\begin{figure}[t]
\centering
\includegraphics[width=0.7\columnwidth]{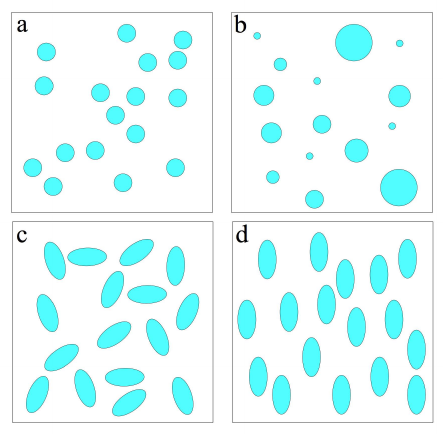}
\caption{
Examples of model systems considered in the text. 
In (a), a system of identical spherical pores is shown, whereas, in (b), the pores have a distribution of sizes. In (c), an approximately isotropic distribution of ellipsoidal pores is sketched and, in (d), the pores are coherently oriented.
 Systems (a)--(c) are macroscopically isotropic, system (d) is not. 
Systems (c) and (d) are {\it microscopically anisotropic}. 
 Ensemble heterogeneity is only seen in systems (b) (size) and (c) (orientation). 
Here, spins contributing to the signal are assumed to only reside within the pores.
}
\label{fig:microanis}
\end{figure}

\subsection{Extra information relative to SDE at $\O(q^4)$ and beyond. Microscopic anisotropy}
\label{sec:mde-q4}

\nin
At larger values of the diffusion weighting, double diffusion encoding was shown from the beginning to have the ability to characterize microscopic anisotropy ($\mu$A) in systems which are macroscopically isotropic, Fig.~\ref{fig:microanis}, see panel (c). Thus, in an early application of the sequence by Cory \etal \cite{RN226}, DDE was used to quantitatively measure the eccentricity of yeast cells, which was shown to be directly related to the difference in signals acquired with parallel and perpendicular diffusion wave vectors. This has since been explored by many authors, e.g., in phantoms \cite{RN684,RN683}, {\it ex vivo} tissues \cite{RN779,shemesh2011}, and {\it in vivo} 
\cite{RN204,RN547,YangDDE2017}. 

The basic sensitivity to anisotropic pores can be understood already from Eq.~(\ref{SneqSS1}) in the long diffusion time and long mixing time limit 
\be \label{S=chichi}	
S_2({\q_1},{\q_2},{t_1},{t_2},\tau ) = \overline {{{\left| {\chi_\alpha ({\q_1})} \right|}^2}{{\left| {\chi_\alpha ({\q_2})} \right|}^2}}  
\ee
%
where $\chi_\alpha (\q)$ is the Fourier transform of the pore structure function, defined as $\chi_\alpha(\r)\equiv 1/V_\alpha$ inside a pore and 0 otherwise \cite{RN208}. 

For spherical pores, the structure function is isotropic, hence $\chi_\alpha(\q)$ does not depend on the direction of $\q$. For anisotropic pores, say ellipsoids, the anisotropic structure functions $\chi_\alpha(\q)$ in Eq.~(\ref{S=chichi}) ensure that the result depends on the directions of ${\q_1}$ and ${\q_2}$. This is because diffusion in the two directions is generally correlated when pores are nonspherical. 

If the overall system is macroscopically isotropic, i.e., the orientations of the individual pores are randomly distributed (Fig.~\ref{fig:microanis}c), the signal will be unaffected by rotations of the sample or of the laboratory system of reference, but the dependence on the relative angles between ${\q_1}$ and ${\q_2}$ will survive the pore averaging in Eq.~(\ref{S=chichi}). Mathematically speaking, this is because the two terms in the product 
$|{\chi_\alpha}({\q_1})|^2 |{\chi_\alpha}({\q_2})|^2$ 
are not independent for a given pore, and hence the average of the product is different from the product of averages. 

A convenient measure of the eccentricity of the pore space (microscopic anisotropy) can therefore be found from the difference of  DDE signals acquired with parallel and perpendicular wave vectors $\q_1$ and $\q_2$. In the presence of macroscopic and microscopic anisotropy, the signal will depend on the orientations of {\it both} $\q_1$ and $\q_2$, and microscopic diffusion anisotropy can no longer be extracted simply from the difference between parallel and perpendicular diffusion wave vectors. 
The rotationally invariant way to circumvent this (cf.\ Sec.~\ref{sec:rotinv} above) is to {\it powder average} the signal, 
analogously to how the $K_0$ invariant was introduced in Sec.~\ref{sec:isoavg}
(although, technically, the averaging here is over the SO(3) group instead of a 2-sphere),
and practical recipes for doing this were proposed in refs.~\cite{RN779,RN1166,RN1678}. Microscopic diffusion anisotropy can then be defined as the difference between (log of) the powder averaged signals acquired with parallel and perpendicular diffusion wave vectors \cite{RN779}.

Mitra's original results were since generalized to arbitrary diffusion and pulse timings by several authors \cite{RN779,RN1166,RN195}. Specifically, it was shown that in the long mixing time regime and for arbitrary diffusion times, the signal can be written as
\bea \non	
\ln S_2({\q_1},{\q_2}) &=&  - ({q_{1i}}{q_{1j}} + {q_{2i}}{q_{2j}}){D_{ij}(t)}t 
\\ \non
&&+ \frac{\overline{D}^2}{6} \, ({q_{1i}}{q_{1j}}{q_{1k}}{q_{1l}} + {q_{2i}}{q_{2j}}{q_{2k}}{q_{2l}}){W_{ijkl}(t)} 
\\ 
&& + \frac1{4} \, {q_{1i}}{q_{1j}}{q_{2k}}{q_{2l}}{Z_{ijkl}(t)}
\label{S-q4}
\eea
where $W$ is the kurtosis tensor as defined in \cite{DKI} from the cumulant expansion (\ref{cum}) of the SDE propagator, 
whereas $Z$ is a rank-4 tensor, unique to DDE, defined as
\be \label{Z}
Z_{ijkl} = 4{t^2}\left( {\overline {D_{ij}^\alpha D_{kl}^\alpha}  - \overline {D_{ij}^\alpha} \,\overline {D_{kl}^\alpha} } \right) .
\ee
The tensors $D^\alpha_{ij}(t)$ refer to the microscopic $t$-dependent diffusion tensors characterizing diffusion within the pores, 
and the SDE-measured overall diffusion tensor $D_{ij}(t)$ entering the first line of Eq.~(\ref{S-q4})  is an average over all pores, 
$D_{ij}(t) = \overline{D_{ij}^\alpha(t)} \equiv \sum_\alpha w_\alpha D_{ij}^\alpha(t)$.

The new tensor $Z$, Eq.~(\ref{Z}), accessible with DDE (and inaccessible with SDE), is  proportional to the covariance tensor of microscopic diffusion tensors. Microscopic diffusion anisotropy, {\it defined} as the difference between log of the powder averaged signals acquired with parallel and perpendicular diffusion wave vectors, can then be expressed as (see \cite{RN779}) 
\bea \non
\varepsilon  &=& \frac{1}{{60}} \lb {3{Z_{ijij}} - {Z_{iijj}} + 2{t^2}(3{D_{ij}}{D_{ij}} - {D_{ii}}{D_{jj}})} \rb
\\ &=& {{t^2} \over 15} \left( {3\overline {D_{ij}^\alpha D_{ij}^\alpha}  - \overline {D_{ii}^\alpha D_{jj}^\alpha} } \right)
 = \frac{3}5 \, t^2 \, \overline{{\mathop{\rm var}\,} \{\sigma _\alpha\}} \,.
 \label{def-epsilon}
\eea
In the last equality, the set $\sigma_\alpha \equiv \{ \sigma_{\alpha,i} \}_{i=1}^3$ denote the eigenvalues of ${D^\alpha}$, and
\be 
{\mathop{\rm var}\,} \{\sigma _\alpha\} \equiv \frac13 \sum_{i=1}^3 \sigma_{\alpha,i}^2 - \lb\frac13 \sum_{i=1}^3 \sigma_{\alpha,i}\rb^2 .
\ee
With the above definition, $3{\,\rm tr\,} (D^\alpha)^2 - ({\rm tr\,} D^\alpha)^2 = 9 {\,\mathop{\rm var}\,} \{\sigma _\alpha\}$
in Eq.~(\ref{def-epsilon}). 
The anisotropy metric $\varepsilon$ has dimensions of [length]$^4$. These somewhat awkward dimensions have a historical root in DDE eccentricity measurements \cite{RN226}. 
While $\varepsilon \geq 0$, in practice it is often estimated from the difference of signals, which can become negative due to noise.

As an example, for randomly oriented (and identical) axially symmetric domains, such as fibers with (time-dependent) diffusivities ${D_\parallel }(t)$ and ${D_ \bot }(t)$, microscopic diffusion anisotropy becomes
\be \label{varepsilon}
\varepsilon  = {2 \over 15} \, t^2 \lb  {{D_\bot }(t) - {D_\parallel }(t)} \rb^2 \,.
\ee
If the domains are different, the corresponding Eq.~(\ref{varepsilon}) should be further averaged over them, cf.\ the
$\overline{{\mathop{\rm var}\,} \{\sigma _\alpha\}}$ term in Eq.~(\ref{def-epsilon}).
Microscopic diffusion anisotropy hence depends explicitly on diffusion time, but tends to the geometric measures of pore shape anisotropy as the diffusion time increases, 
since for any confined region of size $a$, $D(t) \sim a^2/t$, and $t$ asymptotically drops out from Eq.~(\ref{varepsilon}).
From the time dependence of the microscopic diffusion anisotropy, non-Gaussian effects of the individual compartments can be revealed by the time dependence of the compartmental (microscopic) diffusion tensors.


Practically, the anisotropy metric (\ref{def-epsilon}) can be estimated from knowledge of the full $Z$ tensor, or by the difference of the powder averaged log signals with parallel and perpendicular diffusion wave vectors. It has an advantage of being additive (cf.\ the pore average in Eq.~(\ref{def-epsilon})): if several distinct types of pore populations are present in the sample (e.g.,\ a distribution of $D_ \bot $ and $D_\parallel$ in Eq.~\ref{varepsilon}),  $\varepsilon $ simply becomes the  volume-weighted mean
over the corresponding $\varepsilon$ from each of the populations. This is an advantage since it eases the interpretation; however, the disadvantage is the dependence on size of the pore in addition to its anisotropy. This has the additional consequence that  $\varepsilon$ is strongly biased by the larger pores: 
Since $w_\alpha \sim a^3$ and $\varepsilon_\alpha \sim a^4$ for a pore of size $a$, the population averaged eccentricity scales as
 $\varepsilon \sim \overline{a^7}/\overline{a^3}$, heavily preferring the tail of the pore size distribution, --- and hence susceptible to the mesoscopic fluctuations introduced in Sec.~\ref{sec:meso-fluct} above, cf. Eq.~(\ref{r-neu}). 
 
To factor out the pore sizes, normalized dimensionless measures of microscopic diffusion anisotropy were introduced \cite{RN779,RN1257,RN1257,RN1166}, such as the microscopic fractional anisotropy, $\mu$FA: 
\bea \non
{\rm{\mu FA}} &\equiv& \sqrt {\frac{3}{2}\frac{{{{\left( {{\sigma _1} - \bar \sigma } \right)}^2} + {{\left( {{\sigma _2} - \bar \sigma } \right)}^2} + {{\left( {{\sigma _3} - \bar \sigma } \right)}^2}}}{{\sigma _1^2 + \sigma _2^2 + \sigma _3^2}}} 
\\
 &=& \sqrt {\frac{\varepsilon }{\varepsilon  + \frac35 \, t^2 \lp \frac13\, {\rm{tr \, }} D \rp^2}} \,.
\label{muFA}
\eea
In the previous example with axially symmetric domains, microscopic fractional anisotropy
\be
\mu {\rm{FA}} = \sqrt{\frac{2}{3}} \frac{{{{\left| {{D_\parallel } - {D_ \bot }} \right|}}}}{\sqrt{{D_\parallel }^2 + 2{D_ \bot }^2}} \,,  
\ee
whereas fractional anisotropy FA is modulated also by the fiber orientation distribution function \cite{RN612,RN779}, 
and only recovers $\mu$FA when the fibers are all coherently aligned. 
\news Another metric which has been suggested to be of biological importance \cite{szczepankiewicz2015,DIVIDE}, is the variance in isotropic diffusivity, 
 \be
 V_I \equiv \overline{(D_{ii}^\alpha/3)^2} - \overline{(D_{ii}^\alpha/3)}^2 = \frac1{36t^2} Z_{iijj}
\ee
 which can also be inferred from the $Z$ tensor \cite{Topgaard2017,Jespersen2018}. When diffusion within the individual pores is Gaussian, other methods such as the so-called magic angle spinning of the q-vector (q-MAS) \cite{Eriksson2013} can also be used to estimate the diffusion tensor covariance \cite{Westin2016,Topgaard2017}.
 \keep

\subsection{Concluding remarks on MDE}

\nin 
As we can see, MDE can potentially provide unique extra information relative to SDE. However, this information content only starts at the level of $\O(q^4)$, Eq.~(\ref{S-q4}) (in addition to the standard SDE $\O(q^4)$ terms), and hence to claim the true novelty of the information, it has to be properly identified relative to the SDE measurements with similar scan parameters (timings and gradients). 
Clarifying the advantages of MDE is practically essential in the view of much reduced SNR due to a notable increase of the echo time needed for the multiple gradients to play out. 

Overall, the main advantage of MDE so far seems to lie in its ability to detect and quantify microscopic diffusion anisotropy. 
In particular, the advantage of the DDE metrics of microscopic diffusion anisotropy (e.g., Eq.~(\ref{def-epsilon})) is that they do not rely on concrete assumptions regarding pore shapes. 

Of course, if a detailed model of microstructure is available, e.g., of the Standard Model form (Section~\ref{sec:gauss}),  
microscopic diffusion anisotropy is directly accessible in terms of model parameters (which can in principle be determined using SDE).  
However, even in this case, MDE adds value by providing an ``orthogonal" way of being sensitive to model parameters --- and, like in the SM case above, 
Sec.~\ref{sec:branch}, it can provide rotationally invariant independent relations between parameters, which can help lift the parameter estimation degeneracies \cite{coelho2017}.

So far, the existing MDE models have been calculated in the limit of either Gaussian diffusion in all compartments, or in the $t_i\to\infty$ limit (e.g., closed pores). Importantly, the {\it transient} effects, cf.\ Section \ref{sec:Dt}, have not yet been properly accounted for in the MDE framework. In particular, the structural disorder-induced power-law tails in $\Dinst(t)$, or, equivalently, in $\D(\omega)$, such as the ones originating due to disordered axonal packing in the extra-axonal space \cite{burcaw2015,fieremans2016}, will contribute to the ``irreducible" MDE effects (that go beyond the product of a few SDE signals). Taking such transient processes into account seems {\it a priori} as crucial for the interpretation of MDE measurements, as it has been for the SDE --- e.g., in the context of recent re-interpretation of the axonal diameter mapping results (cf. Sec.~\ref{sec:long-t-app} above). 
The relevant coarse-graining formalism of the effective medium theory \cite{EMT,mesopnas,burcaw2015} seems perfectly suitable for the task --- but it has not yet been developed.

%% file: Section_V.tex

\section{Outlook and open questions} 
\label{sec:outlook}

\epigraph{There is nothing more practical than a good theory}{L. Boltzmann}


\subsection{To model or not to model?}

\nin
We are writing this Review at a  transformational moment, when our field of quantitative dMRI is experiencing a revolution due to unprecedented quality of hardware and novel acquisition methods, enabling us to observe very subtle physical effects, even in human subjects and potentially in patients. 

Interpreting these effects in terms of the tissue microarchitecture is highly nontrivial; it is safe to say that the theoretical challenge has been so far greatly underappreciated. This, however, may swing the pendulum the other way, towards an ``anti-modeling" point of view: Since, according to a widespread refrain, ``biology is so much more complicated that anything physicists have ever studied", there is little hope for the quantitative understanding of such effects, and  the best we can do is to stay at the level of  ``representations"  (cf. Sec.~\ref{sec:model-represent}) and to draw empirical correlations between parameters of such representations (e.g. mean diffusivity or fractional anisotropy) and the clinical disease scores. 

One of the messages of our Review is that the whole history of Physics in the 20$^{\rm th}$ century offers the case for optimism. The quote from a nuclear physicist (before Section \ref{sec:intro}) has been a universal refrain for our sustained ability to understand nature's complexity, step-by-step, from the origins of elementary particles to the vast scopes of the Universe. The essence of the effective theory way of thinking is that one certainly does {\it not} need to understand everything about the world in order to understand  some corner of its parameter space really well. We certainly {\it can} quantify tissue microarchitecture without uncovering the origins of the human conscience or mapping full details of the brain's biochemical machinery. Too often, ``biology is way more complex than all your models" has been merely an excuse not to develop better models. 

We reject this excuse \cite{manifesto}. We believe that having appropriate theoretical description of diffusion in tissues at the mesoscopic scale is not a luxury at this point --- rather, this is an indispensable scientific {\it method of investigation} into pathological processes 2-3 orders of magnitude below nominally achievable resolution of MRI in any foreseeable future --- or, in fact, ever, since the MRI resolution is stringently bounded by physical and physiological limitations that have been largely reached by now. The parallels with  super-resolution microscopy \cite{betzig2015nobel} are quite obvious; that discipline took a century to develop, based on employing models and prior information. Our task is harder but, arguably, can lead to even more impactful advances.

With that in mind, let us outline  10 
exciting unresolved problems, focussing on which, to the best of our understanding, will propel our field forward.

\subsection{Ten problems for mesoscopic dMRI}
\label{sec:unresolved}

\renewcommand{\theenumi}{\bf \arabic{enumi}}
\begin{enumerate}

\item {\bf Apparent vs. real diffusion metrics:} What are the confounding effects of mesoscopically varying $R_1(\r)$, $R_2(\r)$, and $\Omega(\r)$ in the mesoscopic Bloch-Torrey equation (\ref{BT}) on the observed diffusion metrics, in the spirit of refs.~\cite{zhong1991,kiselev2004}? Can we develop a multi-modal mesoscopic imaging framework able to self-consistently quantify all these mesoscopic quantities and disentangle their effects in the apparent diffusion coefficients and higher-order metrics in each tissue compartment?

\item {\bf Relation between time-dependent $D_\perp(t)$ and its tortuosity limit $\Dinf$, and the geometric parameters of realistic axonal packings:} As the time dependence of the diffusion transverse to fiber tracts is dominated by the extra-axonal water (Section \ref{sec:Dt}), the natural question is what structural changes (e.g. demyelination, axonal loss) can affect this time dependence, as well as $\Dinf$. 
This is a difficult yet clinically impactful inverse problem \cite{axlossdemyel}, whose approximate solution, relying on the ideas of coarse-graining and renormalization, has so far only been obtained in the $t\to\infty$ limit \cite{shrink-remove}. 

\item {\bf Origin of structural disorder along the neurites:} What causes the time dependence along the fibers or in the gray matter? Is it varicosities, beads, synaptic boutons, undulations, or something else? Which of these structural units' changes in pathology can be detectable?

\item {\bf Parameter estimation challenge for the Standard Model:}  How many Gaussian compartments do we have to include? For increasing the precision, it looks like we need orthogonal measurements, such as MDE (e.g., isotropic diffusion weighting), and varying echo  time. What is an optimal clinically feasible measurement protocol?

\item {\bf Time-dependent rotationally-invariant framework:} Combining the ideas of Sections \ref{sec:Dt} and \ref{sec:gauss} can lead to describing each fiber fascile in terms of the non-Gaussian propagators (inside and outside the neurites) with the corresponding time-dependent diffusion, kurtosis, etc, cumulant tensors; such fascicles then naturally combine into the SM-like signal based on the fiber ODF in a voxel. This difficult parameter estimation problem may offer the all-encompassing description of diffusion process measurable with dMRI in the brain. 

\item {\bf Permeability/exchange time for the neurites:} How well we can approximate compartments as non-exchanging? At which time scales this assumption breaks? The answer most likely will be different for gray and white matter, and maybe even for different brain regions.  

\item {\bf Standard Model for GM:} Can we apply SM as introduced in Section~\ref{sec:gauss} to gray matter {\it in vivo} at clinical diffusion times, or should we modify the compartments? Do we have to include exchange, and if yes, at which level? 

\item {\bf EMT for MDE:} Development of the effective medium theory framework \cite{EMT} for the ``disorder-averaging" involved in the multiple diffusion encoding signal (Section \ref{sec:mpfg}). Which physical effects, from the EMT standpoint, are best captured using MDE, or are completely absent in the SDE?

\item {\bf Signal vs. noise:} As the saying goes, ``noise is signal". The fundamental question is to separate the thermal noise, imaging artifacts, as well as the genuine differences between parameters in voxels belonging to the same region of interest. Recently developed random matrix theory-based approaches \cite{MPnoise,veraart2016} offer an exciting prospect. 

\item{\bf Mesoscopic fluctuations and biological variability:} How different are the mesoscopic tissue parameters within a given region of interest? Their differences provide the ``natural" minimal width for the parameter distributions within an ROI, in the limit of infinite SNR. Sometimes, relatively small differences in the mesoscopic parameters can translate into large differences of the dMRI metrics; 
\mpar{R2.2}
\new the heavy sensitivity of the signal from water inside  axons to the tail \keep of the axonal diameter distribution \cite{burcaw2015} (Secs.~\ref{sec:meso-fluct}, \ref{sec:mde-q4}) is an example of an effect of so-called  ``mesoscopic fluctuations" pioneered within condensed matter physics \cite{imry-book}. Studying these fluctuations can provide fundamental insights on the optimality and robustness of the organization of neuronal tissue microarchitecture, as well as offer practical limits on our detection capabilities.

\end{enumerate}

%% file: app.tex


\section{Causality and  analytical properties in the frequency domain} 
\label{sec:app-G}

\nin
As stated after Eq.~\eq{DE}, the diffusion propagator $\G_{t;\r,\r_0}$ is the density of particles released at a point $\r_0$ at the moment $t=0$. While the diffusion equation is often formulated for positive times only, $t>0$, with an explicit initial condition,  writing Eq.~\eq{DE} for {\it all times} $t$ is more convenient due to reasons that will become clear below. Being the response to an instant point source of particles, the diffusion propagator helps finding the particle density for an arbitrary source $f(t,\r)$ (``particle injection''), using the linearity of diffusion equation,
\be \label{DEf}
\lb \partial_t - \partial_\r  D(\r) \partial_\r \rb \rho(t,\r)  = f(t,\r) \,. 
\ee
The solution takes the form of a $t$- and $\r$-convolution,   
\be \label{rho=Gf}
\rho(t,\r) = \int\! \d t_0 \d^d\r_0 \,\G_{t-t_0;\r,\r_0}  f(t_0,\r_0) \,, 
\ee
which is straightforward to prove by acting with the bracketed operator from Eq.~\eq{DEf} on $\G_{t;\r,\r_0} $ under the integral. This solution explains the notion of {\em causality} that implies that {\it the response $\rho(t,\r)$  follows the source, $f(t_0,\r_0)$, and  it cannot  precede it}. 
This means that $\G_{t-t_0;\r,\r_0} \equiv 0$ for $t<t_0$. 
This is guaranteed by the proportionality of  $\G_{t;\r,\r_0} $ to the step function $\theta(t)$, as stated after Eq.~\eq{def-Gcal}. Synonymous to causality is the notion that $\G_{t;\r,\r_0} $ is a {\em retarded} propagator, which implies that any perturbation, $f$, of the system propagates into the future, which is opposed to the  formally possible {\em advanced} propagator for which perturbations propagate into the past. 

Likewise, quantities $D(t)$, $\Dinst(t)$ and $\D(t)$, entering Eqs.~\eq{Dinst=D}--\eq{Dw=Dt}, are retarded, since they are identically zero for $t<0$. 
In particular, the retarded velocity autocorrelator (\ref{Dt=vv}) has a physical meaning of a  response of the 
current (sometimes called flux) ${\bf J}(t,\r)$ of diffusing particles to that of a lump of particle density $\rho(t,\r)$ (the generalized Fick's law), 
cf. ref.~\cite{EMT}, 
\be 
{\bf J}(t,\r)=-\int\! \d t_0 \, \D(t-t_0) \partial_\r \rho(t_0,\r)  + \O(\partial_\r^3 \rho)\,.
\ee
Equivalently, in the Fourier domain, the convolution becomes \newe a \keep multiplication, cf. Eq.~(\ref{Dw}): 
\be 
{\bf J}_{\w,\r}=-\D(\w)\partial_\r \rho_{\w,\r}  + \O(\partial_\r^3 \rho)\,.
\ee
In physics, the above retarded response functions are known as particular cases of the general linear response theory and the fluctuation-dissipation theorem \cite{kubo,Landau5}. 

\begin{figure}[tp]
\includegraphics[width=0.9\columnwidth]{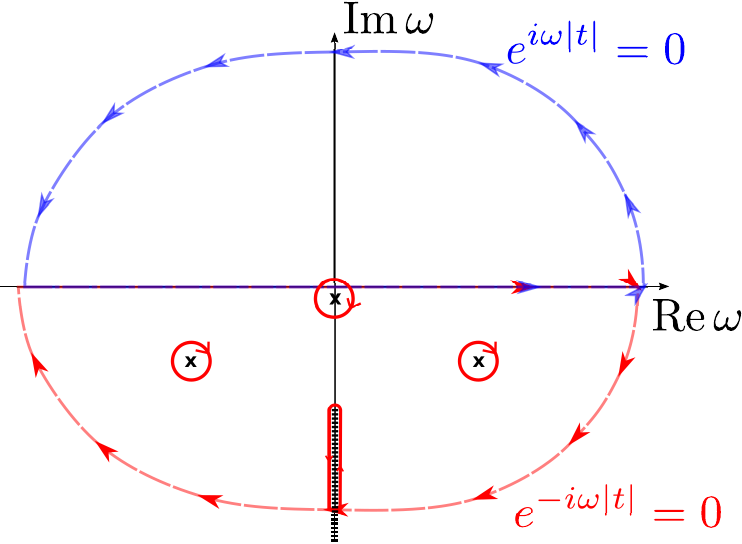}
\caption{Analytical structure of a causal (retarded) response function on the complex plane of $\w$. When calculating the inverse Fourier transform such as Eq.~(\ref{D=iFT}), the original integration contour  over the real axis can be closed in the infinite semicircle with  $\Im \w>0$ (light blue dashed line) when $t<0$, according to the Jordan's lemma. Causality then requires that no singularities are present in the upper half of the complex plane, in which case the integration contour can be shrunk to a point. 
For $t>0$, the contour can be closed where $\Im \w<0$ (light red dashed line). This contour can be shrunk to encircle the singularities of the transformed function (red solid lines). The shown examples include a simple pole as in Eq.~(\ref{DE0_FT}), two other poles and a branch cut along a part of the imaginary axis, to illustrate a few typical options. 
}
\label{fig:omega}
\end{figure}

Our main goal in this Appendix is to investigate how causality, i.e., the retarded character of any response function that identically vanishes for $t<0$, manifests itself in the frequency domain. 
We will now show that causality imposes a strict constraint on the analytical properties of its Fourier transform  
in the complex plane of $\w$. 
Namely, {\it a retarded response must be an analytic function, i.e., it must  have no singularities (e.g., poles or branch points), 
 in the upper half-plane} $\Im \w \ge 0$, Fig.~\ref{fig:omega} \cite{Landau5}. 

To show that, consider the inverse Fourier transform back to the time domain (having $\D(\w)$ as an example): 
\be\label{D=iFT}
\D(t) = \int\!  {\d\w\over 2\pi} \, e^{-i\w t}\, \D(\w)  \,,
\ee
and demand that the resulting $\D(t)\equiv 0$ for $t<0$.

The integration in Eq.~(\ref{D=iFT}) is performed along the $\Re \w$ axis of the complex plane of $\w$ (i.e., over all the frequencies). 
Because of the Fourier exponential, one has to close the integration contour along an infinite semi-circle on which $e^{-i\w t}\to 0$ (Jordan's lemma),%
\footnote{In Eq.~(\ref{D=iFT}), the sign in the exponential $e^{-i\w t}$ is chosen in the tradition of physics; 
the opposite sign would invoke the interchanging of the upper and lower halves of the $\w$ plane.}
and then shrink this contour to single out contributions of all singularities, according to Cauchy's theorem. 
Equivalently, we can view the Fourier integration as proceeding along the equator of the Riemann sphere (topologically equivalent to the complex plane with an added point at infinity); in this case, the fact that the $\Re \w$ axis corresponds to a closed contour is more obvious. Cauchy theorem again applies, and the Jordan lemma dictates in which hemisphere --- top or bottom --- of the Riemann sphere the contour should be shrunk from the equator to encompass the singularities. 

 For negative times, $t<0$, $e^{-i\w t}$ diverges when $\Im \w \to -\infty$, and vanishes when $\Im \w \to +\infty$, which 
 dictates closing the integration contour in the upper half of the $\w$ plane, Fig.~\ref{fig:omega}. 
 For  causality to hold, there must be no singularities in the upper half-plane, in which case the integration contour is constricted to a point, yielding 
 $\D(t)|_{t<0}\equiv 0$. 
 All singularities of $\D(\w)$ must then be present in the lower half of the $\w$ plane, where the contour is closed for $t>0$.  
 The presence of singularities is necessary, since a function without any singularities on a Riemann sphere is a constant.


Let us illustrate the above general considerations by analyzing the analytical structure of the retarded propagator 
of a uniform diffusion equation. Eq.~(\ref{def-Gcal}) with a constant $D(\r) = D_0$ takes the form 
\be \label{DE0}
\lb \partial_t - D_0 \partial_\r^2 \rb \Gbare_{t;\r}  = \delta(t)\delta(\r) \,, 
\ee
where we selected $\r_0=0$ due to translation invariance. 
The above equation in the Fourier domain 
$\Gbare(t,\r) = \int\! {{\rm d}\w\over 2\pi}{{\rm d}^d\q\over (2\pi)^d}\, e^{i\q\r-i\w t} \, \Gbare_{\w,\q}$
becomes algebraic, as the differential operators $\partial_t \to -i\w$ and $\partial_\r \to i\q$ become diagonal: 
\be \non
\lb -i\w + D_0 q^2 \rb \Gbare_{\w;\q}  = 1 \,, 
\ee
 with the solution of a Lorentzian form  
\be \label{DE0_FT}
 \quad \Gbare_{\w;\q} = {1\over -i\w + D_0 q^2} \,.
\ee
This solution preserves causality, since its only singularity, at $\w=-iD_0 q^2$, resides in the lower half-plane of $\w$. 
For $t>0$, closing the integration contour in the lower half of the complex plane and using the residue theorem gives the Gaussian propagator in the $qt$ representation, Eq.~(\ref{Gbare}). 

The above consideration shows that causality is tightly related to the integration in the time domain.  We now inspect this relation closely, first without explicit reference to diffusion propagator, and then applying it to Eqs.~\eq{Dinst=Dw} and \eq{D=Dw}. While the differentiation in the time domain of any function $f(t)$ corresponds to the multiplication with $-i\w$ in the Fourier domain, 
$\partial_t f(t) \Leftrightarrow -i\w f(\w)$, 
the inverse of the differential operator --- i.e., the factor $1/(-i\w)$ --- corresponds to an indefinite integration (the antiderivative) in the time domain. However, infinitesimal shifts of the pole at $\w=0$ result in different integration limits of definite time integrals. Considering both possible shift directions, the same integration technique as above gives the Fourier transformation: 
\be\label{iFT_theta} 
\int\! {\d\w\over 2\pi} \, {e^{-i\w t} \over -i(\w \pm i\varepsilon)} = \pm \theta(\pm t)\,,
\ee
where $\theta(t)$ is the unit step function, and $\varepsilon \to +0$. For an arbitrary function $f(\w)$, which is integrable on the real axis of $\w$, the product $f(\w)/(-i\w\pm \varepsilon)$ is Fourier-transformed according to the convolution theorem, 
\bea\label{retarded}
{f(\w)\over -i\w +  \varepsilon} &\Leftrightarrow& \int\! \d t'\, f(t') \theta(t-t') = \int_{-\infty}^t \! \d t'\, f(t')  \label{intf+} \,,  \\
\label{advanced}
{f(\w)\over -i\w -  \varepsilon}  &\Leftrightarrow& -\int\! \d t'\, f(t') \theta(t'-t) = -\int_t^{\infty} \! \d t'\, f(t') \label{intf-} . \qquad
\eea
Note that differentiating both Eqs.~(\ref{retarded}) and (\ref{advanced}) with respect to $t$ yields back $f(t)$. 
Eqs.~(\ref{retarded}) and (\ref{advanced}) show that while the addition of $\varepsilon$ is unimportant for the Fourier transform of derivatives, it is crucial for inverting differential operators. 
Shifting the pole of $f(\w)/(-i\w)$ downwards from the real axis results in the causal integration (\ref{retarded}), for which the resulted integral up to any time moment $t$ depends on the integrand in the past, $t'<t$. The opposite shift results in the dependence on the future, $t'>t$. 
(The results differ by $f|_{\w=0} = \int\! \d t\, f(t)$, and coincide if $f|_{\w=0} =0$, when $f(\w)/\w$ is not singular.) 
Obviously, the first choice is adequate for the majority of solutions to equations describing the time evolution of physical quantities such as Eq.~\eq{def-Gcal}, or Eqs.~\eq{Dinst=Dw} and \eq{D=Dw}.
The notation $\varepsilon$ with $\varepsilon \to +0$ is often abbreviated to simply $+0$ as it is done in Eqs.~\eq{Dinst=Dw} and \eq{D=Dw}.

\begin{widetext}
\section{OG with a finite number of pulses} 
\label{sec:app-ogse}

\nin
Consider  the   OG gradient wave form $g(t) = g_0 \cos(\w_0 t - \phi)$ with arbitrary initial phase $\phi$ and  $N$ oscillations, such that the total gradient duration 
$T = N\cdot 2\pi/\w_0$. The corresponding 
\be
g(\w) = {g_0\over 2}\lb e^{-i\phi} \cdot {e^{i(\w+\w_0) T}-1 \over i(\w+\w_0)} +  e^{i\phi} \cdot {e^{i(\w-\w_0) T}-1 \over i(\w-\w_0)} \rb ,
\quad e^{\pm i\w_0 T} = 1\,, 
\ee
results in $q_\w = g(\w)/(-i\w)$, such that the wave form acts as the following ``filter" for $\D(\w)$ in Eq.~(\ref{GPA}): 
\be \label{qq}
q_{-\w} q_\w 
= {g_0^2 (1-\cos\w T)\over 2\w^2} \lb {1\over (\w-\w_0)^2} + {1\over (\w+\w_0)^2} + {2\cos2\phi \over (\w-\w_0)(\w+\w_0)}\rb. 
\ee
As  $q_\w$ and Eq.~(\ref{qq}) are not singular when $\w\to0$ and $\w\to\pm\w_0$, we do not need to specify how the zeroes of denominators 
in Eq.~(\ref{qq}) are shifted. 
Hence, one can directly substitute Eq.~(\ref{qq}) into  Eq.~(\ref{GPA}) and integrate with any $\D(\w)$ along the real axis. 

However, to reveal the analytical structure, we find it useful to 
shift the frequency poles inside the square brackets by an infinitesimal positive imaginary part below the real axis, $\w\to\w + i0$ in all the denominators, cf. Appendix~\ref{sec:app-G}.  
As $\D(\w)$ is also analytic in the upper half-plane of the complex $\w$, Appendix~\ref{sec:app-G}, the whole integrand in Eq.~(\ref{GPA}) remains analytic there. Hence, in the prefactor $1-\cos\w T$, the terms $1$ and $-\frac12 e^{i\w T}$ can be dropped for $T>0$, as they identically vanish when closing the  contour in the upper half plane --- we now see that their role was merely to maintain the  $T\to -T$ symmetry of Eq.~(\ref{GPA}) with $q_{-\w}q_\w$ from Eq.~(\ref{qq}), which is forgone by this procedure.  
Hence, $1-\cos\w T \to -\frac12 e^{-i\w T}$, yielding the causal expression 
\be \label{S-sing}
\ln S = {g_0^2\over 4} \int\! {\d\w\over2\pi}\, {\D(\w)\, e^{-i\w T} \over \w_+^2}
\lb {1\over (\w_+-\w_0)^2} + {1\over (\w_++\w_0)^2} + {2-4\sin^2\phi \over (\w_+-\w_0)(\w_++\w_0)}\rb, \quad \w_+ = \w + i0 \,, 
\ee
which starts to mimic the functional form of Eq.~(\ref{D=Dw}) --- and that's the goal! 
Dropping $1$ and $-\frac12 e^{i\w T}$ made the integrand in Eq.~(\ref{S-sing}) singular; 
however, the prescription how to go around its poles regularizes the result,
which we  obtain by a transformation into a sum of simple fractions. This yields
the general relation between the intrinsic $\D(\w)$ and  OG with $N=\w_0 T/2\pi$ pulses: 
\be \label{Sgen-filter}
\ln S = {g_0^2\over 2\w_0^2} \int\! {\d\w\over2\pi}\, \D(\w)\, e^{-i\w T}
\lb \frac12 \lp {1 \over (\w_+-\w_0)^2} + {1\over (\w_++\w_0)^2}\rp + {2\sin^2\phi\over \w_+^2} 
- {1+2\sin^2\phi\over2\w_0} \lp {1\over \w_+-\w_0} - {1\over \w_++\w_0}\rp \rb.
\ee

We can now  see that the singularities in Eq.~(\ref{Sgen-filter}) occur separately at $\w=\pm\w_0$, and  at $\w=0$ for finite $\phi$ (when the gradient is not a pure $\cos$ wave form). Hence, we expect the response of $\D(\w)$ on these two frequencies. 
This response has the contributions of three distinct physical origins. 
The first two terms (in the braces) in Eq.~(\ref{Sgen-filter}) yield the ``pure OG" effect, that of the $\cos$ wave form with $\phi=0$ in the limit $N\to\infty$. The second term describes the $\w=0$ singularity due to the finite time-average component in $q(t)$, 
$\bar q \equiv q_{\w\to 0}/T = (g_0/\w_0)\sin\phi$ present for $\phi\not= 0$, 
which is similar to the PG measurement. Not surprisingly,  it yields the $b_{\rm PG} D(T) \propto D(T)\sin^2\phi$ 
contribution via the {\it exact} relation (\ref{D=Dw}), i.e., the narrow-pulse $D(t)$ with the diffusion time $t=T$ weighted by the 
PG $b$-value contribution $b_{\rm PG} = \bar q^2 T = (g_0/\w_0)^2 T \sin^2\phi$, 
which vanishes for the $\cos$ wave form (cf. Eq.~(7) in ref.~\cite{sv-og}). 
Finally, the last term, representing a finite OG-linewidth effect, is small as $\sim 1/N$, as we will now show. 

While the remaining calculations can be done using the correspondence of $e^{-i(\w-\w_0) T}/(\w-\w_0)$ in the limit $T\to\infty$ with the delta-function, we proceed via a more transparent transformation to the time domain. We use the relations
\be \non
\int\!{\d\w\over2\pi}\, {e^{-i(\w-\w_0)T} \over \w-\w_0+i0} \, \D(\w) = -i \int_0^T\! \d t\, e^{i\w_0 t}\, \D(t) \,,
\quad
\int\!{\d\w\over2\pi}\, {e^{-i(\w-\w_0)T} \over (\w-\w_0+i0)^2} \, \D(\w) = \int_0^T\! \d t\, e^{i\w_0 t}\, (t-T) \D(t) 
\ee
valid for any retarded response functions $\D(t)$ and $\D(\w)$ related by the Fourier transformation (derived based on the property that a convolution in $\w$ is a product in $t$, similar to how Eq.~(\ref{retarded}) was derived). We can see that 
causality means that only $t<T$ are integrated over, as expected.  As a result, we transform Eq.~(\ref{Sgen-filter}) into 
\be \label{Sgen-D}
-\ln S  =  b\cdot \lb {\int_0^T\! \d t\lp1-\frac{t}{T} \rp \D(t) \cos\w_0 t   +  2D(T) \sin^2\phi \over 1 + 2\sin^2\phi}  + 
{1\over 2\pi N} \int_0^T\! \d t\, \D(t)\sin\w_0 t \rb , 
\quad b = {\pi g_0^2 N\over \w_0^3} \, (1+2\sin^2\phi) \,,
\ee
where the $b$-value is calculated \cite{sv-og,sukstanskii2013} assuming constant $\D(\w) \equiv D_0$, such that 
$\D(t) = D_0  \delta(t-\epsilon)|_{\epsilon\to +0}$ \cite{EMT}. 
We can see that the total $b = b|_{\phi=0} + b_{\rm PG}$ is a sum of the pure OG and the pure PG contributions, since $\sin\w_0 t$ is orthogonal 
to a constant. 

In the $N\gg 1$ limit of a clear separation of time scales $2\pi/\w_0 \ll T$, 
we can extend the upper integration limit $T\to\infty$; using $t \cos\w_0 t = \partial_{\w_0} \sin\w_0 t$ in the first term of Eq.~(\ref{Sgen-D}), we can separate the main contribution and the $\sim 1/N$ correction:
\be \label{Sgen-Re-Im}
-\ln S_{N\gg 1} \simeq b\cdot \lb
{\Re \D(\w_0) + 2D(T) \sin^2\phi   \over 1 + 2\sin^2\phi}
+ \frac1{2\pi N} {\lp 1+2\sin^2\phi - \w_0 \partial_{\w_0}\rp \Im \D(\w_0)  
\over 1 + 2\sin^2\phi}  \rb, 
\quad \D(\w_0) = \int_0^\infty\! \d t\, \D(t) \, e^{i\w_0 t}\,.
\ee
Here, the term $\propto D(T)$ is exact due to Eq.~(\ref{D=Dw}), while  setting $\w=\w_0$ in  $\D(\w)$ in the other two terms is precise up to $\sim \w_0/N$. 
We can see that the first term in the square brackets of Eq.~(\ref{Sgen-Re-Im}) is a leading effect (it is not small when $N\to\infty$); it gives the balance of the $\w=\w_0$ and $\w=0$ contributions with the ``filter" weights defined by the phase $\phi$ in agreement with the general property of the second-order cumulant term, discussed after Eq.~(\ref{GPA}). The second term, $\sim \Im \D(\w_0)/N$, is suppressed as $\sim 1/N$. In the limit $N\to\infty$, the imaginary part of $\D(\w)$ does not contribute to the OG measurement \cite{EMT}. 

There exists a case in which the $1/N$ term is comparable with the first one, namely, of a {\it closed pore} with a characteristic size $\sim a$ in the limit of low frequencies (long-times), $\w_0 a^2/D_0 \ll 1$. In this limit, $D(T)\sim a^2/T$ happens to be  parametrically as small as the second term. More precisely,  the second term $\sim 1/2\pi N$ of Eq.~(\ref{Sgen-Re-Im}) exactly cancels the $2D(T)\sin^2\phi$ contribution to the first term, and the net result is determined by 
$\Re \D(\w_0)\sim \w_0^2$. This was first noticed by Sukstanskii \cite{sukstanskii2013} via the time-domain calculation for a one-dimensional impermeable box. Such behavior, however, is completely general. Indeed, for a pore of arbitrary shape, the dispersive diffusivity is a sum over the eigenmodes of the Laplace operator, 
\be \label{Dw-beta}
\D(\w) = D_0 \sum_k C_k\, {-i\w  \over \beta_k^2 D_0/a^2-i\w } \quad \Rightarrow \quad
\Re \D(\w)|_{\w\to0} \simeq  \sum_k {C_k\over \beta_k^4} \cdot {\w^2 a^4\over D_0} 
\ee   
with $C_k$ and $\beta_k$ determined by the pore geometry \cite{callaghan-book,grebenkov-rmp}; 
to the order $\O(\w)$,  $\D(\w) \simeq -i\w \mu a^2$, $D(T) = \mu a^2/T$, where the dimensionless parameter $\mu = \sum_k  (C_k/\beta_k^2)$, 
and so both the $D(T)$ and the $\Im \D(\w_0)$ contributions to Eq.~(\ref{Sgen-Re-Im}) cancel each other exactly. 
The remaining quantity $\Re \D(\w_0) \sim \w_0^2 a^4/D_0$, the leading effect, vanishes very fast (quadratically) in the low-frequency limit  \cite{Stepisnik1993,callaghan-book,dunn-bergman,grebenkov-rmp,Xu2009,stepisnik2007,burcaw2015,nilsson2017limit} 
(this scaling was first observed by Stepisnik \etal \cite{stepisnik2007} in porous media).  
For the relevant case of a cylinder of radius $a$, $C_k = 2/(\beta_k^2-1)$ and the sum 
$\sum_k  {2\over \beta_k^4(\beta_k^2 -1)} = {7\over 96}$ \cite{neuman1974}. 
The estimate (\ref{Dw-beta}) tells that OG (with any phase $\phi$) is  less efficient than PGSE in creating adequate diffusion weighting if one wants to measure  sizes of small fully confining compartments, cf. text after Eq.~(\ref{Dw-cyl}).

\section{Probing the $S/V$ limit with finite-$N$ OG} 
\label{sec:app-SV}

\nin
Let us now use the general finite-$N$ relation (\ref{Sgen-filter}) for the $S/V$ model (\ref{Dt-Mitra}); this setup is practically relevant to study cell density in brain tumors \cite{Reynaud2016,pomace}. 
The intrinsic $\D(\w)$ in this limit can be obtained performing Fourier transform of the outcome of Eq.~(\ref{Dw=Dt}) \cite{sv-og}:
\be \label{Dw-Mitra-complex}
\D(\w) \simeq  D_0 \lp 1 - {S \sqrt{D_0} \over Vd} \cdot {e^{i\pi/4} \over \sqrt{\w}}\rp 
\quad\Leftrightarrow\quad 
\D(t) \simeq  D_0 \theta(t)\lp \delta(t) - {S \sqrt{D_0} \over Vd} \cdot {1 \over \sqrt{\pi t}}\rp. 
\ee
Substituting Eq.~(\ref{Dw-Mitra-complex}) into Eq.~(\ref{Sgen-D}), using 
\be
\int_0^T\! \frac{\d t}{\sqrt{\pi t}}  e^{i\w_0 t} = \sqrt{2\over \w_0} \lb {\cal C}(2\sqrt{N}) + i{\cal S}(2\sqrt{N})\rb , 
\quad
\int_0^T\! \frac{\d t}{\sqrt{\pi t}} {t\over T} \cos \w_0 t = - \frac1{4\pi N} \sqrt{2\over \w_0}\, {\cal S}(2\sqrt{N}) \,, 
\ee
where the Fresnel integrals are defined in a standard way,
\be 
{\cal S}(x) = \int_0^x\! \d u \, \sin {\pi u^2\over 2} \,, \quad {\cal C}(x) = \int_0^x\! \d u \, \cos {\pi u^2\over 2} \,,
\ee
 we obtain
\be \label{Mitra-N}
-\ln S  = b\cdot  {D_0 \lb 1 - c(\phi,N)\cdot {S \over Vd\sqrt{2}} \sqrt{D_0 \over \w_0} \rb  + 2D(T) \sin^2\phi  \over 1+2\sin^2\phi} \,, 
\ee
where $b$ is given in Eq.~(\ref{Sgen-Re-Im}), and the finite-$N$ correction factor 
\be \label{Mitra-c}
c(\phi,N) = 2{\cal C}(2\sqrt{N}) + {3+4\sin^2\phi \over 2\pi N} {\cal S}(2\sqrt{N}) \,.
\ee
Here, the $1/\w^2$ term in Eq.~(\ref{Sgen-filter}) yields the exact $D(T)$ according to Eq.~(\ref{D=Dw}) as discussed in Appendix~\ref{sec:app-ogse}.
In the ``ideal OG" limit $N\to \infty$, using ${\cal C}(\infty)=\frac12$, we obtain $c(\phi,N) \to 1$ for any $\phi$, such that Eq.~(\ref{Mitra-N}) yields Eq.~(7) of ref.~\cite{sv-og}. When, additionally, $\phi=0$, we obtain $-\ln S = b\cdot \Re \D(\w)$, 
cf. Eq.~(\ref{Dw-Mitra}) in the main text. 

We emphasize that performing the calculation in the frequency domain allows us to separate the time scales and identify the contributions to Eq.~(\ref{Mitra-N}) of two distinct physical origins. The $D(T)$ term is completely general --- i.e., the applicability of Eqs.~(\ref{Mitra-N})--(\ref{Mitra-c}) only requires the {\it period} of the oscillation to be short enough so that the model (\ref{Dw-Mitra-complex}) applies; the whole gradient train $T$ can be long and (practically, always) falls out of the short-time $S/V$ limit. Often times, at that point one can set $D(T)\approx D_\infty$. As discussed in Appendix \ref{sec:app-ogse} , the $D(T)$ term appears because at finite $\phi$, the OG wave form can be thought of as a pure $\cos$ wave form and a PG with diffusion time $t=T$
\cite{sv-og}. Hence, Eq.~(\ref{Mitra-N}) allows one to probe the $S/V$ ratio by keeping the period short, yet the total gradient train as long as needed, to accumulate the diffusion weighting $b\propto N$, Eq.~(\ref{Sgen-Re-Im}). 

If, additionally, the {\it whole OG train} $T$ falls into the short-time limit, under a more stringent condition $(S/V) \sqrt{D_0 T} \ll 1$, then $D(T)$ is given by Eq.~(\ref{Dt-Mitra}). In this limit, Eq.~(\ref{Mitra-N}) yields 
\be \label{suks}
-\ln S = bD_0 \lb 1 - \tilde c(\phi,N) \cdot {S \over Vd\sqrt{2}} \sqrt{D_0 \over \w_0} \rb, \quad
\tilde c(\phi,N) = {c(\phi,N) + \frac{16}3 \sqrt{N}\, \sin^2\phi   \over 1+2\sin^2\phi } \,,
\ee 
where the re-defined correction factor $\tilde c(\phi,N)$  corresponds to Eq.~(14) of ref.~\cite{sukstanskii2013}, where the problem of finite-$N$ correction in the $S/V$ limit was first considered. It was noted there, that $\tilde c(\phi,N)$ nominally diverges for $N\to \infty$ as $\sqrt{N}$. It is clear that this divergence occurs due to the $\sqrt{T}$ scaling from $D(T)$, and it eventually gets cut off when $N$ becomes so large than $D(T)$ falls out of the validity regime of Eq.~(\ref{Dt-Mitra}). Hence, this spurious divergence is a result of defining $\tilde c(\phi,N)$  in ref.~\cite{sukstanskii2013} by forcing Eq.~(\ref{suks}) to mimic the form of Eq.~(\ref{Dt-Mitra}), instead of  separating the physics at the two time scales, $2\pi/\w_0$ and $T$. The separation of scales identified in Eq.~(\ref{Mitra-N}) based on the general expression (\ref{Sgen-filter}) extends the validity of OG in the $S/V$ limit far beyond the claim of ref.~\cite{sukstanskii2013}, ``the high-frequency regime can be achieved only when the {\it total} diffusion time is smaller than the characteristic diffusion time" 
(implying $(S/V) \sqrt{D_0 T} \ll 1$), onto the practically relevant domain $(S/V) \sqrt{D_0/\w_0} \ll 1$, for any $\phi$.
Note that for pure $\cos$ gradient,  $\tilde c(0,N) = c(0,N)$, due to the absence of the $\w=0$ singularity in Eq.~(\ref{Sgen-filter}), and Eq.~(\ref{suks}) agrees with Eqs.~(\ref{Mitra-N}) and (\ref{Mitra-c}). 

\newpage
\end{widetext}